\documentclass{JHEP3}

\usepackage{graphicx}
\usepackage{fancyhdr}
\usepackage{amsmath}
\usepackage{dsfont}

\newcommand{\del}{\partial}
\newcommand{\vev}[1]{\langle #1 \rangle}

\title{On the Effective Action of Confining Strings}
\preprint{WIS/04/09-MAR-DPP}

\author{Ofer~Aharony and Eyal~Karzbrun\\\\
Department of Particle Physics, Weizmann Institute of Science,
Rehovot 76100, Israel\\ \\
{\tt E-mail:
Ofer.Aharony@weizmann.ac.il, Eyal.Karzbrun@weizmann.ac.il} }

\abstract{
We study the low-energy effective action on confining strings (in
the fundamental representation) in $SU(N)$ gauge theories in $D$
space-time dimensions. We write this action in terms of the physical transverse
fluctuations of the string. We show that for any $D$, the
four-derivative terms in the effective action must exactly match the
ones in the Nambu-Goto action, generalizing a result of L\"uscher
and Weisz for $D=3$. We then analyze the six-derivative terms, and
we show that some of these terms are constrained. For $D=3$ this
uniquely determines the effective action for closed strings
to this order, while for $D>3$
one term is not uniquely determined by our considerations.
This implies that for $D=3$ the energy levels of a closed string of
length $L$ agree with the Nambu-Goto result at least up to
order $1/L^5$. For any $D$ we find that
the partition function of a long string on a torus is
unaffected by the free coefficient, so it is always equal to
the Nambu-Goto partition function up to six-derivative order. For a
closed string of length $L$, this means that for $D>3$ its energy can,
in principle, deviate from the Nambu-Goto result at order $1/L^5$, but such
deviations must always cancel in the computation of the partition
function (so that the sum of the deviations of all states at each
energy level must vanish). In particular there is no correction at
this order to the ground state energy of a winding string. Next, we
compute the effective action up to six-derivative order for the
special case of confining strings in weakly-curved holographic
backgrounds, at one-loop order (leading order in the curvature). Our
computation is general, and applies in particular to backgrounds
like the Witten background, the Maldacena-Nu\~nez background, and
the Klebanov-Strassler background. We show that this effective
action obeys all of the constraints we derive, and in fact it
precisely agrees with the Nambu-Goto action (the single allowed
deviation does not appear).
}

\begin{document}


\section{Introduction}\label{sec:introduction}

The confining string is a basic object in confining $SU(N)$ gauge
theories, in particular when there is no matter in representations
of non-zero $N$-ality, such that this string is stable. Like any
other solitonic object, it is interesting to study the low-energy
effective action on this string (at energies much lower than the QCD
scale), in order to understand its low-energy fluctuations and the
light excitations of long strings. This study is particularly
interesting since the confining string in large $N$ gauge theories
is believed to be a weakly coupled fundamental string moving in some
background (with a string coupling of order $1/N$ \cite{bib:largeN}). When this
background is known, we can use it to compute the low-energy
effective action on the string. For most interesting confining
theories the corresponding string background is not yet known, and
one can hope that studying (say, by lattice simulations) the
low-energy effective action on a confining string could teach us
about the properties of this background, and give clues for its
construction.

The simplest effective action for a confining string in $D$
space-time dimensions is the
Nambu-Goto action, which is simply the string tension $T$ times the
area of the string
worldsheet. A priori there is no reason why the effective action on
confining strings should take this simple form, but lattice
simulations for pure Yang-Mills theories in $D=3$ and $D=4$ show
(as we will review below)
that the effective action is very well approximated by the
Nambu-Goto form, with only very small deviations. Our goal in this
paper will be to understand why this is the case, and to estimate at
what order deviations from the Nambu-Goto action are expected to
occur.

Two main approaches to constrain the effective action of
a confining string have been studied in the literature. The
Polchinski-Strominger approach \cite{bib:constraints1,
bib:constraints2, bib:constraints3} uses a conformal gauge on the
worldsheet, in which the degrees of freedom in the effective action
are the $D$ embedding coordinates of the string. In this approach
the constraints on the effective action come by requiring that it
must have the correct (critical) central charge, and it was shown
that this implies that the four-derivative effective action must
agree with the Nambu-Goto form. However, it seems difficult to
extend this approach to higher orders. In this paper we follow the
second approach \cite{bib:lsw,bib:luscherterm,bib:constraints4}, writing the effective
action in static gauge, such that the degrees of freedom are only
the $(D-2)$ transverse fluctuations of the string worldsheet.
L\"uscher and Weisz argued in \cite{bib:constraints4} that by computing
the partition function of long winding strings, and expressing it as
a sum over string states, one can constrain the effective action;
they showed that the partition function on the annulus constrains
the four-derivative terms in $D=3$ to be of Nambu-Goto form, but
that for $D>3$ there seems to be one undetermined coefficient.
Essentially, the information that goes into this approach is
\cite{bib:constraints5} that the action should non-linearly realize
the Lorentz symmetry rotating the direction that the string
propagates in and the transverse directions.

In the first part of this paper we generalize the approach of
L\"uscher and Weisz in two directions. First, we compute the
four-derivative partition function of a long winding string on the
torus, and we show that this constrains the four-derivative terms in
the effective action to be of Nambu-Goto form for any $D$. Then, we
extend the computation to the six-derivative terms, computing the
partition function on the torus and on the annulus. For general
$D$ we show that the
considerations of L\"uscher and Weisz allow us to determine two of
the three free coefficients at six-derivative order, but that one
coefficient remains unfixed. Strangely, it turns out that this
free coefficient does not affect the partition function of the
long string on the torus, so that if there are corrections to energy
levels at six-derivative order (order $1/L^5$ for a string of length
$L$) they must cancel exactly in the partition function.
In particular, our results imply that the ground state energy of a
closed winding string is exactly given by its Nambu-Goto form up to
order $1/L^5$, and can deviate from this form only starting at order
$1/L^7$. For $D=3$ we show that the effective action is uniquely determined up
to six-derivative order, so that the previous sentence applies to all
states of closed winding strings. The computations of the partition functions require a
regularization of the effective action, and we use (following
\cite{bib:DF,bib:constraints4}) a zeta function regularization, which gives finite
results, independent of the UV cutoff.

In most confining gauge theories we do not know how to compute
directly the effective action on the confining string, and we can
only do it numerically by lattice simulations. However, in the past
decade a new class of confining gauge theories was discovered, whose
dual string theory description lives in a weakly curved background.
In such backgrounds we can compute explicitly the effective action
on the confining string, and we perform this computation to
six-derivative order in the second part of our paper. More
precisely, we compute the leading dependence of the terms in the
six-derivative effective action on the curvature of the background
(which typically maps to some negative power of the 't Hooft
coupling). There are several motivations for this computation :
\begin{itemize}
\item It is the first example (as far as we know) of a direct computation of the effective
action on a confining string.
\item We show that the effective action we compute obeys all the constraints discussed in the
previous paragraph, thus enabling us to test both the form of these
constraints and our computation of the effective action.
\item Our computation allows us to check whether the term in the effective action that is
allowed to deviate from the Nambu-Goto form is actually present or
not. This is important since there may be additional constraints on
the effective action that may set this term to zero. We find that,
at the leading order that we work in, there are actually no deviations from
the Nambu-Goto action.
\item Some of the backgrounds we study are continuously related (by changing a dimensionless
parameter) to pure Yang-Mills and pure super Yang-Mills theories in
$D=3$ and $D=4$, and we expect that the qualitative form of the
effective action will not change when we change the parameters.
\end{itemize}

We begin in section \ref{sec:generalities} with general comments on
the effective action on confining strings, and with a review of the
known results. In section \ref{sec:effectiveaction} we generalize
the computations of \cite{bib:constraints4} to the torus partition
function and to the next order in the derivative expansion. In
section \ref{sec:superstrings} we write down the worldsheet action
for strings in weakly curved confining backgrounds, and the Feynman
rules that follow from it. Our discussion in this section is
general, and in the following section \ref{sec:examples} we discuss
in detail some of the examples to which our considerations apply. In
section \ref{sec:corfun} we use this worldsheet action to compute
the effective action on the corresponding confining strings, at
leading order in the space-time curvature. We end in section
\ref{sec:conclusions} with our conclusions. Two
appendices contain some technical details. In appendix \ref{sec:appendixA} we present the
computations used in section \ref{sec:effectiveaction}, and in appendix \ref{sec:appendixB} we review our
conventions for sections \ref{sec:superstrings}-\ref{sec:corfun}.

\section{General features of the effective action of a confining string}
\label{sec:generalities}

\subsection{Generalities}

In this paper we consider confining gauge theories in which the
confining string is absolutely stable. For $SU(N)$ gauge theories,
this means that there cannot be any dynamical fields in
representations with non-zero $N$-ality, such as the fundamental
representation. Of course, in the large $N$ limit the confining
string becomes stable even in the presence of dynamical fields in
the fundamental representation. For finite $N$, in the presence of
such fields, the string can break.

In this situation it makes sense to ask about the low-energy
effective action on a long, straight confining string. A string-like object in a
$D$-dimensional gauge theory breaks $(D-2)$ translation symmetries,
so there should be $(D-2)$ massless Nambu-Goldstone bosons on the
worldsheet, whose expectation values are simply the transverse
positions of the string. In a generic confining theory we do not
expect any additional massless fields on the string worldsheet, so
the effective action will involve only these massless fields. In
theories with additional symmetries there may be additional massless
fields on the worldsheet. For instance, in supersymmetric gauge
theories, the confining string typically breaks all the
supersymmetry, so it should have additional massless fermions on its
worldvolume; for instance the confining string in the $D=4$ ${\cal
N}=1$ supersymmetric Yang-Mills theory should have $4$ massless
Majorana-Weyl fermions on its worldvolume. In this paper we will
ignore the possibility of having such additional fields, though we
expect that they will not change most of our conclusions. It would
be interesting to generalize our analysis to include additional
massless fields arising from additional symmetries.

The effective theory of the Nambu-Goldstone bosons is independent of
their expectation value, so all
interactions involve derivatives of all the fields. Thus, it is
necessarily a free field theory at low energies, but it could
involve higher derivative corrections. In addition to these massless
fields, we expect to have for any confining string additional
(bosonic and fermionic) degrees of freedom on the worldsheet at some
scale $m$; in a gauge theory characterized by a single scale
$\Lambda$ (like pure Yang-Mills theories) we expect $m \sim \Lambda$
to be of the same order as the square root of the string tension,
while in gauge theories with dimensionless parameters there may be
some separation between the scales. The theory on the worldsheet at
the scale $m$ may be weakly coupled, in which case we can describe
the additional degrees of freedom as massive particles, or it could
be strongly coupled, in which case we have no such description. The
latter is more likely in a theory with a single scale, in which the
width of any particle-like state is governed by the same scale as
its mass. In either case we expect the effective action to be valid
only below the scale $m$, where it should break down.

For a generic string-like soliton there is no reason to believe that
any effective action makes sense above the typical dynamical scale
$\Lambda$ of the field theory. However, the situation of the
confining string in large $N$ gauge theories is different, since we
believe \cite{bib:largeN} that such gauge theories are equivalent to
weakly coupled string theories, and in such theories there is a
well-defined action on the worldsheet that is valid at all energy
scales. (Indeed, the quantization of this action should include the
full information about the large $N$ gauge theory.) In such a
situation we can think of the low-energy effective action of the
massless fields as coming from the exact string worldsheet action,
when we integrate out all the massive degrees of freedom on the
worldsheet. Note that the action of a fundamental string has a
diffeomorphism symmetry, and, depending on the formalism, it may
also contain a worldsheet metric as a dynamical variable.
Often in string theory we use the conformal gauge, in which the worldsheet action is
conformally invariant and there is no mass scale. It is important to
emphasize that the effective action on a long string arises in a
different gauge, in which we
gauge-fix the diffeomorphism symmetry such that two of the
worldsheet coordinates are identified with space-time coordinates
(the ``static gauge''). In this gauge the action has a mass scale,
and we typically get
(as we will see in various examples) a theory of massive and
massless fields. The low-energy effective action discussed above
arises when integrating out these massive fields. For strings in
flat space, the effective action in this gauge is precisely the
Nambu-Goto action, which has only massless fields but includes an
infinite tower of higher derivative corrections to their action.
This is a special case where the effective action should make sense
at all energies; of course, we know that such a string theory is
only consistent for a superstring in $D=10$. Confining strings arise
from superstrings in curved backgrounds, and then some of the fields
on the worldsheet are massive at some scale $m$, and the low-energy effective action is
more complicated.

As described above, we expect the effective action on a confining
string to depend on the derivatives of $(D-2)$ scalar fields, which
we will denote by $X^i$ ($i=2,\cdots,D-1$). Apriori the action $S =
\int d^2\sigma {\cal L}(\sigma)$ should include the most general
terms consistent with the $SO(D-2)$ rotation symmetry. At
0-derivative order there is a term in the action density
proportional to the effective string tension $T$,
\begin{equation}\label{actionzero}
{\cal L}_0 = -T.
\end{equation}
At 2-derivative order there is a single possible term
\begin{equation}
{\cal L}_2 = -\frac{1}{2} \del^{\alpha} X \cdot \del_{\alpha} X,
\end{equation}
whose coefficient we can always normalize in this way. Here
${\alpha}=0,1$ goes over the worldsheet coordinates, and we use the
notation $X\cdot X \equiv X^{i}X^j\delta_{ij}$ . At 4-derivative order
there are generally two independent terms (ignoring terms
proportional to $\del^2 X^i$ which can be eliminated by field
redefinitions),
\begin{equation}\label{actionfour}
{\cal L}_4 = c_2 (\del^{\alpha} X \cdot \del_{\alpha} X)
(\del^{\beta} X
             \cdot \del_{\beta} X) + c_3 (\del^{\alpha} X \cdot \del^{\beta} X)
             (\del_{\alpha} X \cdot \del_{\beta} X).
             \end{equation}
The notation that we use here follows
\cite{bib:constraints4}, up to a different normalization of $c_2$
and $c_3$. In the special case of $D=3$, there is only one field $X$
and the two terms in \eqref{actionfour} are identical. At six-derivative order, there are
several terms that apparently cannot be eliminated by
field redefinitions :
\begin{eqnarray}\label{actionsix}
{\cal L}_6 &=& {\cal L}_{6,4} + {\cal L}_{6,6}, \\ \nonumber {\cal
L}_{6,4} &=& c_4 (\del_{\alpha} \del_{\beta} X \cdot \del^{\alpha}
\del^{\beta} X) (\del_{\gamma} X \cdot \del^{\gamma} X) +
c_5 (\del_{\alpha} X \cdot \del_{\beta} X) (\del_{\gamma} X \cdot
\del_{\alpha} \del_{\beta} \del_{\gamma} X), \\ \nonumber {\cal
L}_{6,6} &=& c_6 (\del_{\alpha} X \cdot \del^{\alpha} X)^3 + c_7
(\del_{\alpha} X \cdot \del^{\alpha} X) (\del_{\beta} X \cdot
\del_{\gamma} X) (\del^{\beta} X \cdot \del^{\gamma} X) + \\
\nonumber && c_8 (\del_{\alpha} X \cdot \del_{\beta} X)
(\del^{\alpha} X \cdot \del_{\gamma} X) (\del^{\beta} X \cdot
\del^{\gamma} X).
\end{eqnarray}
The $c_5$ term is naively non-trivial, but in fact since our action
lives in a two-dimensional space, one can show that it is actually
proportional to the equation of motion (up to integrations by parts);
this is most easily seen by using light-cone coordinates, where the
leading order equation of motion is $\del_+ \del_- X^i = 0$.
Thus, we will ignore this term from here on. Similarly, in two
dimensions the $c_8$
term can be shown to be equal to a
linear combination of the $c_6$ and $c_7$ terms\footnote{We thank
F. Gliozzi for pointing this out to us.}, so we will ignore it
as well. For the special case of $D=3$
the $c_4$ term is equivalent to the $c_5$ term so it is also trivial, and
there is only one independent term in ${\cal L}_{6,6}$.

The effective action we wrote here is for a string with no
boundaries, and then only terms with an even number of derivatives
are allowed. In many cases it is interesting to consider also
confining strings with boundaries; for instance, such a situation
arises in the computation of Wilson loops (which are boundaries for
a confining string worldsheet), including the computation of the
force between external quarks and anti-quarks. In the presence of
boundaries, there could be additional terms in the effective action
which are localized on the boundary (and may involve an even or an
odd number of derivatives); in particular, in the analysis
above we did not write down terms which differ by an integration by
parts, so if we make a different choice for the form of the terms we
write we will generate some boundary terms. However, it is important
to emphasize that the same confining string could have different
types of boundaries; for instance the string could end either on a
Wilson loop or on a domain wall, and it is not obvious that the
boundary terms should be the same for different boundaries. In this
paper we focus on the closed string effective action, and on the
corrections to the closed string spectrum, so we ignore the boundary
terms (which only affect the open string spectrum). In some of our
computations we will use worldsheets with boundaries, and we will
then assume that there are no boundary terms; this seems reasonable
for a string ending on a domain wall described by a D-brane, though
it is not necessarily true for strings ending on Wilson loops.
This assumption does not influence our results
concerning the closed string effective action.

What can we compute using the effective action ? Obviously, we can
use the tree-level effective action to compute any dynamical
processes below the scale $m$ where the action is expected to break
down. However, it would be nice to be able to use the effective
action also for loop computations, such as the computation of the
partition function of a long string whose worldsheet is (say) a
torus (this includes the corrections to the energies of winding
closed string states), or loop corrections to scattering amplitudes
on the worldsheet. Generic loop computations lead to divergences, so
the answer depends on the physics at the cutoff scale and additional
information is required (beyond the effective action) to obtain
finite answers. However, in special cases loop computations may give
finite results, in which case we can trust them. We will see that
this happens in many cases when we use the effective action on a
superstring; presumably this is because in a different gauge (the
conformal gauge) this action is finite, so it should lead to
well-defined results. In other cases, like the computation of the
partition function of the low-energy effective action on a torus, we will
encounter divergences. Of course, such divergences arise already in
the partition function of a free field theory.
We will regulate these divergences, as in
\cite{bib:DF}, using a zeta function regularization. This
regularization satisfies some nice physical properties (described in
\cite{bib:DF}) which make it effectively independent of the physics
at the cutoff, so we expect it to give correct answers for the
physics below the scale $m$ (which should be independent of the
cutoff).

\subsection{Constraints on the effective action of a confining
string}

Two types of constraints on the effective action have been
considered in the literature. One constraint, originally
analyzed by L\"uscher and Weisz \cite{bib:constraints4}, arises from
the fact that the partition function of the string wrapped on
various surfaces must have an interpretation in terms of the
propagation of physical string states along these surfaces. This
constraint is relevant for strings which have a limit in which they
are weakly coupled (such as confining strings in $SU(N)$ gauge
theories), since in that case the single-string states do not mix
(in the weak coupling limit) with any other states, so the partition
function has an interpretation involving purely the propagation
of single-string states.

A specific case of this, which was considered in
\cite{bib:constraints4}, involves worldsheets with the topology of
an annulus. Suppose that we consider a confining string in a
Euclidean space, in which one of the directions is compactified on a
circle of circumference $L$ (say, $X^0 \equiv X^0 + L$). We can now
consider a string whose worldsheet is wrapped once around this
circle, and which has two boundaries separated by a distance $R$ in
another spatial direction (say, boundaries at $X^1=0$ and at
$X^1=R$). For a confining string one example of this is the
correlation function of two Wilson loops, wrapped on the circle and
separated by a distance $R$. The partition function
$Z^{annulus}(L,R)$ on such a worldsheet has two interpretations. On
one hand, we can view $X^0$ as the ``time'' direction, and then the
diagram is a one-loop vacuum diagram for an open string of length
$R$, which can be expressed as
\begin{equation}
\label{openannulus} Z^{annulus}(L,R) = \sum_{k} e^{-E_{k}^{o}(R) L},
\end{equation}
where the sum is over all open string states $k$ of length $R$, with
vanishing transverse
 momentum (since the ends of the open string are fixed),
 which have energies $E_{k}^{o}(R)$. Note that these energies depend only on $R$ and not
 on $L$, since we interpret $Z$ as a statistical mechanical partition function; when we have
 fermionic states for the string this requires that we put anti-periodic boundary conditions
 for the fermions in the $X^0$ direction, otherwise we have an extra factor of $(-1)^F$.
 For confining strings in pure Yang-Mills theories we do not expect to have any
 fermionic states so this is not relevant.

On the other hand, we can view $X^1$ as the ``time'' direction.
Then, we have a closed string state (winding on a circle of
circumference $L$), which is created at $X^1=0$ from some ``boundary
state'' and is annihilated at $X^1=R$. The closed strings can have
any transverse momentum, since they propagate between two points
with vanishing transverse separation. In other words, if we allow
for some transverse separation between the two boundaries and
integrate over it, we would sum over only zero transverse momentum
closed strings, so we have
\begin{equation}
\int dx_{\perp} Z^{annulus}(L,\sqrt{R^2+x_{\perp}^2}) = \sum_{n}
|v_n(L)|^2 e^{-E_n^{c}(L) R},
\end{equation}
where the sum is over all closed string states $n$ with zero transverse
momentum, $v_n(L)$ are the overlaps of these states with the boundary
state, and $E_n^{c}(L)$ are their energies. It was shown in
\cite{bib:constraints4} that this implies that
\begin{equation}
\label{closedannulus} Z^{annulus}(L,R) = \sum_n |v_n(L)|^2 2 R
\left( \frac{E_n^{c}(L)}{2\pi R} \right)^{(D-1)/2}
K_{(D-3)/2}(E_n^{c}(L) R),
\end{equation}
where $K_{\nu}(x)$ is a Bessel function. The same equation may be
derived from a string theoretic computation of the partition
function for a string wrapping the annulus along the lines of
\cite{bib:partfunc} (see also \cite{Billo:2005iv}), 
allowing arbitrary energies for the states in the closed string channel.

For very large $L$ and $R$, the higher derivative corrections to the
effective action are negligible, and the energy levels will be those
of a free string in $D$ dimensions; this implies that the closed
string energy levels are all of the form
\begin{equation}\label{energies}
E_n^{c,L} = E_n^{c,0}(L) + E_n^{c,1}(L) + \cdots = T L + \frac{4\pi}{L}
\left[ -{1\over 24} (D-2) + N_n \right] + \cdots
\end{equation}
if the state $n$ arises at excitation level $N_n \in {\bf Z}$ (this
is actually the excitation level for both the right-moving and the
left-moving states on the worldsheet; the two are equal for any
state that has an overlap with the boundary state). The first term
is the classical string energy, and the second is the well-known
L\"uscher term \cite{bib:luscherterm}. We expect the higher
derivative corrections to the action to give corrections to \eqref{energies}
which, on
dimensional grounds, begin at order $1/L^3$; in particular, in a
flat-space string theory, which is well-described by the Nambu-Goto
action, the exact formula for the energy levels of zero-momentum
states is given by
\begin{equation}
E_n^{c,NG}(L) = \sqrt{(T L)^2 + 8\pi T \left[ -\frac{1}{24} (D-2) +
N_n \right]},
\end{equation}
but we do not expect this equation to be exact for general confining
strings\footnote{In particular, we expect the ground state energy of
the winding string to go to zero at $L=1/T_H$ where $T_H$ is the
Hagedorn temperature, and we expect this temperature in general to
be smaller than the one of the Nambu-Goto string, $T_H^{NG} =
\sqrt{3 T / (D-2) \pi}$. This is confirmed by lattice simulations
\cite{Bringoltz:2005xx}.}. Similarly, for large $R$ the open string energy levels
take the form
\begin{equation}
\label{eq:spectrumopen}
E_k^{o}(R) = E_k^{o,0}(R) + E_k^{o,1}(R) + \cdots = T R + {\pi\over
R} \left[-{1\over 24}(D-2) + N_k \right] + \cdots,
\end{equation}
for levels $N_k \in {\bf Z}$, with corrections starting at order
$1/R^3$. Note that the form of the effective action guarantees that
closed string energy levels have an expansion involving purely odd
powers of $1/L$. For open strings the same is true if there are no
boundary terms, but for strings stretched between a quark and an
anti-quark there could be boundary terms which introduce also
even powers of $1/R$.

The partition function on the annulus that we compute must be
consistent with the two forms \eqref{openannulus} and
\eqref{closedannulus} above. The partition function for large $L,R$
in both channels may be expanded in a power series in
inverse powers of $L$ and $R$ (multiplying the exponential terms),
which is really an expansion in $E^1, E^2, \cdots$ (where we take
$E^k(R)$ to scale as $1/R^{2k-1}$). The comparison to the effective
field theory partition function turns out to be simplest if we
expand around an expression in which we write both $E^0$ and $E^1$
in the exponent, but expand just around $E^0$ in other places,
since this is what we will find in the effective field theory
partition function in the free field approximation. In the ``open
string channel'' we then have at the leading non-trivial order
\begin{equation}
Z^{annulus}(L,R) = \sum_{k} e^{-(E_{k}^{o,0}(R)+E_k^{o,1}(R)) L} (1
- E_k^{o,2}(R) L + \cdots).
\end{equation}
So, in this channel, the leading order correction to the partition
function should look like $L$ times a sum over corrections to
energies (scaling as $1/R^3$) times exponentials. In the ``closed
string channel'' we obtain to leading non-trivial order (up to two
inverse powers of lengths)
\begin{eqnarray}\label{closedcorr}
Z^{annulus}&&(L,R) = \sum_n |v_n(L)|^2 \left( {E_n^{c,0}(L)\over
{2\pi R}} \right)^{(D-2)/2} e^{-(E_n^{c,0}(L)+E_n^{c,1}(L)) R} \cdot
\nonumber \\ && \left(1 - E_n^{c,2}(L) R + {{D-2}\over 2}
{E_n^{c,1}(L)\over E_n^{c,0}(L)} + \cdots \right) \cdot \left[ 1 +
{{(D-2)(D-4)}\over {8 E_n^{c,0}(L) R}} + \cdots \right].
\end{eqnarray}
We will
analyze this expansion in detail in the next section.
The leading correction here is more complicated, involving terms
scaling as $R/L^3$, $1/L^2$ and $1/LR$. A $1/L^2$ contribution can
also arise by expanding $v_n(L)$ in inverse powers of $L$.

The method of \cite{bib:constraints4} to constrain the partition
function is to first compute the partition
function coming from the effective action described in the
previous subsection, and then to try to match it to the equations in
the previous paragraph (for some corrections to the energy levels
and to $v_n(L)$). Of course, the partition function should only
match for states whose excitation energies are much smaller than the
scale $m$ where the effective action breaks down, but this is true
for any state for large enough $L$ and $R$. In
\cite{bib:constraints4} this matching was performed for the
4-derivative terms, by expanding the partition function to leading
order in $c_2$, $c_3$, and writing it using exponentials either of
$L/R$ or of $R/L$ (modular transformation properties of the
resulting partition function can be used to relate the two). We will review this
computation in section \ref{sec:effectiveaction}. In the ``open
string channel'' this gave reasonable results for any $c_2, c_3$
(with some specific corrections to open string energy levels at
order $1/R^3$, linear in $c_2$ and $c_3$). However, in the ``closed
string channel'', the corrections of order $1/L R$, coming from the
last parentheses in \eqref{closedcorr}, matched only when
\begin{equation}\label{lwconstraint}
(D-2) c_2 + c_3 = {{D-4}\over {8 T}}.
\end{equation}
Thus, they concluded that the effective action is only consistent
when this constraint is satisfied. When it is satisfied, the
correction takes the form \eqref{closedcorr} and one can compute the
leading corrections to the energy levels (and to $v_n(L)$) in both
the open and closed string channels. Note that the Nambu-Goto action
(whose quantum open string spectrum was computed in
\cite{bib:arvis}), with
 $c_2^{NG} = 1 / 8 T$ and $c_3^{NG} = - 1 / 4 T$, satisfies this constraint, as it has to since one can check that it leads to
a partition function on the annulus which is consistent in both
channels (to all orders). For $D=3$, since there is only a single
four-derivative coefficient, the constraint \eqref{lwconstraint}
implies that the four-derivative action must agree with the
Nambu-Goto action. For $D>3$ there is one free coefficient
remaining, so it seems that the action (and the energy levels) need
not agree with Nambu-Goto at the four-derivative order (namely, for
corrections to energies of order $1/R^3$ or $1/L^3$).
In section \ref{sec:effectiveaction} we will extend the
considerations of this paragraph to the next order in the derivative
expansion.

In \cite{bib:constraints4} they also showed that no boundary terms
can contribute up to two-derivative order; here, as
mentioned above, we will not discuss the boundary terms, but we will
just take them to zero (assuming that there is some consistent
boundary of the confining string which gives this). It would be
interesting to generalize our analysis to obtain constraints on
boundary terms involving higher derivatives than considered
in \cite{bib:constraints4}.

A similar analysis may be performed by considering a string wrapping
a torus in space-time, which we will take for simplicity to involve
two orthogonal periodic coordinates, $X^0 \equiv X^0+L$ and
$X^1\equiv X^1+R$. Obviously, the partition function in this case
must be invariant under $L \leftrightarrow R$ (modular invariant),
and this is automatically satisfied for any effective action.
However, it must also have an interpretation as the sum over closed
string states winding the $X^1$ circle, propagating for a time $L$,
or the other way around. By similar arguments to the ones above (or
by a simple generalization of the computation of the torus partition
function in \cite{bib:partfunc}, see also \cite{Billo:2006zg}), 
this partition function must take
the form (up to an unimportant constant depending on the radii and
on the transverse volume)
\begin{equation}
\label{closedtorus} Z^{torus}(L,R) = \sum_n R \left(
{E_n^{c}(L)\over R} \right)^{(D-1)/2} K_{(D-1)/2}(E_n^{c}(L) R),
\end{equation}
where the states summed over here are all the closed string states
winding once around the circle with zero transverse momentum
(including states with different numbers of left and right-moving
excitations on the worldsheet, which carry a non-zero momentum
around the worldsheet). Again, we can expand this in inverse powers
of $L$ and $R$, and compare to the expressions that we obtain from
the low-energy effective action. Here we have less freedom (since
there are no unknown coefficients $v_n(L)$), so we will find more
constraints. We will see in section \ref{sec:effectiveaction} that
at the leading non-trivial order this will allow us to uniquely
determine $c_2$ and $c_3$ for any $D$, such that they must equal the
Nambu-Goto values, and we will extend the analysis to the next order
as well.

Before we conclude this section, it is important to stress the
assumptions that go into the constraints discussed above on the
string effective action. One assumption is that the string has a
limit in which it is weakly coupled. If the string is not weakly
coupled, there is no physical observable that gets contributions
only from single-string states as we assumed above, since there is
generically mixing between single-string and multi-string states. It
is not clear how this mixing affects the energy levels of winding
states; it would be interesting to analyze this. Thus, for confining
$SU(N)$ strings, we expect the effective action on the worldsheet to
obey the constraints above (since there is a large $N$ limit where
the string theory is weakly coupled), but for finite $N$ the energy
levels obtain corrections from mixing so there may be deviations
from the large $N$ predictions derived above. A second assumption,
which goes into the way we regularize the loop diagrams in the
worldsheet effective action, is that there is no dependence of the
results on any UV cutoff scale. This assumption is presumably true
for solitonic strings that correspond to standard weakly coupled
string theories, since in such theories there is a gauge where the
worldsheet theory is conformal, and there is no dependence on any
high energy scale. However, generic solitonic strings may not
correspond to any weakly coupled string theory, and for such strings
it is not clear that physics at high energies decouples from the
low-energy physics captured by the effective action. Thus, our
predictions for the form of the effective action apply to confining
strings (in the fundamental representation) in large $N$ gauge
theories, but a priori it is not clear whether they should hold more
generally or not.

Even with these assumptions it seems that we are getting constraints
for ``free'', but we are really getting them by using the symmetries
of the problem. Specifically, our theories have a $SO(D-1,1)$
Lorentz symmetry, but the effective action on the string is only
explicitly invariant under an $SO(D-2)$ subgroup. The derivations of
the expressions \eqref{closedannulus} and \eqref{closedtorus} use
the rotation symmetry between the transverse coordinates and the
coordinate which the string propagates in, so our constraints really
test this $SO(D-2,1)$ symmetry (this was explicitly shown in
\cite{bib:constraints5} for the annulus). It would be interesting to
derive the constraints above directly from Ward identities for the
broken symmetries, and to check whether additional constraints may
be derived by using the full Poincar\'e symmetry.
Note that the Nambu-Goto string, whose full
spectrum is not Poincar\'e-invariant (for $D \neq 26$, since the
massive states do not all lie in representations of $SO(D-1)$),
still satisfies all the constraints that we discuss here.

A different type of constraint on the effective action was
considered by Polchinski and Strominger in \cite{bib:constraints1}.
They used the fact that the confining string is expected to be (at
least in the large $N$ limit \cite{bib:largeN}) a fundamental string
in some curved background. This implies that one can go to a
``conformal gauge'' in which the theory on the worldsheet of the
string must be a conformal theory with a specific central charge
($c=26$ if it is a bosonic string, or $c=15$ if it is a
superstring). They attempted to write down such a conformal theory
for the $D$ scalar fields describing the position of the string in
${\bf R}^D$ (since they use a conformal gauge they cannot go to
static gauge, so the action involves all the coordinates). The
action they wrote down is singular (it involves negative powers of
$(\del X \cdot \del X)$), but becomes non-singular when expanded
around a long string configuration of the type described above; we
can interpret this by saying that their action may be derived by
integrating out all the other degrees of freedom on the string, and
this integrating out is useful (gives a non-singular action) when
expanding around a long string configuration in which these
additional degrees of freedom are heavy. In this formalism an
expansion similar to, but not identical to, the derivative expansion
described above was obtained in \cite{bib:constraints1}. They showed
that the leading correction to the free action in this expansion is
uniquely determined, and it was later shown in
\cite{bib:constraints2, bib:constraints3} that, for any $D$, this
implies that the leading correction to the string ground state
energy (at order $1/L^3$) is the same as in the Nambu-Goto action.
Of course, this, as well as the claim that the leading correction to
the effective action is unique (and equal to Nambu-Goto), is
consistent with the constraints we described above.

It is not clear to us precisely how to relate the effective action
in the Polchinski-Strominger formalism to the one in the
L\"uscher-Weisz formalism -- it would be very interesting to
understand this. In particular the Polchinski-Strominger formalism
seems to involve the full Poincar\'e symmetry (since it is only
valid for a string with the critical central charge) which is not
used in the L\"uscher-Weisz formalism. It is not completely obvious
to us if the Polchinski-Strominger formalism is valid or not
(namely, if integrating out the other fields on the string
indeed always gives an effective action of the form that they
assume). If it is valid, it would be interesting to extend it to the
next order, in order to see if this leads to more or to less
constraints on the corrections to the effective action than the ones
that we derive from the L\"uscher-Weisz formalism. This is not
clear, since the assumptions that underlie the two formalisms do not
seem to be identical.

\subsection{The effective string action in weakly curved holographic
backgrounds}

As already mentioned above, we believe that the confining string (in
the fundamental representation) in a large $N$ gauge theory is
equivalent to a weakly coupled fundamental string in some
background, by a generalization of the AdS/CFT correspondence
\cite{bib:adscft1, bib:adscft2, bib:adscft3}. In principle, given
such a background, we can write down the string action for a long
string configuration in static gauge\footnote{In the last 12 years it has
been realized that also non-confining gauge theories have a string theory
description. However, in these theories there is no classical long closed
string configuration of the type we analyze in this paper, and no expansion
in the inverse string tension.}.
As mentioned above, we expect
that all degrees of freedom except the $(D-2)$ Nambu-Goldstone
bosons will have some mass $m$ in this background, and, thus, we can
integrate them out to obtain the effective action below the scale
$m$. Recall that already for a string in flat space, the Nambu-Goto
action is an effective action expanded in derivatives divided by the
string scale $M_s \sim \sqrt{T}$ so we expect it to break down (or
become strongly coupled) at this scale. For a generic gauge theory,
with no dimensionless parameters, we expect $m \sim M_s$. This means
that the effective action at the scale $m$ will generically be
strongly coupled, and it is not clear how to integrate out the
degrees of freedom at this scale.

However, for a special class of gauge theories, the dual string
background is weakly curved (such a background for a superstring is
necessarily ten dimensional). Several examples of this class were
discovered in the last decade \cite{bib:backgrounds1,
bib:backgrounds2, bib:backgrounds3, bib:backgrounds4}. The fact that
the background is weakly curved means that the string tension is
much larger than the typical curvature scale of the background; the
latter determines the masses of the additional fields on the
worldsheet, so that in such a case we have $m^2 \ll T$.
The effective theory on the worldsheet at the scale $m$ is then
weakly coupled -- the dimensionless coupling constant is $m^2 / T$
-- and we can perturbatively integrate out the massive fields to
obtain the low-energy effective action of the massless fields. We
will do this in detail in section \ref{sec:corfun}. In these
backgrounds we naively expect the effective action to be equal to
the Nambu-Goto action, with corrections that are a power series in
$m^2 / T$ (coming from loops of massive string states). Note that
usually when massive states are integrated out, the corrections to
the effective action go as a negative power of their mass, while
here the corrections go as a positive power of the mass. This is
because the massless limit corresponds exactly to a string in flat
space (the Nambu-Goto action), so the deviations from this limit
must go as a positive power of the mass; in practice we will see
that this arises because the couplings of the additional fields to
the massless fields will contain powers of $m$.

As mentioned above, in typical gauge theories (like pure Yang-Mills
theory) we do not expect $m$ to be small, so we cannot perform such
a power series expansion of the corrections. Our goal will be to see
which terms in the action deviate from the Nambu-Goto form in the
regime of small $m$ and which do not; it seems plausible that any terms that
are allowed to deviate, will do so already at leading order, and
thus we conjecture that the same terms should deviate from
Nambu-Goto also in general gauge theories (though in general
we expect any allowed deviations to be of order one in string
units). We will find that indeed our one-loop effective action will
take the most general form allowed by the constraints discussed in
the previous subsection, up to one additional constraint that it
satisfies.

A special example of such a computation of corrections to the
action, which was already analyzed in the literature,
is the computation of the effective tension of
the string; the classical tension receives corrections at one-loop
from integrating out the massive fields, and it was found in
\cite{bib:tension} that the corrected tension takes the form
\begin{equation}
\label{tensioncorr} T' = T + {1\over {8\pi}} \left( \sum_{fermions\
F} m_F^2 \log(m_F^2) - \sum_{bosons\ B} m_B^2 \log(m_B^2) \right),
\end{equation}
where $T$ is the classical tension, and $m_B$ and $m_F$ are the
masses of the bosonic and fermionic degrees of freedom on the
worldsheet, respectively. Note that we generally expect to have
$(10-D)$ massive scalars on the worldsheet (since we are dealing with
a string in a background which is ten dimensional), and $8$ massive
fermions (since this is the number of physical fermions living on a
superstring), though in some cases some of these fields may be
massless (and do not contribute to \eqref{tensioncorr}). Note also
that the logarithms appearing in this formula are really of the form
$\log(m^2/\Lambda^2)$ where $\Lambda$ is the UV cutoff, so in order
for \eqref{tensioncorr} to be finite, it must be the case that
\begin{equation}
\label{massequal} \sum_{bosons\ B} m_B^2 = \sum_{fermions\ F} m_F^2.
\end{equation}
This is true in all known holographic backgrounds, and it seems to
be necessary to obtain finite (cutoff-independent) results for
various worldsheet computations (as we expect, since the theory in
conformal gauge is manifestly independent of the cutoff). We will
assume from here on that equation \eqref{massequal} holds.

In all known examples of confining string backgrounds, we actually
find more massless scalars than expected on the worldsheet. This is
because in these backgrounds the confining string lives in some
``IR'' region of space, and in this region there is some
$p$-dimensional sphere that the string is localized on (and there is
an exact or approximate $SO(p+1)$ symmetry), which generally carries
some non-zero flux which stabilizes the background. The string is
localized on this sphere, meaning that it has $p$ additional
massless fields $e^j$ on its worldvolume corresponding to its
position on the sphere. Generally, upon integrating out the massive
fields (including the fermions), the effective action of these
fields looks like a sigma model on $S^p$. So, we expect that
perturbatively the $e^j$ fields will remain massless, but that
non-perturbatively this sigma model will develop a mass gap, at some
scale $\tilde \Lambda$ which is exponentially smaller than the scale $m$
(the radius of the sphere is of order $1/m$, so we expect $\tilde\Lambda
\sim m \exp(-C T / m^2)$ for some constant $C$). In such a case, in
the effective action for length scales between $1/\tilde\Lambda$ and $1/m$
we should include the $e$ fields as well, and only in the effective
action at length scales above $1/\tilde\Lambda$ we can integrate them out.
However, the field theory of the $e$'s is strongly coupled, so this
will lead to various corrections to the action depending on the
scale $\tilde\Lambda$ that we do not know how to compute.

Our attitude will be to ignore the $e$ fields in our computations,
and only to integrate out the fields which obtain a mass at the
curvature scale $m$. For the effective action between the length
scales $1/\tilde\Lambda$ and $1/m$, this means that we will reliably
compute the terms in the effective action that depend only on the
$X^i$, but that the full effective action will also contain the $e$
fields and various couplings involving them (which we will not
compute, though they can easily be computed). In the effective
action at very low energies (below $\tilde\Lambda$), there will be
additional terms coming from integrating out also the $e$'s, which
we do not know how to compute. However, these terms will depend on
$\tilde\Lambda$ rather than on the scale $m$, so they will be
parameterically separated from the terms that we do compute. Namely,
we compute all the terms in the effective action that depend on the
scale $m$ and not on other scales. Our main interest
will be in seeing for which terms in the effective action deviations
from Nambu-Goto exist; if we find a deviation from Nambu-Goto
associated with the scale $m$ then we can be sure that this
deviation remains non-vanishing also after adding the contributions
from the $e$ fields, since their contributions involve different
scales.

Since we are only interested in effects depending on $m$, we can
also ignore other contributions to the effective action which are
independent of $m$, such as effects coming from ghost loops and
loops of the worldsheet metric (at least at the one-loop order that
we will be working in). Since we ignore metric loops, it will not
matter whether we work in the Polyakov or in the Nambu-Goto
formalism. However, the difference between the two formalisms may be
important at higher orders.

Note that in the holographic description, we can introduce a
boundary to the string worldsheet in two ways. One possibility is to
compute a Wilson loop (or a correlation function of Wilson loops);
in this case the fundamental string worldsheet ends on the boundary
of the higher dimensional space (it goes to infinity in the radial
direction) at the position of the Wilson loop \cite{bib:wilsonloop1,
bib:wilsonloop2}. The worldsheet in
this case does not sit just in the ``IR'' region of space, so the
action in this case may well have boundary terms which depend on
what happens throughout the holographic dual space-time. The second
possibility is that the string can end on a D-brane; D-branes can
describe various objects in a confining string theory, including
dynamical quarks and domain walls. In this case, if the D-brane
extends into the ``IR'' region of space so that the open string can
be completely localized in this region, we do not expect to get any
boundary terms on the worldsheet, since there are no such terms for
a string ending on a D-brane.

\subsection{A special class of holographic backgrounds}
\label{specialclass}

In general weakly curved holographic backgrounds, we obtain small
corrections to the Nambu-Goto action as described above. In these
backgrounds the confining string is described as a fundamental
string moving in some curved space with various $p$-form fields
(typically including Ramond-Ramond (R-R) fields) turned on. We expect some such
background (weakly curved or not) to correspond to any large $N$
gauge theory in the 't Hooft limit. However, there is also a class
of large $N$ confining gauge theories whose confining string is not
well-described as a weakly coupled fundamental string as above;
these do not arise from a 't Hooft large $N$ limit but from a
different type of limit. As an example of this, consider the
background of $N$ D5-branes compactified on a two-sphere, in the
limit where the theory on the D5-branes is decoupled from the string
theory in the bulk; the gravity solution describing this background
was found by Chamseddine and Volkov in \cite{bib:backgrounds5} and
interpreted in this way by Maldacena and Nu\~nez in
\cite{bib:backgrounds2}. In the decoupling limit the $5+1$
dimensional effective Yang-Mills coupling $g_6$ is kept fixed, and
there is a dimensionless parameter corresponding to the size of the
two-sphere in units of the 't Hooft coupling $g_6^2 N$. (Of course,
the six dimensional Yang-Mills theory requires some UV completion,
and this is provided in this decoupling limit by a ``little string
theory''.) In the limit where this size is small, the background is
weakly coupled in an S-dual frame to the original frame, such that
the confining string is best described by a D-string moving in a weakly
curved background \footnote{There is also a range of values of the
size for which the IR region of the background is weakly coupled in
the original frame, and there the confining string is described by a
fundamental string, but the coupling becomes strong away from the IR
region. For this range of values the confining string behaves in the
generic way that we expect for large $N$ gauge theories, see section
\ref{ssec:examplesMN}.}. At first
sight this case seems very similar to the case of a fundamental
string; the effective action on the D-string, which is the DBI
action, is essentially the same as the Nambu-Goto action, and again
there are some fields on the long D-string worldsheet that are
massive and that may be integrated out. However, the fact that the
D-string action has an inverse string coupling $1/g_s$ multiplying
it, means that corrections from loops of the massive states are down
by powers of $g_s$ compared to the classical action. Additional
corrections of the same order come from string diagrams of higher
genus, which have not been computed (recall that the DBI
action just captures the string disk diagrams). So, in this case the
effective action on the confining string worldsheet, at leading
order in $g_s$, is precisely the same as the Nambu-Goto action, with
no deviations at all. At the next order in $g_s$ there will be
deviations, but computing them requires higher genus diagrams so it
is rather complicated. This case is thus rather different from the
standard confining string case discussed in the previous subsection,
where we expect any allowed
deviations from Nambu-Goto to appear already at leading order in
$g_s \sim 1/N$.

\subsection{A brief review of lattice results}

The effective action of confining strings on the lattice has been
studied in the last few years with increasing accuracy (see \cite{Kuti:2005xg}
for a review of results before the ones we explicitly discuss below). In
particular, for the three dimensional pure Yang-Mills theory, very
precise results have been obtained for the spectrum of closed
confining strings in large $N$ $SU(N)$ gauge theories (in the
fundamental representation) winding on a circle of circumference $L$
\cite{bib:lat1, bib:lat2}. For the ground state energy, it has been
found that it agrees with the Nambu-Goto result at order $1/L$ (the
L\"uscher term) and is consistent with it at order $1/L^3$,
and that if there is a deviation at
order $1/L^5$ its coefficient is very small (so that the deviation
may well be at a higher order in $1/L$). Excited states
seem to again agree with Nambu-Goto at orders $1/L$ and
$1/L^3$, but to deviate more from it at higher orders, perhaps
already at order $1/L^5$ (the deviations seem larger than those of
the ground state, but the lattice data is not precise enough to
determine at which order it occurs).
In $3+1$ dimensions the simulations of large $N$ gauge theories again
find agreement with Nambu-Goto for large $L$, but they are
not yet precise enough to tell at what order deviations from
Nambu-Goto arise.
Simulations of interfaces in the $2+1$-dimensional Ising model similarly
show good agreement with Nambu-Goto (see, for instance, \cite{Caselle:2007yc} and
references therein), and again they are not yet precise enough to tell at
what order deviations from Nambu-Goto arise.

Recently, $2+1$ dimensional confining strings in higher
representations (``$k$-strings'') were also studied \cite{Bringoltz:2008nd,Athenodorou:2008cj} in the
large $N$ limit and compared to Nambu-Goto, and it was found that
they exhibit larger deviations from Nambu-Goto (which may already
start at order $1/L^5$) for
all states, including the ground state. However, it is subtle
to interpret these results, for the following reason. In the large
$N$ limit (with fixed $k$), the binding energy of $k$ fundamental
strings to form a $k$-string goes to zero as $1/N^{\alpha}$; some
theoretical arguments suggest that $\alpha=2$, while other arguments
(see, for instance, \cite{KorthalsAltes:2005ph}) and lattice results
suggest that $\alpha=1$. This means that
in the large $N$ limit there are (at least) $(k-1)(D-2)$ light modes
on the worldvolume of a $k$-string, whose mass goes to zero in the
large $N$ limit as $1/N^{\alpha/2}$.
The general constraints above concerning the form of the effective
action are valid only at length scales larger than the inverse of
any mass of a worldsheet field (and also large enough so that there is
no mixing of the $k$-string states with states of $k$ fundamental strings);
thus, they only apply for
length scales bigger than $N^{\alpha/2} / \sqrt{T}$, and it is not clear if these
scales are accessed by the simulations yet (with enough precision to extract
the $1/L^5$ corrections to energy levels).
It would be interesting to analyze $k$-strings at longer scales, to
see if they obey our constraints at these scales.

In another recent paper \cite{bib:lat3}, the energy of the ground
state of a confining string in a $2+1$ dimensional system arising
from a continuum limit of percolation was numerically computed, and
found to agree again with Nambu-Goto at orders $1/L$ and $1/L^3$,
but to significantly deviate from it at order $1/L^5$ (though it is
not clear if there are enough data points within the range where the
$1/L$ expansion can be trusted in order to be able to reliably
compute this). Since this model does not have any obvious large $N$
limit where it corresponds to a weakly coupled string theory, it is
not clear if the general arguments above apply to it or not; as we
will see below, the results of \cite{bib:lat3} seem to contradict
the general predictions for an effective action of a string which
has a weakly coupled limit.


\section{The effective theory on a confining string}
\label{sec:effectiveaction}
In this section we compute the partition function for the general
effective action \eqref{actionzero}-\eqref{actionsix}
at six derivative order, both on the annulus and on
the torus. The computation is perturbative, and the partition
function is a power series in $(\sqrt{T}L)^{-1}$ and
$(\sqrt{T}R)^{-1}$. We obtain expressions such as the Dedekind
function and its derivatives. This allows us to expand the results
both in exponents of $L/R$ (times powers) and in exponents of $R/L$
(times powers).
On the annulus
this corresponds to the closed and open string channels, while on the
torus both limits correspond to closed string partition
functions. We require the partition function to have the form
(\ref{closedannulus}) for the annulus and (\ref{closedtorus}) for
the torus, when expanded in the long string limit. From
this we can extract the open and closed string spectrum.
Most importantly, we discover that the coefficients $c_{2},\cdots,c_7$
are not arbitrary.

In each subsection we start by writing the general form of the
partition function, given in section \ref{sec:generalities}, which is relevant at that
order. We then compute, perturbatively, the partition function,
using the action $S=S_0+S_2+S_4+S_6$, and compare the two.

\subsection{Partition function at $O(T^{0})$}
To leading order, the expressions \eqref{openannulus}, (\ref{closedannulus}) and
(\ref{closedtorus}) take the following forms :
\begin{eqnarray}\label{eq:lus_corelator_1}
\nonumber
Z^{ann.}&=&\sum_{k=0}^{\infty}\omega_{k}e^{-(E^{o,0}_{k}+E^{o,1}_{k})L}=
\sum_{n=0,2,4,\cdots}^{\infty}|v_n^0|^2 \left(\frac{T
L}{2\pi R }\right)^{\frac{1}{2}(D-2)}e^{-(E^{c,0}_n+E^{c,1}_n)R}\textrm{ ,
}
\\  Z^{tor.}&=&\sum_{n=0}^{\infty}\sqrt{\frac{\pi}{2}}\left(\frac{T L}{R}\right)^{\frac{1}{2}(D-2)}\omega_n^{tor.} e^{-(E^{c,0}_n+E^{c,1}_n)R}\textrm{ . }
\end{eqnarray}
The summations are over energy levels $n=N_L+N_R$, where $N_{L(R)}$
are the number of left(right)-moving excitations on the worldsheet.
$\omega_n$ is a weight factor corresponding to the number of states
at each energy level (we join together the contributions from different
states with the same energies, anticipating a degeneracy
in the leading order).
The annulus boundary carries zero momentum in
the compact direction, and so the annulus partition function
contains only states with an equal number of left-moving and right-moving
excitations,
which explains the summation over even $n$ in our equations. We use
the closed (open) energy level expansion
$E_n^{c(o)}=E_n^{c(o),0}+E_n^{c(o),1}+\cdots$, and
$v_n(L)=v_n^0(L)+v_n^1(L)+\cdots$ are the overlaps between the boundary
state and the closed string states at level $n$. More precisely,
$|v_{n}|^2=\sum_{i\in n}|v_{n,i}|^2$, where $v_{n,i}$ is the overlap
with a specific state $i$ at level $n$.

From the worldsheet effective action point of view,
we derive the partition function at this order from the free action
$S=S_0+S_2$. It is convenient to write the partition function using
the definitions
\begin{eqnarray}\label{partdefs}
\nonumber q&\equiv&e^{2\pi i \tau} \quad\textrm{,}\quad \tilde{q}\equiv e^{2\pi
i \tilde{\tau}} \quad ,
\\ \tau_{ann.}&=&i\frac{L}{2R} \quad\textrm{,}\quad \tilde{\tau}_{ann.}\equiv-\frac{1}{\tau_{ann.}}=i\frac{2R}{L}
\quad\textrm{,}\quad\tau_{tor.}=i\frac{L}{R} \quad\textrm{,}\quad
\tilde{\tau}_{tor.}=i\frac{R}{L} \quad.
\end{eqnarray}
The partition function (up to constants including the
transverse volume in the torus case) is \cite{bib:DF}:
\begin{eqnarray}\label{eq:lus_0th}
\nonumber
Z_0^{ann.}&=&e^{-TLR}\eta(q^{ann.})^{2-D}=\sum_{k=0}^{\infty}\omega_{k}e^{-TLR-\frac{\pi
L}{R}[-\frac{1}{24}(D-2)+k]}
\\ \nonumber&=&\left(\frac{2R}{L}\right)^{\frac{2-D}{2}}\sum_{n=0,2,4,\cdots}^{\infty}\omega_n^{ann.}e^{-TLR-\frac{4\pi
R}{L}[-\frac{1}{24}(D-2)+\frac{n}{2}]}\textrm{ , }
\\
Z_0^{tor.}&=&R^{2-D}e^{-TLR}\eta(\tilde{q}^{tor.})^{2(2-D)}=R^{2-D}\sum_{n=0}^{\infty}\omega_n^{tor.}e^{-TLR-\frac{4\pi
R}{L}[-\frac{1}{24}(D-2)+\frac{n}{2}]}\textrm{ . }
\end{eqnarray}
Here $\omega_n$ are weight factors, proportional to the number of
states at each energy level $n$, and we also used
the modular properties of the Dedekind eta function,
\begin{eqnarray}\label{etafunc}
\eta(q)\equiv q^{1/24}\Pi_{n=1}^{\infty}(1-q^n) ,\quad
\eta(q)=\sqrt{\frac{i}{\tau}}\eta(\tilde{q}) \textrm{ .}
\end{eqnarray}
By matching (\ref{eq:lus_0th}) and (\ref{eq:lus_corelator_1}) we
find the energies and overlap functions to first order. For the
annulus,
\begin{eqnarray}\label{eq:lus_energy1}
\nonumber E_n^{c,0}&=&T L,\quad
E_n^{c,1}=\frac{4\pi}{L}[-\frac{1}{24}(D-2)+\frac{n}{2}],\quad
\\|v_n^0|^2&=&\omega_n^{ann.}\left(\frac{\pi}{T}\right)^{\frac{1}{2}(D-2)} \textrm{ , }
\end{eqnarray}
and the open string energies are given in (\ref{eq:spectrumopen}).
The same closed string energies appear in the torus partition function, so that the
annulus and torus computations are consistent.

\subsection{Partition function at $O(T^{-1})$}

We carry on with the expansion of the partition function to order
$O(T^{-1})$. We explicitly use our results from the previous
subsection. By expanding equations
(\ref{closedannulus}) and (\ref{closedtorus}) we expect to obtain
expressions of the form
\begin{eqnarray}\label{eq:lus_corelator_3}
\nonumber Z^{ann.}&=&\sum_{n=0,2,4,\cdots}^{\infty}|v_n^0|^2\left(\frac{T
L}{2\pi R
}\right)^{\frac{1}{2}(D-2)}e^{-(E^{c,0}_n+E^{c,1}_n)R}\{1-[E^{c,2}_n]_{ann.}R+\frac{|v_n^1|^2}{|v_n^0|^2}
\\ \nonumber&&+\frac{2\pi(D-2)}{TL^2}[-\frac{1}{24}(D-2)+\frac{n}{2}]+\frac{(D-2)(D-4)}{8 T L R}+\cdots\}\textrm{ , }
\\ \nonumber Z^{tor.}&=&\sum_{n=0}^{\infty}\sqrt{\frac{\pi}{2}}\left(\frac{T L}{R}\right)^{\frac{1}{2}(D-2)}e^{-(E^{c,0}_n+E^{c,1}_n)R}\omega_n^{tor.}\{1-[E^{c,2}_n]_{tor.}R+\frac{D(D-2)}{8TLR}
\\ &&+\frac{2\pi(D-2)}{TL^2}[-\frac{1}{24}(D-2)+\frac{n}{2}]+\cdots\}\textrm{ . }
\end{eqnarray}
The notation $[E^{c,2}_n]$ indicates there is an averaging over all
states at level $n$, for the torus with equal weight and for
the annulus with weight $|v_{n,i}|^2$.

\begin{figure}[!htb]
\centering
\includegraphics[width=3cm,height=2cm]{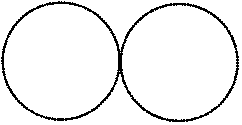}
\caption{The 2-loop contribution to the partition function at
$O(T^{-1})$.\label{fig:Z2}}
\end{figure}

We now compute the partition function of the action
$S=S_0+S_2+S_4$.
There are two 2-loop bubble diagrams (see figure \ref{fig:Z2}), with the
vertices $c_2$ and $c_3$. Two possible contractions in each diagram lead to expressions
proportional to $(D-2)^2$ and $(D-2)$. The computation of the diagrams gives
\begin{eqnarray}
\vev{S_4}_0&=& c_2[(D-2)^2\times I_1+2(D-2)\times  I_2]
\\ \nonumber
&&+c_3[(D-2)\times I_1+((D-2)^2+(D-2))\times I_2],\quad
\end{eqnarray}
with
\begin{eqnarray}
I_1&=&\int
d^2\sigma\partial_{\alpha}\partial^{\alpha'}G\partial_{\beta}\partial^{\beta'}G,\quad
I_2=\int
d^2\sigma\partial_{\alpha}\partial_{\beta'}G\partial^{\alpha}\partial^{\beta'}G\textrm{
. }
\end{eqnarray}
Here $G=\lim_{\sigma\rightarrow\sigma'}G(\sigma,\sigma')$ is the
propagator of the free field $X^i$ in coordinate space. We compute the diagrams in detail in
appendix \ref{sec:appendixA}, as originally done in \cite{bib:constraints4, bib:DF}.
The computation is rather straightforward, apart from the need to carefully
use a consistent zeta function regularization. We find the following result for the
annulus, expressed using the Eisenstein series $E_n(q)$ and their derivatives
$H_{n,k}(q)$, all defined in section \ref{ssec:modfun} :
\begin{eqnarray}\label{ioneann}
I_1^{ann.}&=&\frac{2\pi^2L}{R^3}H_{2,2}(q^{ann.})=-\frac{1}{L
R}+\frac{2\pi}{3L^2}E_2(\tilde{q}^{ann.})+\frac{32\pi^2R}{L^3}H_{2,2}(\tilde{q}^{ann.})\textrm{
, }
\\ \nonumber I_2^{ann.}&=&\frac{\pi^2L}{R^3}[\frac{2}{(24)^2}E_2^2(q^{ann.})+H_{2,2}(q^{ann.})]=\frac{16\pi^2R}{L^3}[\frac{2}{(24)^2}E_2(\tilde{q}^{ann.})^2+H_{2,2}(\tilde{q}^{ann.})]\textrm{ .}
\end{eqnarray}
For the torus we find (a similar computation for the Nambu-Goto action was performed in
\cite{Billo:2006zg})
\begin{eqnarray}\label{ionetor}
\nonumber I_1^{tor.}&=&\frac{1}{R L}\textrm{ , }
\\ I_2^{tor.}&=&\frac{\pi^2R}{18L^3}E_2^2(\tilde{q}^{tor.})-\frac{\pi}{3L^2}E_2(\tilde{q}^{tor.})+\frac{1}{R L} \textrm{ . }
\end{eqnarray}

The partition function at
this order is given by,
\begin{eqnarray}\label{eq:Z_2}
\nonumber Z(q)&=&Z_0(q)(1-\vev{S_4}_0)
\\ \nonumber &\propto& \sum_{n=0}^{\infty}\omega_n e^{-R(E_n^{c,0}+E^{c,1}_n)}\{1-I_1[(D-2)^2c_2+(D-2)c_3]
\\&&\qquad \qquad -I_2[(D-2)(D-1)c_3+2(D-2)c_2]\}\textrm{ . }
\end{eqnarray}
In the annulus case we see that
the corrections to the open string partition function are all energy
corrections proportional to $L/R^3$, as expected from (\ref{openannulus}). Plugging our
results \eqref{ioneann}, \eqref{ionetor} in (\ref{eq:Z_2}) gives the following partition functions
for closed strings,
\begin{eqnarray}
\nonumber
Z^{ann.}&=&\left(\frac{L}{2R}\right)^{\frac{1}{2}(D-2)}\sum_{n=0,2,4,\cdots}^{\infty}
\omega_n^{ann.}e^{-R(E_n^{c,0}+E^{c,1}_n)} \cdot
\\ \nonumber&&\times\{1-\frac{\pi^2R}{18L^3}E_2(\tilde{q}^{ann.})^2[(D-2)(D-1)c_3+2(D-2)c_2]
\\ \nonumber&&+\left(\frac{1}{LR}-\frac{2\pi}{3L^2}E_2(\tilde{q}^{ann.})\right)[(D-2)^2c_2+(D-2)c_3]
\\ \nonumber&&-\frac{16\pi^2 R}{L^3}H_{2,2}(\tilde{q}^{ann.})[(D-2)(D+1)c_3+2(D-2)(D-1)c_2]\},
\\ \nonumber Z^{tor.}&=&R^{2-D}\sum_{n=0}^{\infty}\omega_n^{tor.}e^{-R(E_n^{c,0}+E^{c,1}_n)} \cdot \{1+\left(\frac{\pi}{3L^2}E_2(\tilde{q}^{tor.})-\frac{\pi^2R}{18L^3}E_2^2(\tilde{q}^{tor.})\right)
\\ &&\times[(D-2)(D-1)c_3+2(D-2)c_2]-\frac{1}{RL}(D-2)D(c_3+c_2)\}.
\end{eqnarray}
We can now compare our result to (\ref{eq:lus_corelator_3}). By
looking at the ground state ($n=0$) we find the following constraint
from the $\frac{1}{LR}$ term in the annulus partition function \cite{bib:constraints4} :
\begin{eqnarray}
\frac{D-4}{8T}&=&(D-2)c_2+c_3\textrm{ , }
\end{eqnarray}
and we also find,
\begin{eqnarray}
{[E_0^{c,2}]}_{ann.}&=&\frac{\pi^2}{18L^3}(D-2)[(D-1)c_3+2c_2]
\textrm{ , } \frac{|v_0^1|^2}{|v_0^0|^2}=-\frac{2\pi}{3L^2}(D-2)((D-2)c_2+c_3)\textrm{ . }
\end{eqnarray}
Comparing other states does not give us additional constraints, as shown
in \cite{bib:constraints4}.

The result from the torus is consistent
with the results above, and we also get one additional constraint due to the fact
that there are no unknown overlap functions, so we can compare both
the $\frac{1}{LR}$ and the $\frac{1}{L^2}$ terms:
\begin{eqnarray}
\nonumber c_2+c_3&=&-\frac{1}{8T} \textrm{ , }
-\frac{D-2}{4T}=(D-1)c_3+2c_2 \textrm{ , }
\\ {[E_0^{c,2}]}_{tor.}&=&\frac{\pi^2}{18L^3}(D-2)[(D-1)c_3+2c_2]=-\frac{\pi^2}{72TL^3}(D-2)^2 \textrm{ . }
\end{eqnarray}
There are two independent constraints in total, which completely fix
the effective action at this order to be the Nambu-Goto action for any $D$:
\begin{eqnarray}
c_2=c_2^{NG}=\frac{1}{8T} \textrm{ , }
c_3=c_3^{NG}=-\frac{1}{4T}\textrm{ . }
\end{eqnarray}
In particular, this implies that the partition function (or any
other physical observable) is constrained to be the one given by the
Nambu-Goto action to this order. Again one can check that higher $n$'s do
not give additional constraints, but just give formulas for
$[E_n^{c,2}]_{tor.}$ for each level $n$.

\subsection{Partition function at $O(T^{-2})$}

At order $O(T^{-2})$ there are numerous contributions to the
partition function. We explicitly write only the unknown parameters,
namely $[E^{c,3}_n]$ and $|v^2_n(L)|^2$. All other terms at this
order, such as $[(E_n^{c,2})^2]$, were already determined in the
previous subsection to be the same as in the Nambu-Goto partition function. We then have,
\begin{eqnarray}\label{eq:lus_corelator_5}
\nonumber Z^{ann.}&=&\left(\frac{T L}{2\pi R
}\right)^{\frac{1}{2}(D-2)}\sum_{n=0,2,4,\cdots}^{\infty}|v_n^0|^2e^{-R(E_n^{c,0}+E^{c,1}_n)}
\{\cdots-{[E_n^{c,3}]}_{ann.}R+\frac{|v_n^2(L)|^2}{|v_n^0|^2}+\cdots\}\textrm{
, }
\\ Z^{tor.}&=&\sqrt{\frac{\pi}{2}}\left(\frac{T L}{R }\right)^{\frac{1}{2}(D-2)}\sum_{n=0}^{\infty}e^{-R(E_n^{c,0}+E^{c,1}_n)} \omega_n^{tor.}
\{\cdots-{[E_n^{c,3}]}_{tor.}R+\cdots\}\textrm{ . }
\end{eqnarray}

As in the previous subsections we compute the partition function to
six derivative order, given by the following action,
\begin{eqnarray}
S=S_0+S_2+S_4+S_6\textrm{ . }
\end{eqnarray}
The partition function is then,
\begin{eqnarray}\label{partsixd}
Z(q)&=&Z_0(q)(1-\vev{S_4}_0+\frac{1}{2}\vev{(S_4)^2}_0-\vev{S_6}_0) \textrm{ . }
\end{eqnarray}
Diagrammatically, the $c_{4}$ contributions to $\vev{S_6}$ are two-loop diagrams
 similar to the ones of the previous subsection, while the $c_{6,7,8}$ contributions
 to $\vev{S_6}$, and $\vev{S_4^2}$, are three-loop diagrams
(see Figure 2). We do not compute $\vev{S_4^2}$ explicitly since we know from the
previous subsection that this contribution is constrained to equal
its form in the Nambu-Goto action. In particular, we know it will match all
terms (such as $[(E_n^{c,2})^2]$) which we did not write in
(\ref{eq:lus_corelator_5}), and which do not get contributions from
$S_6$.
\begin{figure}[!htb]
\centering
\includegraphics[width=10cm,height=3cm]{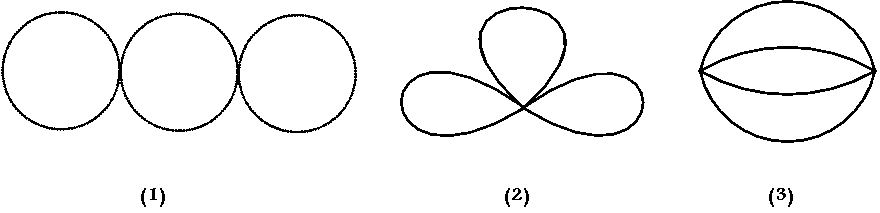}
\caption{The 3-loop contribution to the partition function. Diagrams
(1) and (3) are the two contributions to $\vev{S_4^2}$ and diagram (2) is
a single vertex diagram appearing in $\vev{S_6}$.\label{fig:Z3}}
\end{figure}

The computation of $\vev{S_6}$, using the diagrams of figures \ref{fig:Z2} and \ref{fig:Z3}, gives
\begin{eqnarray}\label{eq:Z_3}
\vev{S_6}&=&c_4[(D-2)^2I_3+2(D-2)I_4]
+c_6[(D-2)^3 I_6+6(D-2)^2I_7+8(D-2) I_8]
\\ \nonumber &&+c_7[((D-2)^3+(D-2)^2+4(D-2))I_7+4(D-1)(D-2)I_8+(D-2)^2I_6],
\end{eqnarray}
where
\begin{eqnarray}
\nonumber I_3&=& \int
d^2\sigma\partial_{\alpha}\partial^{\alpha'}\partial_{\beta}\partial^{\beta'}G\partial_{\gamma}\partial^{\gamma'}G,\quad
  I_4=\int d^2\sigma\partial_{\alpha}\partial_{\beta}\partial_{\gamma}'G\partial^{\alpha}\partial^{\beta}\partial^{\gamma'}G,\quad
\\ \nonumber
 I_6&=&\int d^2\sigma \partial_{\alpha}\partial^{\alpha'}G\partial_{\beta}\partial^{\beta'}G\partial_{\gamma}\partial^{\gamma'}G,\quad
\\ I_7&=&\int d^2\sigma\partial^{\alpha}\partial'_{\alpha}G\partial^{\beta}\partial_{\gamma}'G\partial_{\beta}'\partial^{\gamma}G,\quad
  I_8=\int d^2\sigma\partial^{\alpha}\partial'_{\beta}G\partial^{\beta}\partial_{\gamma}'G\partial^{\gamma}\partial_{\alpha}'G\textrm{ . }
\end{eqnarray}
In appendix \ref{sec:appendixA} we obtain the following results for the annulus,
\begin{eqnarray}\label{eq:intann3}
\nonumber
I_3^{ann.}&=&-4\frac{\pi^4L}{R^5}H_{2,4}(q^{ann.})=-4\frac{\pi^4L}{R^5}\left(\frac{4R^5}{15\pi
L^5}E_4(\tilde{q}^{ann.})-\frac{64R^6}{L^6}H_{2,4}(\tilde{q}^{ann.})\right)\textrm{
, }
\\ \nonumber I_4^{ann.}&=&-\frac{1}{2}I_3^{ann.},\quad
I_6^{ann.}=\frac{3\pi^3L}{2R^5}F(q^{ann.})=\frac{\pi^3L}{2R^5}\left[\frac{ R^5}{\pi L^5}-\frac{4R^4}{\pi^2
L^4}+\frac{4R^3}{\pi^3L^3}+O(\tilde{q})\right],
\\  I_7^{ann.}&=&\frac{1}{2}I_6^{ann.},\quad
I_8^{ann.}=\frac{1}{4}I_6^{ann.},
\end{eqnarray}
where $F(q)$ is defined in (\ref{eq:modfun1}). There is an
uncertainty $O(\tilde{q})$ in $I_{6,7,8}$ because we computed the
$\tilde{q}$ expansion of $F(q)$ numerically, as explained
in detail in section \ref{numericalF}.

For the torus we find
\begin{eqnarray}\label{eq:inttor3}
\nonumber I_3^{tor.}&=&I_4^{tor.}
=0\textrm{ , }
\\ \nonumber I_6^{tor.}&=&-\frac{1}{L^2R^2} \textrm{ , } I_7^{tor.}=-\frac{2\pi}{3RL^3}E_2({\tilde q}^{tor.})-\frac{\pi^2}{18L^4}E^2_2({\tilde q}^{tor.})-\frac{1}{L^2R^2}\textrm{ , }
\\  I_8^{tor.}&=&-\frac{\pi}{RL^3}E_2({\tilde q}^{tor.})-\frac{\pi^2}{12L^4}E^2_2({\tilde q}^{tor.})-\frac{1}{L^2R^2}\textrm{ . }
\end{eqnarray}
Using (\ref{eq:intann3}) and (\ref{eq:inttor3}) in (\ref{eq:Z_3}) we
find
\begin{eqnarray}\label{eq:S_6}
\nonumber\vev{S_6}^{ann.}&=&-4(D-2)(D-3)c_4\left[\frac{4\pi^3}{15L^4}E_4(\tilde{q}^{ann.})-\frac{64\pi^4
R}{L^5}H_{2,4}(\tilde{q}^{ann.})\right]
\\ \nonumber&&+[(D-2)^3(4c_6+2c_7)+(D-2)^2(12c_6+10c_7)
\\ \nonumber&&+(D-2)(8c_6+12c_7)]\times\left[\frac{\pi^2}{8 L^4}-\frac{\pi}{2 R L^3}+\frac{1}{2 R^2L^2}+O(\tilde{q})\right]\textrm{ , }
\\ \nonumber\vev{S_6}^{tor.}&=&-\frac{1}{L^2R^2}(c_6+c_7)[(D-2)^3+6(D-2)^2+8(D-2)]
\\ \nonumber&&-\left(\frac{\pi}{3RL^3}E_2({\tilde q}^{tor.})+\frac{\pi^2}{36L^4}E^2_2({\tilde q}^{tor.})\right)[2 c_7 (D-2)^3
\\ && +(D-2)^2(12c_6+14c_7)+(D-2)(24c_6+20c_7)]\textrm{ . }
\end{eqnarray}
Putting this result in \eqref{partsixd} and comparing to (\ref{eq:lus_corelator_5}), we are able to
find the deviations of the spectrum and overlap functions from the Nambu-Goto (NG) case. We
define for convenience: $\Delta E_n=E_n-E_n^{NG}$, $\Delta
c_i=c_i-c_i^{NG}$, where  $E_n^{NG}$ and $c_i^{NG}$ refer to the NG
spectrum and coefficients. By comparing the $\frac{1}{R^2L^2}$
term in the annulus partition function in (\ref{eq:S_6}) and
(\ref{eq:lus_corelator_5}), we find the following constraint :
\begin{eqnarray}\label{eq:compare2ann}
\nonumber
0 &=& (D-2)^3(4\Delta c_6+2\Delta  c_7)
+(D-2)^2(12\Delta c_6+10\Delta c_7)
\\ &&
+(D-2)(8\Delta c_6+12\Delta c_7)\ .
\end{eqnarray}
This constraint is enough to exclude any correction to the annulus partition
function coming from $\Delta c_{6,7}$. The coefficient $c_4$ is
not constrained.
To compute the spectrum
${[E_n^{c,3}]}_{ann.}$ we compare the $\frac{R}{L^5}$ terms. We see
that there is a deviation from Nambu-Goto only when $D>3$ and $c_4\neq0$. Since the
expansion of $H_{2,4}$ does not contain any constant term, we find that there are
deviations from the Nambu-Goto spectrum only for the excited states $n>0$.
Note that this does not teach us about deviations of each state $i$
from the Nambu-Goto form (in its energy or overlap functions), but just that the
sum over the states at each energy level should give the results that we
stated.

For the torus we find, by comparing the $\frac{1}{L^2R^2}$ and $\frac{1}{L^3R}$ terms and using the consistency of the Nambu-Goto expressions :
\begin{eqnarray}\label{eq:compare2tor}
\nonumber{[\Delta E_n^{c,3}]}_{tor.}&=&0 ,
\\ \nonumber 0&=&(\Delta c_6+ \Delta c_7)[(D-2)^3+6(D-2)^2+8(D-2)],
\\ \nonumber0&=&(D-2)^3(2\Delta c_7)+(D-2)^2(12\Delta c_6+14 \Delta c_7)
\\ &&+(D-2)(24\Delta c_6+20\Delta c_7)\quad.
\end{eqnarray}
On the torus, the partition function is exactly the Nambu-Goto partition function at this order,
and there are no contributions at all coming from $\Delta c_{4,6,7}$.
This result is different from what we got for the annulus, but there
is no contradiction since in the torus there are contributions from all
states, while in the annulus only some of the states contribute.
Therefore our results imply a relation between the sums of the
corrections to the energies of the states with the different
possible momenta at each level. For instance, at the level $n=2$,
there is one possible value $(N_L,N_R)=(1,1)$ for the annulus and 3
possible values for the torus $(N_L,N_R)=(0,2),(2,0),(1,1)$. Thus,
at this level we have the relation $\Delta E_{(1,1)}=-2\Delta
E_{(2,0)}=-2\Delta E_{(0,2)}$. If $D=3$ or $c_4=0$ then we obtain
from the annulus the additional relation $\Delta E_{(1,1)}=0$, and the
corrections to the energies of the $(2,0)$,$(0,2)$ states also sum
to zero.

The two constraints we find on $c_{6,7}$ from the torus are linearly independent of
each other, but not of the annulus constraint, so there is a consistent solution.
For $D=3$, there is
only one independent term which is constrained to have the Nambu-Goto
coefficient. For $D>3$ the general solution to the two constraints
turns out to be independent of $D$,
\begin{eqnarray}\label{eq:const3}
c_6=\frac{1}{16T^2},\quad
c_7=-\frac{1}{8T^2},
\end{eqnarray}
so that these coefficients also exactly agree with their Nambu-Goto values.

We summarize our results at this order:

For $D=3$ there is a single parameter in the effective action,
$c_6$ which is constrained to be the Nambu-Goto
coefficient. Thus, the effective action to this order is the same as
the Nambu-Goto action, and all energy levels should agree with the Nambu-Goto
levels up to order $1/L^5$.

For $D>3$, there are three independent terms : $c_4$ which is
unconstrained, and $c_{6,7}$ which have two constraints and agree with
their Nambu-Goto values.
The annulus partition function is
generally affected by $c_4$, but the ground state energy has no
deviations from Nambu-Goto at this order. The torus partition function at
this order is not affected by $c_4$, and it is
always equal to the Nambu-Goto partition function.

\section{Superstrings in confining backgrounds}
\label{sec:superstrings}

String theory in flat space-time is described, in the conformal
gauge, by a free worldsheet theory of massless degrees of freedom
(d.o.f), corresponding to the worldsheet fluctuations. In the
general, non-flat, case there are background fields with non-trivial
vacuum expectation values. These fields couple to the worldsheet
d.o.f and create interaction terms, so that the worldsheet theory is
no longer free.

In this section we will derive the worldsheet action for an infinite
confining string, in a specific class of holographic backgrounds. In
these backgrounds, the confining string sits at the minimal value of
a radial coordinate. As we will see, in the static gauge, the
worldsheet action in such a configuration will contain massive and
massless modes. We will then be able, in later sections, to define
an effective action which includes the massless fields only. This
action can be considered on different topologies, as in the previous section.

We begin with a description of the possible background fields and
heuristically describe their coupling to the worldsheet theory. This
discussion should make clear how general our analysis is, and which
backgrounds it fails to describe (examples of backgrounds which are
included in our analysis will be presented in the next section).
Then, we present the superstring action in these
backgrounds, which was derived in \cite{bib:action1, bib:action2}, and
we derive its Feynman rules. To avoid confusion, we
summarize our notations in appendix \ref{sec:appendixB}.

\subsection{The backgrounds}

The class of confining backgrounds which we consider (a sub-class of the general
confining backgrounds discussed in \cite{Kinar:1998vq}) contains a
cycle $S^{p'}$ (with the topology of a sphere) that vanishes smoothly at the minimal value of the
radial coordinate, $\rho_0$. At $\rho_0$ the warp factor, $f(\rho)$,
is minimal and so the string is forced to sit at this point. We can
write the radial direction and the $S^{p'}$ coordinates together in Cartesian
coordinates, $\rho\in {\bf R}^{p'+1}$, and expand the metric around the
minimal point $\rho_0$. Up to the order we will need in
$\Delta\rho=\rho-\rho_0$, the metric of the $X$ and $\rho$ coordinates is given by
(in the string frame)
\begin{eqnarray}
\frac{ds^2}{2\pi\alpha'}&=&f(\rho)dX^2+d\rho^2=f(\rho_0)
\left(1+\frac{f''(\rho_0)}{2f(\rho_0)}\Delta\rho^2\right)dX^2+d\rho^2.
\end{eqnarray}
The string is stretched along the $X$ directions, where the field theory lives.
The effective
string tension in $\alpha'$ units is $T=f(\rho_0)$. The coefficient
of the $\Delta\rho^2dX^2$ term is proportional to the curvature
$\cal{R}$ at the minimal point
,$\frac{f''(\rho_0)}{f(\rho_0)}\propto \frac{\cal{R}}{T}$, which is
positive. The metric should be smooth, so the warp factor is a
function of $\Delta\rho^2$ and there are no odd powers appearing in
the expansion.

Expanding the metric perturbatively near the minimal point is valid
when the curvature is small compared with the string tension,
${\cal{R}}<<T$. This limit often corresponds to a large 't Hooft
coupling limit in the dual field theories. In the static gauge, the
quadratic part of the warp factor will create worldsheet masses (and
interaction terms) for the $\rho$ coordinates. This is rather
intuitive, as the curvature suppresses excitations in the $\rho$
directions, and in the language of the worldsheet theory causes
these fields to become heavy.

There is usually also another compact $p$ dimensional subspace, with
a radius which scales like a power of $N$. As discussed in section
2, we will not take these coordinates into consideration. However,
these compact subspaces are stabilized by some flux which they carry,
and this flux will appear in our computations. Since the geometry is not
flat, it is convenient to choose for these coordinates an appropriate vielbein ($e^a\cdot
e^b=\delta^{ab}$) to work with. In this frame the volume form is
always proportional to the antisymmetric tensor, and is not
coordinate dependent.

The metric which we will use is thus (rescaling the $X$ coordinates,
renaming the $\rho$ coordinates $Y$, and renaming the curvature  term
$m^2_B$)
\begin{eqnarray}\label{eq:metric}
\frac{ds^2}{2\pi\alpha'}&=&(1+\frac{m^2_{B}}{2T}Y^2_B)\sum_{\xi=0}^{D-1}
dX^{\xi}dX_{\xi}+\sum_{B=D}^{D+N_B-1}dY_BdY_B+\frac{1}{2\pi\alpha'}\sum_{a=D+N_B}^{9}de^ade^a.
\end{eqnarray}
Here $N_B=p'+1$ is the number of massive scalars on the worldsheet.
The ten space-time coordinates, $Z^{\mu}$ ($\mu=0,\cdots,9$), include a
${\bf R}^D$ part, which is spanned by the $X$ coordinates and is
multiplied by the warp factor, and also the $Y$ and $e^{a}$
coordinates. Our $X$ and $Y$ coordinates are dimensionless. In our
computation we will integrate out at one-loop order the $Y$
coordinates and the massive fermions. This means that in our action
we only need to keep terms up to quadratic order in these fields.

We choose the static gauge fixing
$X^{\alpha}=\sqrt{T}\sigma^{\alpha}$ ($\alpha=0,1$), where $T$ is the
string tension. In
this gauge $m^2_{B}$ become the masses of the radial coordinates
$Y_B$, where $B$ is an index running over all these fields. Note
that in this gauge, the range of the dimensionless $X$ on (say) the
torus is $0\leq X^0\leq L\sqrt{T}$, $0\leq X^1\leq R\sqrt{T}$.

The second possible bosonic background is the NS-NS 2-form field. We
assume this field is not polarized in the ${\bf R}^D$ directions, and
therefore has no interaction with the $X$ coordinates to one loop order.
For $D>3$, this assumption follows from Lorentz invariance.

The background generally includes also various R-R field
strengths which couple to the worldsheet fermions through
the covariant derivative including terms
proportional to
$\bar{\Theta}F_{\mu_1\cdots\mu_p}\Gamma^{\mu_1\cdots\mu_p}\Theta$. For
each $p$-form we define
$\tilde{\Gamma}_{p}\equiv
\frac{1}{8p!}e^{\phi}F_{\mu_1\cdots\mu_p}\Gamma^{\mu_1\cdots\mu_p}$
(this has units of mass in our conventions). We
work under the assumption that $\tilde{\Gamma}_{p}$ can be expressed
with gamma matrices polarized orthogonally to the ${\bf R}^D$ directions.
There may be non-zero background fields in these directions, such as
the self-dual 5-form in the  original AdS/CFT correspondence
\cite{bib:adscft1}, which is polarized in all ten dimensions.
However, because of Lorentz invariance in the ${\bf R}^D$ directions,
$\tilde{\Gamma}_{p}$ is a sum over gamma matrices which either
contain all $D$ directions (and possibly other directions), or none.
Therefore, the gamma matrices which are polarized in the flat
directions can be expressed as gamma matrices polarized in
orthogonal directions using the chirality operator; e.g.
$\tilde{\Gamma}_{5}=F_{01234}\Gamma^{01234}+F_{56789}\Gamma^{56789}=(F_{01234}\Gamma^{11}+F_{56789})\Gamma^{56789}$.

To conclude, our analysis will include all the possible $p$-form
backgrounds which appear in IIA/B superstring theories, apart from
the $B$-field which, if polarized orthogonally to the field theory
dimensions, does not couple directly to the $X$ scalars, and it can
only give interaction terms which do not contribute in our one-loop
computation.

There may be more then one background $p$-form field present, and
their sum generates a mass and interaction terms for the fermions.
For simplicity we will start with the case of a single $p$-form
background. As we will discover, the fermionic action is identical
for all $p$-forms which we consider (up to some sign differences).
At this point, the addition of several forms will be trivial. In our
convention $\tilde{\Gamma}_{p}$ is a real matrix that has the
following symmetry and commutation properties with
$\Gamma^{i}$ ($i=0,\cdots,D-1$)
\begin{eqnarray}
\begin{array}{cccc}
& \tilde{\Gamma}^T_p=\tilde{\Gamma}_p \qquad & \tilde{\Gamma}_p^T=-\tilde{\Gamma}_p &\\
IIA\ ([\tilde{\Gamma}_p,\Gamma^i]=0)\qquad &p=4 \qquad &  p=2
\\IIB\ (\{\tilde{\Gamma}_p,\Gamma^i\}=0)\qquad &p=1,5 \qquad &p=3
\end{array}
\end{eqnarray}
Each $\tilde{\Gamma}_p$ matrix is symmetric (skew-symmetric), and
therefore by itself diagonalizable (block diagonalizable). As we will
see there are other matrices which multiply $\tilde{\Gamma}_p$, so
the final mass matrix is always skew-symmetric, and we can bring it
to a block diagonalized form, which is the standard form for Dirac mass
terms. For the total mass matrix to be brought into this form, we
need all the matrices which originate in different fluxes to commute
with each other. This occurs in all the examples which we discuss,
and should be the case in any well-defined worldsheet theory. We
stress that although the fermion and scalar masses come from
different background fields, they are always related through the
background equations of motion to give the sum rule (\ref{massequal}). This is
necessary for our results to be finite, as we will see in section
\ref{sec:corfun}. We will discuss separately the type IIA and type
IIB cases, and show that they lead to the same action in the
backgrounds we consider.

\subsection{Type IIA action}

We begin with the type IIA action in its Polyakov form, in the
Green-Schwarz (GS) formalism, to second order in
fermions \cite{bib:action1, bib:action2, bib:actionWTN}:
\begin{eqnarray}
\nonumber S_P&=&-\frac{1}{4\pi\alpha'}\int
d^2\sigma\{\sqrt{-h}h^{\alpha\beta}(\partial_{\alpha}Z^{\mu}\partial_{\beta}Z_{\mu}-2i\partial_{\alpha}Z^{\mu}\bar{\Theta}\Gamma_{\mu}D_{\beta}\Theta)
\\ \nonumber&&~~~~~~~~~~~~~~~~~~~+2i\epsilon^{\alpha\beta}\partial_{\alpha}Z^{\mu}\bar{\Theta}\Gamma^{11}\Gamma_{\mu}D_{\beta}\Theta\},
\\  D_{\alpha}&\equiv&\partial_{\alpha}+\sum_{p}\partial_{\alpha}Z^{\mu}\tilde{\Gamma}_p\Gamma_{\mu}Q_p
,\quad \{Q_{2}=\Gamma^{11} \textrm{ , } Q_{4}=I \}\quad.
\end{eqnarray}
Here the fields $Z^{\mu}$ are contracted with the metric $g_{\mu\nu}$ written
above (\ref{eq:metric}). $\Theta$ is a space-time Majorana fermion
with 32 real degrees of freedom off-shell (the Majorana condition is
taken such that the fermions are real variables). The gamma matrices
obey the general relation
$\{\Gamma_{\mu},\Gamma_{\nu}\}=2g_{\mu\nu}$. The worldsheet
directions are $\alpha,\beta=0,1$. Apart from diffeomorphism and
Weyl invariance, the action is kappa symmetric and reduces to the
familiar GS action in flat space-time. The matrices $Q_p$ are needed
when we write the action in this compact form, without an explicit
summation over two Weyl fermions as in \cite{bib:GSbook}. These
matrices ensure that the matrix sitting between the two fermions is
antisymmetric.

Since we are interested in describing the low-energy effective
action on a long string, the convenient gauge to work in is the
static gauge $X^{\alpha}=\sigma^{\alpha}\sqrt{T}$ ($\alpha=0,1$) \footnote{Note
that in this gauge
we still need to take into account the equation of motion of $X^{\alpha}$ which is not
automatically satisfied, and should be viewed as a constraint. At leading order the
constraint follows from the other equations of motion, so the leading non-trivial
constraint on physical states arises at 5-derivative order and involves four fields. It
would be interesting to find a different formalism in which the constraints are
automatically satisfied.}. This is
useful as we have a dimensionless parameter to expand in, which is
$k^i/\sqrt{T}$, where $k^i$ are the momenta of the fields on the
string. After this gauge fixing, we cannot completely gauge away the
worldsheet metric $h_{\alpha \beta}$. We therefore set the metric to its classical value, using
the equations of motion, and expand around this solution:
\begin{eqnarray}
\nonumber
K_{\alpha\beta}&\equiv&\partial_{\alpha}Z^{\mu}\partial_{\beta}Z_{\mu}-i(\partial_{\alpha}Z^{\mu}\bar{\Theta}\Gamma_{\mu}D_{\beta}\Theta+\alpha\leftrightarrow\beta)\textrm{
,}
\\ \nonumber h_{\alpha\beta}\sqrt{-K}&=&\sqrt{-h}K_{\alpha\beta}\textrm{ ,}
\\ h&\equiv&\det(h_{\alpha\beta}) \textrm{  ,  }  K\equiv \det(K_{\alpha\beta})\textrm{ .}
\end{eqnarray}
Integrating out the metric classically we obtain the following
Nambu-Goto like action
\begin{eqnarray}\label{eq:NG}
\nonumber S_{NG}&=&-\frac{1}{2 \pi \alpha'}\int
d^2\sigma\{\sqrt{-K}+S_2\} \textrm{ ,}
\\  S_2&=&i\epsilon^{\alpha\beta}\partial_{\alpha}Z^{\mu}\bar{\Theta}\Gamma^{11}\Gamma_{\mu}D_{\beta}\Theta\textrm{ .}
\end{eqnarray}
As mentioned earlier, we will not consider effects from loops of the
massless fields in the theory, and so we will ignore the worldsheet
metric fluctuations. In principle we still need to fix the Weyl
gauge symmetry, however this symmetry acts only on the metric $h_{\alpha \beta}$ so this
will not affect our computation. Note that the metric does not have a kinetic term, and
therefore will not contribute to the L\"uscher term.

We split our Majorana fermion into two space-time Weyl-Majorana
fermions:
\begin{eqnarray}\label{eq:split}
\Theta&=&\Theta^1+\Theta^2 \quad\textrm{,}\quad
\Gamma^{11}\Theta^I=(-1)^{I+1}\Theta^I ,\quad (I=1,2) \quad.
\end{eqnarray}
We write the gamma matrices as a product of gamma matrices in the 2 dimensional
worldsheet directions and the transverse 8 directions. In the
following, $\rho_{\alpha}(\gamma_i)$ are $2(8)$ dimensional gamma
matrices in flat space (see appendix \ref{sec:appendixB} for more details), and then we have
\begin{eqnarray}\label{eq:gammadef}
\nonumber \Gamma_{\alpha}&=&\sqrt{2\pi\alpha'}\rho_{\alpha} \otimes
I\quad\textrm{,}\quad
 \Gamma_{i}=\sqrt{2\pi\alpha'}\rho \otimes \gamma_i\quad\textrm{,}\quad
\\ \nonumber \tilde{\Gamma}_p&\equiv&\frac{1}{\sqrt{2\pi\alpha'}}\rho^{p}\otimes\tilde{\gamma}_p,
\\ \rho&\equiv&\rho_0\rho_1 \quad\textrm{,}\quad \gamma^c\equiv\gamma^2\cdots\gamma^9 \quad\textrm{,}\quad \Gamma^{11}\equiv\rho\otimes\gamma^c\quad\textrm{.}
\end{eqnarray}
Here $\rho$ is the worldsheet chirality operator, $\gamma^c$ is the
chirality operator in the transverse 8 dimensions, and $\Gamma^{11}$
is the 10 dimensional chirality operator. In (\ref{eq:gammadef}) we
wrote explicitly the vielbein
$e^{a(i)}_{b(j)}\propto\delta_{a(i),b(j)}\sqrt{2\pi\alpha'}$ (to
leading order in the radial variables $Y$). We fix the kappa
symmetry by identifying the worldsheet chirality of the fermions
with their space-time chirality. This means that both fermions have
positive 8 dimensional chirality in the transverse directions. In
our basis of gamma matrices the kappa fixing then takes the form
\begin{eqnarray}
\rho\Theta^I&=&\Gamma^{11}\Theta^I
\quad\textrm{,}\quad\gamma^c\Theta^I=\Theta^I\quad\textrm{.}\quad
\end{eqnarray}
This eliminates half of the degrees of freedom
carried by the fermions, and our space-time spinors now have a
worldsheet spinor index and an 8-dimensional space-time spinor index.
This kappa fixing is not entirely arbitrary, as we shall now
explain.

Our action (\ref{eq:NG}) is invariant under local fermionic
transformations which are generalizations of the flat-space kappa
symmetry transformation presented in \cite{bib:GSbook},
\begin{eqnarray}
\nonumber \delta_k\Theta&=&2i\Gamma\cdot\Pi_{\alpha}
P^{\alpha\beta}k_{\beta} \quad\textrm{,}\quad
\delta_kX^{\mu}=i\bar{\Theta}\Gamma^{\mu}\delta_k\Theta
\\ P^{\alpha\beta}&\equiv&\frac{1}{2}(K^{\alpha\beta}-\frac{\epsilon^{\alpha\beta}  \Gamma^{11}}{\sqrt{-K}})\quad\textrm{,}\quad
 \Pi_{\alpha}^{\mu}\equiv \partial_{\alpha}X^{\mu}-i\bar{\Theta}\Gamma^{\mu}\partial_{\alpha}\Theta\textrm{ . }
\end{eqnarray}
Here $k_{\alpha}$ is a Majorana fermion in space-time and
$P^{\alpha\beta}$ is a projector operator, affecting only half of
the d.o.f in $\Theta$. When we split our Majorana fermions in
(\ref{eq:split}), each fermion $\Theta^{1,2}$ is only affected by
the part of $k$ which has the opposite space-time chirality. We then
define $k=k^1+k^2$, where $\Gamma^{11}k^i=(-1)^{i}k^i$. In the
static gauge the projection operator $P^{\alpha\beta}$ becomes a
worldsheet chirality projector,
\begin{eqnarray}
\delta_k\Theta^1=-4i\Gamma_-k^1+O(Z^{\mu}) \quad\textrm{,}\quad
\delta_k\Theta^2=-4i\Gamma_+k^2+O(Z^{\mu}) \textrm{ , }
\end{eqnarray}
where $\Gamma_{\pm}=\frac{1}{2}(\Gamma_{0}\pm \Gamma_{1})$ refer to the lightcone
coordinates. Here $O(Z^{\mu})$ refers to corrections which involve other fields.
A simple gauge fixing which completely fixes our gauge freedom
is then \cite{bib:action2}
\begin{eqnarray}
\Gamma^{-}\Theta^1=0 \quad\textrm{,}\quad
\Gamma^{+}\Theta^2=0\textrm{ . }
\end{eqnarray}
One should make sure that the Fadeev-Popov determinant of our full gauge
fixing, namely the kappa symmetry and diffeomorphism gauge fixing, does not vanish.
This determinant will contain a kinetic term for $k_i$, which will
become ghost fields, and the $O(Z^{\mu})$ term in the kappa
transformation will produce interaction terms between the ghosts and the
other coordinates.
We have verified the consistency of our gauge-fixing, but we will not
describe this in detail here.

After taking the static gauge the $X^{\alpha}$ no longer appear in
the action, which involves (up to terms which do not contribute in
our one-loop computation)
\begin{eqnarray}\label{eq:NGaction}
\nonumber
K_{\alpha\beta}&=&-\delta_{\alpha,o}\delta_{\beta,0}2\pi\alpha'T+\delta_{\alpha,1}\delta_{\beta,1}2\pi\alpha'T+\partial_{\alpha}X^{i}\partial_{\beta}X_{i}+\partial_{\alpha}Y^B\partial_{\beta}Y_B
\\ \nonumber&+&i\sqrt{2\pi\alpha'T}(\Theta^1\rho_1(\rho_{\alpha}\partial_{\beta}+\rho_{\beta}\partial_{\alpha})\Theta^1-\Theta^2\rho_1(\rho_{\alpha}\partial_{\beta}+\rho_{\beta}\partial_{\alpha})\Theta^2)
\\ \nonumber&\pm&i\sqrt{2\pi\alpha'}T[\Theta^1\rho_1\{\rho_{\alpha},\rho_{\beta}\}\tilde{\gamma}_p\Theta^2-\Theta^2\rho_1\{\rho_{\alpha},\rho_{\beta}\}\tilde{\gamma}_p^T\Theta^1]
\\ \nonumber&\pm&2i\sqrt{2\pi\alpha'}\partial_{\alpha}X^i\partial_{\beta}X_i[\Theta^1\rho_1\tilde{\gamma}_p\Theta^2-\Theta^2\rho_1\tilde{\gamma}_p^T\Theta^1],
\\ \nonumber S_2&=&2i\sqrt{2\pi\alpha'T}(\Theta^1\partial_+\Theta^1+\Theta^2\partial_-\Theta^2)\pm 2i\sqrt{2\pi\alpha'}T(\Theta^1\tilde{\gamma}_p\Theta^2-\Theta^2\tilde{\gamma}_p^T\Theta^1)
\\ &\pm&\frac{i}{2}\sqrt{2\pi\alpha'}\epsilon^{\alpha\beta}\partial_{\alpha}X^i
\partial_{\beta}X^j(\Theta^1[\gamma_i,\gamma_j]\tilde{\gamma}_p\Theta^2+\Theta^2[\gamma_i,\gamma_j]\tilde{\gamma}_p^T\Theta^1)\quad.
\end{eqnarray}
The upper(lower) sign relates to the 4(2)-form background, and the
relative minus sign between the two backgrounds can be swallowed
into $\tilde{\gamma}_p$. Since our final results will only depend on
physical quantities, such as the masses squared of the
fermions (which are proportional to $\tilde{\gamma}_p^2$), this will
not make a difference. Thus, we can see that the two different possible type
IIA backgrounds give the same form to the action. When deriving
\eqref{eq:NGaction}, we used the worldsheet chirality of the fermions to express
$\rho^0$ as $\rho_1$ which is a positive matrix. One should notice
that there are no interactions linear in $X^i$. This is because of
our gauge choice for kappa symmetry fixing and the assumptions on
$\tilde{\Gamma}_p$ which includes no gamma matrices in the
worldsheet directions. Notice that fermions of opposite chirality
only mix through mass terms proportional to $\tilde{\gamma}_p$,
and so this mixing vanishes in flat
space as it should.

\subsection{Type IIB action}

Here we start directly with the NG like action for the type IIB case
\cite{bib:action1, bib:action2}, given by (\ref{eq:NG})
with
\begin{eqnarray}
\nonumber
K_{\alpha\beta}&\equiv&\partial_{\alpha}Z^{\mu}\partial_{\beta}Z_{\mu}-i(\partial_{\alpha}Z^{\mu}\bar{\Theta}^I\Gamma_{\mu}D_{\beta}^{IJ}\Theta^J+\alpha\leftrightarrow\beta),\quad(I,J=1,2),
\\ \nonumber S_2&=&i\epsilon^{\alpha\beta}\partial_{\alpha}Z^{\mu}\bar{\Theta}^I\rho^{IJ}\Gamma^{11}\Gamma_{\mu}D^{JK}_{\beta}\Theta^K\textrm{ ,}
\\  D^{IJ}_{\alpha}&\equiv&\partial_{\alpha}\delta^{IJ}+\partial_{\alpha}Z^{\mu}\tilde{\Gamma}_p\Gamma_{\mu}Q_p^{IJ}
,\quad  \{Q^{IJ}_{p=1,5}=\rho_0^{IJ} ,\quad
Q^{IJ}_{p=3}=-\rho_1^{IJ}\}.
\end{eqnarray}
In the type IIB case we have two fermions $\Theta^{1,2}$, both with positive
space-time chirality. One technical difference between the type IIA
and type IIB actions, is that the first is initially written for a
single Majorana fermion, while the latter is written in terms of two
fermions. This difference is because in the type IIA case we can use
the space-time chirality operator to distinguish the right and left
movers and so to write down the correct interactions. After
gauge-fixing and using the metric (\ref{eq:metric}), the determinant $K$
and the topological part $S_2$ are the same in the type IIB case as
in (\ref{eq:NGaction}), where the upper sign refers to the 1,5-form cases and the lower sign to the 3-form
case. The kappa symmetry fixing is
restricted to be the same as in the type IIA case, so that here the
worldsheet chirality of the fermions is not the same as their
space-time chirality, but as their 8-dimensional chirality :
\begin{eqnarray}
\Gamma^{11}\Theta^I=\Theta^I ,\quad
\rho\Theta^I=(-1)^{I+1}\Theta^I,\quad
\gamma^c\Theta^I=(-1)^{I+1}\Theta^I.
\end{eqnarray}
Again, we see that the action is the same for all flux sectors.

When we have several background fields, we will still have the same action, but
we should replace $\tilde{\gamma}_p$ by the sum
$\sum_{p}\tilde{\gamma}_p$. As we claimed before, this sum of
matrices can always be block diagonalized. Both for type IIA and for
type IIB we can re-express it in the following way, in terms of
projection operators $\tilde{\gamma}_F$ on mass eigenstates:
\begin{eqnarray}\label{eq:Projector}
 \sum_{p}\tilde{\gamma}_p=\sum_{F}\frac{m_F}{2\sqrt{T}}\tilde{\gamma}_F\quad ,
\quad {\rm tr}[\tilde{\gamma}_F^T\tilde{\gamma}_F]=\frac{1}{8}{\rm tr}[1]=2
,\quad(\tilde{\gamma}_F^T\tilde{\gamma}_F)^2=\tilde{\gamma}_F^T\tilde{\gamma}_F\quad.
\end{eqnarray}
These matrices obey the commutation relation
$[\tilde{\gamma}_F,\gamma^c]=0$ for the type IIA theory, and the
anti-commutation relation $\{\tilde{\gamma}_F,\gamma^c\}=0$ for type
IIB. The sum $\sum_{F}$ is over all massive fermions.
$\tilde{\gamma}_F^T\tilde{\gamma}_F$ is a projection operator in the
16-dimensional spinor space, projecting onto one on-shell fermionic
d.o.f. In this description we have eight worldsheet Dirac fermions with
(generally) different masses. An example of such a decomposition is given in
section \ref{ssec:examplesMN} where the Maldacena-Nu\~nez R-R
background is discussed.

\subsection{The Nambu-Goto determinant}

We now carry on with a derivative expansion of the action \eqref{eq:NG}.
In the rest of the section we will use the Einstein summation rule for the worldsheet coordinates.
We incorporate the space-time metric explicitly, and use $A\cdot A=\sum_{a}A_{a}A_{a}$, with the relevant indices for $X$ and $Y$ coordinates.
Our NG-like action contains a square root of the determinant
$K=K_{00}K_{11}-K_{01}K_{10}$. This can be consistently expanded in
powers of derivatives over the tension, $\frac{k}{\sqrt{T}}\ll1$.
We will perform this expansion up to sixth order in derivatives and up to second order in the massive fields $Y$ and $\Theta$ \footnote{Actually we will not require in this paper the six-derivative
terms with fermion couplings, so we will not write them down.}.
This will be enough to compute the one-loop effective action of the $X$'s by integrating out the massive fields.
We rescale the fermionic fields $\Theta\rightarrow(\frac{2\pi\alpha'}{64 T})^{\frac{1}{4}}\Theta$.
The spinors $\Theta^{1},\Theta^2$ are Weyl spinors, and have only one d.o.f on the worldsheet ($\Theta^1=\left(\begin{array}{c} \theta^1 \\0 \end{array}\right)$,
$\Theta^2=\left(\begin{array}{c} 0 \\ \theta^2 \end{array}\right)$). From here on we write our action in terms of $\theta^{1}$ and $\theta^2$,
which are worldsheet scalars. The elements of $K$ are given by
\begin{eqnarray}
\nonumber \frac{K_{00}}{2
\pi\alpha'}&=&(-T+\partial_0X\cdot\partial_0X)(1+\frac{m^2_{B}}{2T}Y^2_B)+\partial_0
Y\cdot \partial_0 Y
\\ \nonumber&-&\frac{i}{4}(\theta^1\partial_0\theta^1+\theta^2\partial_0\theta^2)+\frac{im_F}{8}(\theta^1\tilde{\gamma}_F\theta^2-\theta^2\tilde{\gamma}_F^T\theta^1)(\frac{1}{T}\partial_0X\cdot\partial_0X-1),
\\ \nonumber
\frac{K_{11}}{2\pi\alpha'}&=&(T+\partial_1X\cdot\partial_1X)(1+\frac{m^2_{B}}{2T}Y^2_B)+\partial_1
Y\cdot \partial_1 Y
\\ \nonumber&+&\frac{i}{4}(\theta^1\partial_1\theta^1-\theta^2\partial_1\theta^2)+\frac{im_F}{8}(\theta^1\tilde{\gamma}_F\theta^2-\theta^2\tilde{\gamma}_F^T\theta^1)(\frac{1}{T}\partial_1X\cdot\partial_1X+1),
\\ \nonumber
\frac{K_{01}}{2\pi\alpha'}&=&\partial_0X\cdot\partial_1X(1+\frac{m^2_{B}}{2T}Y^2_B)+\partial_0
Y\cdot \partial_1 Y +\frac{i}{4}(\theta^1\partial_-\theta^1+\theta^2\partial_+\theta^2)
\\ &+&\frac{im_F}{8T}\partial_0X\cdot\partial_1X(\theta^1\tilde{\gamma}_F\theta^2-\theta^2\tilde{\gamma}_F^T\theta^1),
\end{eqnarray}
and its determinant (to the order we need) by
\begin{eqnarray}
\nonumber
&&\frac{-K}{(2\pi\alpha'T)^2}=1+\frac{1}{T}\partial_{\alpha}X\cdot
\partial^{\alpha}X+\frac{1}{T}\partial_{\alpha}Y\cdot
\partial^{\alpha}Y+\frac{m^2_{B}}{T}Y^2_B+\frac{i}{2T}(\theta^1\partial_+\theta^1+\theta^2\partial_-\theta^2)
\\ &&+\frac{im_F}{4T}(\theta^1\tilde{\gamma}_F\theta^2-\theta^2\tilde{\gamma}_F^T\theta^1)
\\ \nonumber&&+\frac{1}{T^2}\partial_{\alpha}X\cdot\partial^{\alpha}X\{\partial_{\beta}Y\cdot\partial^{\beta}Y+m^2_BY^2_B+\frac{i}{2}(\theta^1\partial_+\theta^1+\theta^2\partial_-\theta^2)+\frac{im_F}{4}(\theta^1\tilde{\gamma}_F\theta^2-\theta^2\tilde{\gamma}_F^T\theta^1)\}
\\ \nonumber&&-\frac{1}{T^2}\partial_{\alpha}X\cdot\partial_{\beta}X\partial^{\alpha}Y\cdot\partial^{\beta}Y-\frac{i}{2T^2}\partial_{\alpha}X\cdot\partial_+X\theta^1\partial^{\alpha}\theta^1-\frac{i}{2T^2}\partial_{\alpha}X\cdot\partial_-X\theta^2\partial^{\alpha}\theta^2
\\ \nonumber &&+\frac{1}{T^3}((\partial^{\alpha}X\cdot\partial_{\alpha}X)^2-\partial_{\alpha}X\cdot\partial_{\beta}X\partial^{\alpha}X\cdot\partial^{\beta}X)\{\frac{T}{2}+\frac{m_B^2}{2}Y^2_B+\frac{i m_F }{8}(\theta^1\tilde{\gamma}_F\theta^2-\theta^2\tilde{\gamma}_F^T\theta^1)\}\ .
\end{eqnarray}
Already at this stage we can see that in the static gauge, we have a canonical kinetic term for the fermions, and part of the interactions became mass terms for the fermions and scalars.

Our first step will be to expand the square root in the action (\ref{eq:NG}) in powers of $-\frac{K}{(2\pi\alpha'T)^2}-1$, to the order we are interested in:
\begin{eqnarray}
\nonumber S&=&-T\int
d^2\sigma\{1+\frac{1}{2}(\frac{-K}{(2\pi\alpha'T)^2}-1)-\frac{1}{8}(\frac{-K}{(2\pi\alpha'T)^2}-1)^2+\frac{1}{16}(\frac{-K}{(2\pi\alpha'T)^2}-1)^3
\\ &-&\frac{5}{128}(\frac{-K}{(2\pi\alpha'T)^2}-1)^4+\frac{1}{2\pi\alpha'T}S_2\}\quad.
\end{eqnarray}
Keeping only terms we are interested in, the powers in $S$ are given by:
\begin{eqnarray}
\nonumber&&(\frac{-K}{(2\pi\alpha'T)^2}-1)^2=\frac{2}{T^2}\partial_{\alpha}X\cdot
\partial^{\alpha}X\{\partial_{\beta}Y\cdot
\partial^{\beta}Y+m^2_BY^2_B
\\ \nonumber&&-\frac{i}{2}(\theta^1\partial_+\theta^1+\theta^2\partial_-\theta^2)+\frac{im_F}{4}(\theta^1\tilde{\gamma}_F\theta^2-\theta^2\tilde{\gamma}_F^T\theta^1)\}
-\frac{2}{T^3}\partial^{\gamma}X\cdot\partial_{\gamma}X\partial_{\alpha}X\cdot\partial_{\beta}X\partial^{\alpha}Y\cdot\partial^{\beta}Y
\\ \nonumber&&-\frac{1}{T^3}\partial^{\gamma}X\cdot\partial_{\gamma}X\{\partial_{\alpha}X\cdot\partial_+X\theta^1\partial^{\alpha}\theta^1+\partial_{\alpha}X\cdot\partial_-X\theta^2\partial^{\alpha}\theta^2\}
+\frac{1}{T^2}(\partial^{\alpha}X\cdot\partial_{\alpha}X)^2
\\
\nonumber&&+\frac{1}{T^3}(3(\partial^{\alpha}X\cdot\partial_{\alpha}X)^2-\partial_{\alpha}X\cdot\partial_{\beta}X\partial^{\alpha}X\cdot\partial^{\beta}X)\{\partial_{\gamma}Y\cdot \partial^{\gamma}Y+m^2_BY^2_B
\\ \nonumber&&+\frac{i}{2}(\theta^1\partial_+\theta^1+\theta^2\partial_-\theta^2)+\frac{im_F}{4}(\theta^1\tilde{\gamma}_F\theta^2-\theta^2\tilde{\gamma}_F^T\theta^1)\}
\\ \nonumber&&+\frac{1}{T^4}((\partial_{\alpha}X\cdot \partial^{\alpha}X)^3-\partial_{\gamma}X\cdot \partial^{\gamma}X\partial_{\alpha}X\cdot \partial_{\beta}X\partial^{\alpha}X\cdot \partial^{\beta}X)(T+\partial_{\delta}Y\cdot\partial^{\delta}Y+2m^2_BY_B^2)
\\ &&-\frac{1}{T^4}((\partial_{\alpha}X\cdot \partial^{\alpha}X)^2-\partial_{\alpha}X\cdot \partial_{\beta}X\partial^{\alpha}X\cdot \partial^{\alpha}X)\partial_{\gamma}X\cdot
\partial_{\delta}X\partial^{\gamma}Y\cdot\partial^{\delta}Y,
\end{eqnarray}

\begin{eqnarray}
\nonumber&&(\frac{-K}{(2\pi\alpha'T)^2}-1)^3=\frac{3}{T^3}(\partial_{\alpha}X\cdot
\partial^{\alpha}X)^2\{\partial_{\beta}Y\cdot
\partial^{\beta}Y+m^2_BY^2_B
\\ \nonumber&&\qquad\qquad\qquad\qquad\qquad\qquad\qquad+\frac{i}{2}(\theta^1\partial_+\theta^1+\theta^2\partial_-\theta^2)+\frac{im_F}{4}(\theta^1\tilde{\gamma}_F\theta^2-\theta^2\tilde{\gamma}_F^T\theta^1)\}
\\ \nonumber&&+\frac{1}{T^4}(\partial_{\alpha}X\cdot \partial^{\alpha}X)^3(T+6\partial_{\beta}Y\cdot \partial^{\beta}Y+6m^2_BY^2_B)-\frac{3}{T^4}(\partial^{\alpha}X\cdot\partial_{\alpha}X)^2\partial_{\beta}X\cdot\partial_{\gamma}X\partial^{\beta}Y\cdot\partial^{\gamma}Y
\\&&-\frac{3}{T^4}(\partial^{\gamma}X\cdot\partial_{\gamma}X\partial_{\alpha}X\cdot\partial_{\beta}X\partial^{\alpha}X\cdot\partial^{\beta}X)(\partial_{\delta}Y\cdot \partial^{\delta}Y+m^2_BY^2_B),
\end{eqnarray}
and
\begin{eqnarray}
(\frac{-K}{(2\pi\alpha'T)^2}-1)^4&=&\frac{4}{T^4}(\partial_{\alpha}X\cdot
\partial^{\alpha}X)^3\{\partial_{\beta}Y\cdot
\partial^{\beta}Y+m^2_BY^2_B\}\quad.
\end{eqnarray}
The full action to order $O((\del X)^6Y^2,(\del X)^4\theta^2)$ is then
\begin{eqnarray}
\label{eq:10daction}
\nonumber S&=&-\int d^2\sigma\{T+\frac{1}{2}\partial_{\alpha}X\cdot
\partial^{\alpha}X+\frac{1}{2}\partial_{\alpha}Y\cdot
\partial^{\alpha}Y+\frac{1}{2}m^2_BY^2_B
\\ \nonumber&+&\frac{i}{2}(\theta^1\partial_+\theta^1+\theta^2\partial_-\theta^2)+\frac{im_F}{4}(\theta^1\tilde{\gamma}_F\theta^2-\theta^2\tilde{\gamma}_F^T\theta^1)
\\
\nonumber&+&\frac{1}{4T}\partial_{\alpha}X\cdot\partial^{\alpha}X[\partial_{\beta}Y\cdot\partial^{\beta}Y+m^2_BY_B^2+\frac{i}{2}(\theta^1\partial_+\theta^1+\theta^2\partial_-\theta^2)+\frac{im_F}{4}(\theta^1\tilde{\gamma}_F\theta^2-\theta^2\tilde{\gamma}_F^T\theta^1)]
\\ \nonumber&-&\frac{1}{2T}\partial_{\alpha}X\cdot\partial_{\beta}X\partial^{\alpha}Y\cdot\partial^{\beta}Y-\frac{i}{4T}\partial_{\alpha}X\cdot\partial_+X\theta^1\partial^{\alpha}\theta^1-\frac{i}{4T}\partial_{\alpha}X\cdot\partial_-X\theta^2\partial^{\alpha}\theta^2
\\ \nonumber&+&\frac{1}{4T^2}\partial^{\gamma}X\cdot\partial_{\gamma}X\partial_{\alpha}X\cdot\partial_{\beta}X\partial^{\alpha}Y\cdot\partial^{\beta}Y
\\
\nonumber&+&\frac{i}{8T^2}\partial^{\gamma}X\cdot\partial_{\gamma}X[\partial_{\alpha}X\cdot\partial_+X\theta^1\partial^{\alpha}\theta^1+\partial_{\alpha}X\cdot\partial_-X\theta^2\partial^{\alpha}\theta^2]
\\ \nonumber&+&\frac{1}{T^2}(\partial^{\alpha}X\cdot\partial_{\alpha}X)^2[\frac{T}{8}-\frac{3}{16}\partial_{\beta}Y\cdot \partial^{\beta}Y+\frac{1}{16}m^2_BY_B^2-\frac{3i}{32}(\theta^1\partial_+\theta^1+\theta^2\partial_-\theta^2)
\\ \nonumber&+&\frac{im_F}{64}(\theta^1\tilde{\gamma}_F\theta^2-\theta^2\tilde{\gamma}_F^T\theta^1)]
\\
\nonumber&+&\frac{1}{T^2}(\partial_{\alpha}X\cdot\partial_{\beta}X\partial^{\alpha}X\cdot\partial^{\beta}X)[-\frac{T}{4}+\frac{1}{8}\partial_{\gamma}Y\cdot \partial^{\gamma}Y-\frac{1}{8}m^2_BY_B^2+\frac{i}{16}(\theta^1\partial_+\theta^1+\theta^2\partial_-\theta^2)
\\ \nonumber&-&\frac{im_F}{32}(\theta^1\tilde{\gamma}_F\theta^2-\theta^2\tilde{\gamma}_F^T\theta^1)]
+\frac{im_F}{32T}\epsilon^{\alpha\beta}\partial_{\alpha}X^i\partial_{\beta}X^j(\theta^1[\gamma_i,\gamma_j]\tilde{\gamma}_F\theta^2+\theta^2[\gamma_i,\gamma_j]\tilde{\gamma}_F^T\theta^1)
\\ \nonumber&+&\frac{1}{32T^3}(\partial_{\alpha}X\cdot \partial^{\alpha}X)^3(-2T+3\partial^{\beta}Y\cdot\partial_{\beta}Y-m^2_BY_B^2) -\frac{1}{16T^3}(\partial_{\alpha}X\cdot \partial^{\alpha}X)^2\partial_{\beta}X\cdot\partial_{\gamma}X\partial^{\beta}Y\cdot\partial^{\gamma}Y
\\ \nonumber&+&\frac{1}{16T^3}\partial_{\alpha}X\cdot \partial^{\alpha}X\partial_{\beta}X\cdot\partial_{\gamma}X\partial^{\beta}X\cdot\partial^{\gamma}X(2T-\partial_{\delta}Y\partial^{\delta}Y+m^2_BY^2_B)
\\ &-&\frac{1}{8T^3}\partial_{\alpha}X\cdot\partial_{\beta}X\partial^{\alpha}X\cdot\partial^{\beta}X\partial_{\gamma}X\cdot \partial_{\delta}X\partial^{\gamma}Y\cdot\partial^{\delta}Y\}\quad.
\end{eqnarray}
Note that in terms such as
$m_F\theta^1\tilde{\gamma}_F\theta^2$ there is an
implicit sum over the massive fermions $F=1,\cdots,N_F$.

\subsection{Feynman rules}

The action \eqref{eq:10daction} above yields the following Feynman rules
in Minkowski space, where each vertex is accompanied by a momentum delta function and
each momentum integral (which we perform in Euclidean space)
should be multiplied by $i$ due to Wick
rotation.

\subsubsection{Propagators}

For each fermion with mass $m_F$ the propagator is
\begin{eqnarray}\label{eq:ferprop}
\nonumber G_{ab}(k)&=&<\theta^a(k)\theta^b(-k)>
\\&=& \frac{4}{p^2+m_F^2}\left[{\left(\begin{array}{cc}
-ip_- & \frac{m_F}{2}\tilde{\gamma}_F \\
 -\frac{m_F}{2}\tilde{\gamma}_F^T &  -ip_+
\end{array}\right)} \frac{1}{2}
{\left(\begin{array}{cc}
1+\gamma^c &  \\
 &  1\pm\gamma^c
\end{array}\right)}\right]_{ab}
\times \tilde{\gamma}_F^T\tilde{\gamma}_F\textrm{ ,}
\end{eqnarray}
where the indices $a,b=1,2$ indicate the entry for the $2\times2$
matrix.
Notice the two projection operators in the fermion propagator. The
first is to project the fermion onto the proper 8 dimensional
chirality, according to the kappa symmetry fixing (upper sign for type IIA
backgrounds and lower sign for type IIB). The second projector
projects onto one d.o.f only, with a specific mass $m_F$.

For a scalar with mass $m_B$ the propagator is
\begin{eqnarray}\label{eq:bosprop}
<Y_a(k)Y_b(-k)>&=&\frac{-i}{k^2+m_B^2}\delta_{ab}\textrm{ .}
\end{eqnarray}

\subsubsection{Interactions}

In our conventions, we put into the vertices we write below the sum over
permutations of $X$ fields but not of other fields. We will
consistently take this into account in the symmetry factors of the
loop diagrams. The vertices including only scalar fields are:
\begin{eqnarray}\label{eq:tree4pt}
\nonumber&&<X_i(k_1)X_j(k_2)X_k(k_3)X_l(k_4)>=\delta_{ij}\delta_{kl}\{-\frac{i}{T}k_1\cdot
k_2k_3\cdot k_4+\frac{i}{T}k_1\cdot k_3k_2\cdot
k_4+\frac{i}{T}k_1\cdot k_4k_2\cdot k_3\}
\\ &&+\delta_{il}\delta_{jk}(k_1\leftrightarrow k_3)+\delta_{ik}\delta_{jl}(k_1\leftrightarrow k_4),
\end{eqnarray}
\begin{eqnarray}\label{eq:tree6pt}
\nonumber&&<X_i(k_1)X_j(k_2)X_k(k_3)X_l(k_4)X_m(k_5)X_n(k_6)>=\delta_{ij}\delta_{kl}\delta_{mn}\{-\frac{3i}{T^2}k_1\cdot
k_2k_3\cdot k_4k_5\cdot k_6
\\ \nonumber&&+\frac{i}{T^2}[k_1\cdot k_2(k_3\cdot k_5 k_4\cdot k_6+k_4\cdot k_5 k_3\cdot k_6)+(1;2)\leftrightarrow(3;4)+(1;2)\leftrightarrow(5;6)]\}
\\ \nonumber&&+\delta_{ij}\delta_{km}\delta_{ln}(k_3\leftrightarrow k_6)+\delta_{ij}\delta_{kn}\delta_{lm}(k_3\leftrightarrow k_5)
+\delta_{kl}\delta_{im}\delta_{jn}(k_1\leftrightarrow
k_6)+\delta_{kl}\delta_{in}\delta_{jm}(k_1\leftrightarrow k_5)
\\ &&+\textrm{ more permutations on }[(i;1),(j;2),(k;3),(l;4),(m;5),(n;6)],
\end{eqnarray}
\begin{eqnarray}
\nonumber&& <X_i(k_1)X_j(k_2)Y_a(k_3)Y_b(k_4)>=\delta_{ij}\delta_{ab}\frac{i}{2T}\{(-k_1\cdot
k_2k_3\cdot k_4+k_1\cdot k_3k_2\cdot k_4+k_1\cdot k_4k_2\cdot k_3)
\\ &&+m_B^2k_1\cdot k_2\},
\end{eqnarray}
\begin{eqnarray}
\nonumber&&
<X_i(k_1)X_j(k_2)X_k(k_3)X_l(k_4)Y_a(k_5)Y_b(k_6)>=\delta_{ab}\delta_{ij}\delta_{kl}\frac{i}{2T^2}\{k_1\cdot
k_2(k_3\cdot k_5k_4\cdot k_6+k_3\cdot k_6k_4\cdot k_5)
\\ \nonumber&&+k_3\cdot k_4(k_1\cdot k_5k_2\cdot k_6+k_1\cdot k_6k_2\cdot k_5)-3k_1\cdot k_2k_3\cdot k_4k_5\cdot k_6+k_1\cdot k_3k_2\cdot k_4k_5\cdot k_6
\\ \nonumber&&+k_1\cdot k_4k_2\cdot k_3k_5\cdot k_6+m_B^2(-k_1\cdot k_2k_3\cdot k_4+k_1\cdot k_3k_2\cdot k_4+k_1\cdot k_4k_2\cdot
k_3)\}
\\ &&+\delta_{ab}\delta_{il}\delta_{jk}(k_1\leftrightarrow k_3)+\delta_{ab}\delta_{ik}\delta_{jl}(k_1\leftrightarrow k_4),
\end{eqnarray}
\begin{eqnarray} \nonumber&&<X_i(k_1)X_j(k_2)X_k(k_3)X_l(k_4)X_m(k_5)X_n(k_6)Y_a(k_7)Y_b(k_8)>=\delta_{ij}\delta_{kl}\delta_{mn}\delta_{ab}
\\ \nonumber&&\times\frac{i}{2T^3}\{k_1\cdot k_2k_3\cdot k_4k_5\cdot k_6(-9k_7\cdot k_8-3m_B^2)
\\ \nonumber&&+[k_3\cdot k_4 k_5\cdot k_6 (k_1\cdot k_7k_2\cdot k_8+k_2\cdot k_7k_1\cdot k_8)+(1;2)\leftrightarrow(3;4)+(1;2)\leftrightarrow(5;6)]
\\ \nonumber&&+(k_7\cdot k_8+m_B^2)[k_1\cdot k_2(k_3\cdot k_5 k_4\cdot k_6+k_3\cdot k_6 k_4\cdot k_5)+(1;2)\leftrightarrow(3;4)+(1;2)\leftrightarrow(5;6)]
\\ \nonumber&&+[(k_5\cdot k_3 k_6\cdot k_4+k_5\cdot k_4k_6\cdot k_3)(k_1\cdot k_7k_2\cdot k_8+k_1\cdot k_8k_2\cdot k_7)+(1;2)\leftrightarrow(3;4)+(1;2)\leftrightarrow(5;6)]\}
\\ &&+\textrm{ permutations on }[(i;1),(j;2),(k;3),(l;4),(m;5),(n;6)]\quad.
\end{eqnarray}
The vertices involving fermions are, using
$k_i\times k_j\equiv \epsilon^{\alpha\beta}k_{i\alpha}k_{j\beta}$,
\begin{eqnarray}
\nonumber &&<X_i(k_1)X_j(k_2)\theta^1(k_3)\theta^1(k_4)>=\delta_{ij}\frac{i}{8T}(-k_1\cdot k_2({k_4}_+-{k_3}_+)+k_2\cdot (k_4-k_3){k_1}_+
\\ &&+k_1\cdot (k_4-k_3){k_2}_+),
\end{eqnarray}
\begin{eqnarray}
\nonumber &&<X_i(k_1)X_j(k_2)\theta^2(k_3)\theta^2(k_4)>=\delta_{ij}\frac{i}{8T}(-k_1\cdot k_2({k_4}_--{k_3}_-)+k_2\cdot (k_4-k_3){k_1}_-
\\ &&+k_1\cdot (k_4-k_3){k_2}_-),
\end{eqnarray}
\begin{eqnarray}
&&<X_i(k_1)X_j(k_2)\theta^1(k_3)\theta^2(k_4)>=-\frac{m_F\tilde{\gamma}_F}{8T}k_1\cdot
k_2-\frac{m_F[\gamma_i,\gamma_j]\tilde{\gamma}_F}{16T}k_1\times k_2,
\\ \nonumber\\
&&<X_i(k_1)X_j(k_2)\theta^2(k_3)\theta^1(k_4)>=\frac{m_F\tilde{\gamma}^T_F}{8T}k_1\cdot
k_2-\frac{m_F[\gamma_i,\gamma_j]\tilde{\gamma}^T_F}{16T}k_1\times k_2,
\end{eqnarray}
\begin{eqnarray}
\nonumber&&
<X_i(k_1)X_j(k_2)X_k(k_3)X_l(k_4)\theta^1(k_5)\theta^1(k_6)>=\delta_{ij}\delta_{kl}\frac{i}{8T^2}\{k_1\cdot k_2k_3\cdot (k_6-k_5){k_{4}}_+
\\ \nonumber&&+k_1\cdot k_2k_4\cdot (k_6-k_5){k_{3}}_++k_3\cdot k_4k_1\cdot (k_6-k_5){k_{2}}_++k_3\cdot k_4k_2\cdot (k_6-k_5){k_{1}}_+
\\ \nonumber&&-3k_1\cdot k_2k_3\cdot k_4(k_{6+}-k_{5+})+k_1\cdot k_3k_2\cdot k_4(k_{6+}-k_{5+}),
+k_1\cdot k_4k_2\cdot k_3(k_{6+}-k_{5+})\}
\\ &&+\delta_{il}\delta_{jk}(k_1\leftrightarrow k_3)+\delta_{ik}\delta_{jl}(k_1\leftrightarrow k_4),
\end{eqnarray}
\begin{eqnarray}
\nonumber&&
<X_i(k_1)X_j(k_2)X_k(k_3)X_l(k_4)\theta^2(k_5)\theta^2(k_6)>=\delta_{ij}\delta_{kl}\frac{i}{8T^2}\{k_1\cdot
k_2k_3\cdot (k_6-k_5){k_{4}}_-
\\ \nonumber&&+k_1\cdot k_2k_4\cdot (k_6-k_5){k_{3}}_-+k_3\cdot k_4k_1\cdot (k_6-k_5){k_{2}}_-+k_3\cdot k_4k_2\cdot (k_6-k_5){k_{1}}_-
\\ \nonumber&&-3k_1\cdot k_2k_3\cdot k_4(k_{6-}-k_{5-})+k_1\cdot k_3k_2\cdot k_4(k_{6-}-k_{5-})
+k_1\cdot k_4k_2\cdot k_3(k_{6-}-k_{5-})\}\\ &&+\delta_{il}\delta_{jk}(k_1\leftrightarrow k_3)+\delta_{ik}\delta_{jl}(k_1\leftrightarrow k_4),
\end{eqnarray}
\begin{eqnarray}
\nonumber&&
<X_i(k_1)X_j(k_2)X_k(k_3)X_l(k_4)\theta^1(k_5)\theta^2(k_6)>=\delta_{ij}\delta_{kl}\frac{m_F\tilde{\gamma}_F}{8T^2}(k_1\cdot
k_2k_3\cdot k_4
\\ &&-k_1\cdot k_3k_2\cdot k_4-k_1\cdot k_4k_2\cdot k_3)+\delta_{il}\delta_{jk}(k_1\leftrightarrow k_3)+\delta_{ik}\delta_{jl}(k_1\leftrightarrow k_4),
\end{eqnarray}
\begin{eqnarray}
\nonumber&&
<X_i(k_1)X_j(k_2)X_k(k_3)X_l(k_4)\theta^2(k_5)\theta^1(k_6)>=-\delta_{ij}\delta_{kl}\frac{m_F\tilde{\gamma}_F^T}{8T^2}(k_1\cdot
k_2k_3\cdot k_4
\\ &&-k_1\cdot k_3k_2\cdot k_4-k_1\cdot k_4k_2\cdot k_3)+\delta_{il}\delta_{jk}(k_1\leftrightarrow k_3)+\delta_{ik}\delta_{jl}(k_1\leftrightarrow k_4)\quad.
\end{eqnarray}

\section{Examples}\label{sec:examples}

In this section we review some known confining backgrounds which have a
dual gauge theory interpretation. All of these examples are special
cases of the general background which we analyze in this paper.
We provide a very short description for each
background, followed by a derivation of the physical parameters of
the background (scalar and fermion masses). In all of these
examples, one can see that the sum rule (\ref{massequal}) holds.

\subsection{Witten background for $D=3$}\label{ssec:examplesWN3}

A confining theory related to pure $SU(N)$ Yang-Mills (YM) theory in
3 dimensions was proposed in \cite{bib:backgrounds1}. The approach
taken there was to start with the AdS$_5$/CFT$_4$ duality and to
compactify the conformal theory on a circle with radius $R_0$,
taking anti-periodic boundary conditions for the fermions. This
breaks supersymmetry explicitly, and the fermions
all have a mass proportional to $1/R_0$.
The 3 dimensional coupling is
$g_3=g^2_4N/R_0$. The pure YM theory is obtained in the limit
$g^2_4N\rightarrow0$, $R_0\rightarrow0$, $g_3$ fixed. However, with
present knowledge, we can only analyze the gravity side at small
curvature, which implies $g^2_4N>>1$. Therefore the theory we
analyze is dual to a strongly coupled 3 dimensional theory (which
becomes four dimensional at the scale $1/R_0$; this turns out to also
be the scale of the mass gap at strong coupling).

On the gravity side, the theory is a type IIB superstring theory.
%
The background is given \cite{bib:backgrounds1} by a metric and a five-form,
\begin{eqnarray}
\nonumber\frac{ds^2}{R^2}&=&(u^2-\frac{u_0^4}{u^2})d\tau^2+(u^2-\frac{u_0^4}{u^2})^{-1}du^2+u^2\sum_{i=0}^{2}dX_i^2+d\Omega_5^2,\quad
\\ F_5&=&16\pi N\alpha'^2\omega_5  ,\quad R^2=\sqrt{4\pi g_s N}\alpha' ,\quad \int_{S^5}\omega_5=\pi^3,\quad e^{\phi}=g_s,
\end{eqnarray}
where $u_0$ is related to the periodicity of the circle coordinate
$\tau$, and $\omega_5$ is the volume form on a 5-sphere with unit
radius. The $\tau$ circle shrinks smoothly at $u = u_0$, ending the space.
In this background, as we will show in the next subsections,
there are six massless scalars, two massive scalars with mass
$\frac{m^2_B}{T}=\sqrt{\frac{16\pi}{g_s N}}$ and eight massive
fermions with mass $\frac{m^2_F}{T}=\sqrt{\frac{\pi}{g_s N}}$.

\subsubsection{Scalar masses}

We take the limit $u \to u_0$ :
\begin{eqnarray}\label{eq:WN3}
\nonumber &&u=u_0(1+\frac{2\pi \alpha'}{R^2}\rho^2+O(\rho^4)),\quad
\\ &&\frac{ds^2}{2\pi\alpha'}=(1+\frac{4\pi\alpha'}{R^2}\rho^2)\sum_{i=0}^{2}dX_i^2+\rho^2d\tau^2+d\rho^2+\frac{R^2}{2\pi\alpha'}d\Omega_5^2,\quad
\end{eqnarray}
where we rescaled $X$ and $\tau$. Comparing the last equation in
(\ref{eq:WN3}) with (\ref{eq:metric}) we find two massive scalars of
mass $\frac{m^2_B}{T}=\frac{8\pi\alpha'}{R^2}=\sqrt{\frac{16\pi}{g_s
N}}$, which are the two radial directions ($\rho$ and $\tau$), and six massless
transverse scalars coming from the $X$ and $S^5$ directions.

\subsubsection{Fermion masses}
The 5-form in this background couples to the fermions, as described
in \cite{bib:action1, bib:action2}, so that the covariant derivative
on the worldsheet is:
\begin{eqnarray}
\nonumber
D_{\alpha}^{IJ}&=&\partial_{\alpha}\delta^{IJ}+\frac{e^{\phi}}{16\cdot
5!}F_{abcde}\Gamma^{abcde}\partial_{\alpha}Z^{\mu}\Gamma_{\mu}Q_5^{IJ}
=\partial_{\alpha}\delta^{IJ}+
\sqrt{2\pi\alpha'} g_s \frac{2\pi
N\alpha'^2}{R^5}\rho\gamma^{56789}\partial_{\alpha}Z\cdot \Gamma
Q_5^{IJ}
\\ \nonumber &=&\partial_{\alpha}\delta^{IJ}+\frac{1}{2}\left(\frac{\pi}{g_s N}\right)^{\frac{1}{4}}\rho\gamma^{56789}\partial_{\alpha}Z\cdot \Gamma Q_5^{IJ}
\\&\equiv&\partial_{\alpha}\delta^{IJ}+\tilde{\Gamma}_5\partial_{\alpha}Z \cdot \Gamma Q_5^{IJ}
=\partial_{\alpha}\delta^{IJ}+\sum_{F=1}^8\frac{m_F}{2\sqrt{T}}\rho\tilde{\gamma}_F\partial_{\alpha}Z\cdot
\Gamma Q_5^{IJ},\quad
\end{eqnarray}
where in $\partial_{\alpha}Z\cdot\Gamma$ we contract with
$\delta_{ab}$. Notice the factor $2$ in the second equality coming
from the fact that we included the dual 5-form. In this notation
$\tilde{\Gamma}_5=\frac{1}{2}(\frac{\pi}{g_s
N})^{\frac{1}{4}}\rho\gamma^{56789}$.  There are 8 massive fermions
with mass $\frac{m_F^2}{T}=\sqrt{\frac{\pi}{g_s N}}$.
We can take $\tilde{\gamma}_F$ to be $\gamma^{56789}$ times
any basis of projection operators commuting with $\gamma^{56789}$.

\subsection{Witten background for $D=4$}\label{ssec:examplesWN4}

A theory related to a pure YM theory in 4 dimensions can be
achieved by methods similar to those in the previous example,
starting from D4-branes \cite{bib:backgrounds1}. The string theory
is a type IIA theory, with a background including the metric, a
4-form on the sphere, and a dilaton which diverges at infinity.
Again, there is a circle which vanishes at a finite radial
coordinate. The background is given by \cite{bib:backgrounds1}
\begin{eqnarray}
\nonumber
\frac{ds^2}{\alpha'}&=&\frac{2\pi\lambda}{3u_0}u
\left(4u^2\sum_{i=0}^{3}dx_i^2+\frac{4}{9u_0^2}u^2(1-\frac{u_0^6}{u^6})d\tau^2+4\frac{du^2}{u^2(1-\frac{u_0^6}{u^6})}+d\Omega_4^2\right),\quad
\\ F_4&=&3\pi N\alpha'^{\frac{3}{2}}\omega_4  ,\quad R^2=\frac{2\pi \lambda}{3}\alpha' ,\quad\int_{S_4}\omega_4=\frac{8\pi^2}{3},\quad
e^{2\phi}=\frac{8\pi \lambda^3u^3}{27u_0^3N^2} \quad,
\end{eqnarray}
where $\lambda$ is related to the four dimensional 't Hooft coupling, and $\omega_4$
is the volume form on the unit 4-sphere.
In this background we find there are six massless scalars, two massive scalars with mass $\frac{m^2_B}{T}=\frac{27}{4\lambda}$ and eight massive fermions with mass $\frac{m^2_F}{T}=\frac{27}{16\lambda}$.

\subsubsection{Scalar masses}

We take the limit $u\to u_0$, and obtain (after rescalings)
\begin{eqnarray}
\nonumber &&u=u_0(1+\frac{9}{8 \lambda}\rho^2+O(\rho^4)),\quad
\\ &&\frac{ds^2}{2\pi\alpha'}=(1+\frac{27}{8\lambda}\rho^2)\sum_{i=0}^{3}dx_i^2+\rho^2d\tau^2+d\rho^2+\frac{R^2}{2\pi\alpha'}d\Omega_4^2\quad.
\end{eqnarray}
Comparing to (\ref{eq:metric}) we find six transverse massless scalars
and two massive scalars with mass $\frac{m^2_B}{T}=\frac{27}{4\lambda}$.

\subsubsection{Fermion masses}
The 4-form in this background couples to the fermions, as described
in \cite{bib:action1, bib:action2}, so that the covariant derivative
on the worldsheet is

\begin{eqnarray}
\nonumber D_{\alpha}&=&\partial_{\alpha}+\frac{1}{8\cdot
4!}e^{\phi}F_{abcd}\Gamma^{abcd}\partial_{\alpha}Z^{\mu}\Gamma_{\mu}
=\partial_{\alpha}+\frac{1}{8}\sqrt{2\pi\alpha'}\sqrt{\frac{8\pi\lambda^3}{27N^2}}\frac{27
N}{4\pi \lambda^2 \sqrt{\alpha'}}
\gamma^{6789}\partial_{\alpha}Z\cdot\Gamma
\\ &=&\partial_{\alpha}+\frac{1}{2}\sqrt{\frac{27}{16\lambda}}\gamma^{6789}\partial_{\alpha}Z\cdot\Gamma \equiv\partial_{\alpha}+\sum_{F=1}^8\frac{m_F}{2\sqrt{T}}\tilde{\gamma}_F\partial_{\alpha}Z\cdot \Gamma \quad.
\end{eqnarray}
Here we used
$\tilde{\Gamma}_4=\frac{1}{2}\sqrt{\frac{27}{16\lambda}}\gamma^{6789}$.
We find 8 massive fermions, each with mass
$\frac{m_F^2}{T}=\frac{27}{16\lambda}$.

\subsection{The Maldacena-Nu\~nez background}\label{ssec:examplesMN}

It was proposed in \cite{bib:backgrounds2} that the gravity solution
found in \cite{bib:backgrounds5} is associated with the theory of $N$ D5-branes on a 2-sphere,
which in a specific limit becomes the four dimensional ${\cal{N}}=1$ supersymmetric YM
theory (SYM).
The UV theory is 6
dimensional and maximally supersymmetric. The spin structure on the
sphere is taken such that only 4 supersymmetries remain, and in the
limit of small 't Hooft coupling the IR theory is the 4 dimensional
${\cal{N}}=1$ SYM. In the weakly curved limit (large 't Hooft
coupling) there is no separation between the SYM theory and the six
dimensional modes. The background consists of a metric, a R-R 3-form,
and a dilaton:
\begin{eqnarray}
\nonumber
\frac{ds^2}{\alpha'}&=&e^{\phi_D}N[\sum_{i=0}^3dx_i^2+d\rho^2+e^{2g(\rho)}d\Omega^2_2+\frac{1}{4}\sum_a(\omega^a-A^a)^2],
\\ \nonumber e^{2\phi_D}&=&e^{2\phi_{D,0}}\frac{\sinh(2\rho)}{2e^{g(\rho)}}=g_s^2(1+\frac{8}{9}\rho^2+O(\rho^4)),
\\ F_3&=&N[-\frac{1}{4}(\omega^1-A^1)\wedge(\omega^2-A^2)\wedge(\omega^3-A^3)+\frac{1}{4}\sum_aF\wedge (\omega^a-A^a)],
\end{eqnarray}
where (see \cite{bib:backgrounds2} for details)
\begin{eqnarray}
\nonumber
a(\rho)&=&\frac{2\rho}{\sinh(2\rho)}=1-\frac{2}{3}\rho^2+O(\rho^4),
\\ \nonumber e^{2g}&=&\rho\coth(2\rho)-\frac{\rho^2}{\sinh^2(2\rho)}-\frac{1}{4}=\rho^2+O(\rho^4),
\\ \nonumber
A&=&\frac{1}{2}[\sigma^1a(\rho)d\theta+\sigma^2a(\rho)\sin(\theta)d\phi+\sigma^3\cos(\theta)d\phi],\quad
F=dA+A\wedge A,
\\ \omega^1+i\omega^2&=&e^{-i\psi}(d\tilde{\theta}+i\sin(\tilde{\theta})d\Phi) ,\quad \omega^3=d\psi+\cos(\tilde{\theta})d\phi\quad.
\end{eqnarray}
There is also a limit of this background where the string coupling becomes
strong so one needs to use an S-duality transformation, after which only
NS-NS fields are turned on. In this limit the confining string is a D-string,
and it belongs in the class of backgrounds discussed in section \ref{specialclass}; we will
not discuss it further here.

In this background there are 3 massive scalars with mass
$\frac{m^2_B}{T}=\frac{16\pi}{9 g_s N}$ and 5 massless transverse
scalars. There are 6 massive fermions with mass
$\frac{m^2_F}{T}=\frac{8\pi}{9g_s N}$ and 2 massless fermions, which
are Goldstinos for the supersymmetries broken by the string.

\subsubsection{Scalar masses}

Carefully taking the IR limit $\rho\rightarrow0$, and using 3
Cartesian coordinates $Y_B$ instead of $\rho$ and $\Omega_2$, the
metric is:
\begin{eqnarray}\label{eq:MNmetric}
\frac{ds^2}{2\pi\alpha'}&=&\left(1+\frac{8\pi Y^2}{9g_sN}\right)\sum_{i=0}^3dX_i^2+\sum_{B=4}^6 dY_B^2+\frac{1}{8\pi}\sum_{a=7}^{9}(\omega^{a-6}-A^{a-6})^2\quad.
\end{eqnarray}
Comparing to (\ref{eq:metric}) we find 5 massless scalars and 3
massive scalars with mass $\frac{m^2_B}{T}=\frac{16\pi}{9 g_s N}$.

\subsubsection{Fermion masses}

The covariant derivative contains the following term :
\begin{eqnarray}\label{eq:MNfermass}
\nonumber  \frac{\sqrt{2\pi
\alpha'}}{8\cdot3!}e^{\phi}F_{\mu\nu\rho}e^{\mu}_ae^{\nu}_be^{\rho}_c\Gamma^{abc}\Gamma
\cdot
\partial_{\alpha}Z&=&-\frac{1}{2}\sqrt{\frac{\pi}{18g_sN}}\rho(\gamma^{457}+\gamma^{468}+\gamma^{569}+3
\gamma^{789})\Gamma\cdot \partial_{\alpha}Z
\\ \nonumber&=&-\frac{1}{2}\sqrt{\frac{8\pi}{9g_sN}}\rho\gamma^{789}(1-P_{I+}P_{II+})\Gamma\cdot \partial_{\alpha}Z,
\\ P_{I\pm}\equiv\frac{1}{2}(1\pm\gamma^{5678}) ,\quad P_{II\pm}&\equiv&\frac{1}{2}(1\pm\gamma^{4589}),
\quad P_{III\pm}\equiv\frac{1}{2}(1\pm\gamma^{4679})\quad.
\end{eqnarray}
The indices on the gamma matrices $\gamma^{ijk}$ are flat space
indices, and their numbers correspond to the directions in the metric
(\ref{eq:MNmetric}). We defined projection operators $P_{I\pm}$,
$P_{II\pm}$ and $P_{III\pm}$, each with half zero eigenvalues,
so that each product of three of them
projects onto one physical d.o.f.
The three projectors commute, so we can block diagonalize them
simultaneously and in this basis our 8 fermions split into 8
sectors, according to the projectors
eigenvalues \{$P_{I+}=(0,1)$, $P_{II+}=(0,1)$, $P_{III+}=(0,1)$\}.
There are two sectors, $\{1,1,0\}$ and $\{1,1,1\}$, for which the
mass matrix vanishes. Thus there are 2 massless fermions, while the
other 6 fermions all have the same mass,
$\frac{m^2_F}{T}=\frac{8\pi}{9g_sN}$.

\subsection{Klebanov-Strassler  background}\label{ssec:examplesKS}

The Klebanov-Strassler background is obtained by considering a
collection of $N$ regular and $M$ fractional D3-branes in the
geometry of a deformed conifold \cite{bib:backgrounds3}. The gravity
solution includes a ${\bf R}^4$ part with a warp factor, and a six
dimensional conifold (including the radial direction). There are R-R
forms $F_3$ and $F_5$, and an NS-NS 2-form $B$.
We refer to \cite{bib:backgrounds3b} for the exact background and
write here only the expansion near the minimal radial coordinate.
We find that there are 5 massless scalars, 3 massive scalars with
mass $\frac{m^2_B}{T}=\frac{4\pi}{3 a_0^{3/2}g_s M}$ (where
$a_0\approx0.718$ is computed in \cite{bib:backgrounds3b}), 2
massless fermions (corresponding to the ${\cal N}=1$ Goldstinos) and
6 massive fermions with mass $\frac{m^2_F}{T}=\frac{2\pi}{3
a_0^{3/2} g_sM}$.

\subsubsection{Scalar masses}

The metric near $\rho=0$ is
\begin{eqnarray}\label{eq:KSmetric}
\nonumber \frac{ds^2}{2\pi\alpha'}&=&\left(1+\frac{6^{1/3}2\pi
a_1}{a_0^{3/2}g_s
M}\rho^2\right)\sum_{i=0}^3dX_i^2+d\rho^2+\frac{\rho^2}{2}((g^1)^2+(g^2)^2)+\frac{\sqrt{a_0}g_sM\alpha'}{6^{1/3}}((g^3)^2+(g^4)^2)
\\ &+&\frac{\sqrt{a_0}g_s M \alpha'}{4 \cdot 6^{1/3}}(g^5)^2\quad.
\end{eqnarray}
Here $\rho$ is the radial direction, and $g^1$ and $g^2$ are the
tangent directions on a 2-sphere which shrinks to zero at $\rho=0$.
$g^{3},g^4,g^5$ are other directions on the sphere which are
massless. Comparing to (\ref{eq:metric}) we find 5 massless scalars
and 3 massive scalars with mass $\frac{m^2_B}{T}=\frac{6^{1/3}4\pi
a_1}{a_0^{3/2}g_s M}=\frac{4\pi}{3 a_0^{3/2}g_s M}$. We used the
value $a_1=\frac{6^{2/3}}{18}$ computed in \cite{bib:backgrounds3b}.

\subsubsection{Fermion masses}

Out of the 4 background fields, there are two which do not vanish at
the minimal radial coordinate, $F_3$ and $H_3$. As we stated in section \ref{sec:superstrings},
if $H_3$ is not polarized along the worldsheet then it does not
contribute to the fermion mass terms, which is the case here.
Therefore, the only contribution comes from the R-R 3-form, whose
value at the minimal radial coordinate is
\begin{eqnarray}\label{eq:KSfermass}
\nonumber
F_{\mu\nu\rho}e^{\mu}_ae^{\nu}_be^{\rho}_c\Gamma^{abc}&=&\frac{4}{\sqrt{3}a_0^{3/4}
M^{1/2}\alpha'^{1/2}g_s^{3/2}}\rho(3\gamma^{345}+\gamma^{125}+\gamma^{\rho13}+\gamma^{\rho24})
\\  &=&\frac{4}{\sqrt{3}a_0^{3/4}M^{1/2}\alpha'^{1/2}g_s^{3/2}}\rho\gamma^{345}({1-P_{I+}P_{II+}})\quad.
\end{eqnarray}
The covariant derivative is therefore
\begin{eqnarray}
\nonumber \frac{\sqrt{2\pi
\alpha'}}{8\cdot3!}e^{\phi}F_{\mu\nu\rho}e^{\mu}_ae^{\nu}_be^{\rho}_c\Gamma^{abc}\Gamma
\cdot
\partial_{\alpha}Z&=&\frac{1}{2}\sqrt{\frac{2 \pi} {3 a_0^{3/2} g_s M}}\rho\gamma^{345}(1-P_{I+}P_{II+})\Gamma\cdot \partial_{\alpha}Z\quad,
\\ P_{I\pm}\equiv\frac{1}{2}(1\pm\gamma^{1234}) ,\quad P_{II\pm}&\equiv&\frac{1}{2}(1\pm\gamma^{\rho145}),
\quad P_{III\pm}\equiv\frac{1}{2}(1\pm\gamma^{\rho235})\quad.
\end{eqnarray}
The indices on the gamma matrices $\gamma^{ijk}$ are flat space
indices and correspond to the directions in the metric
(\ref{eq:KSmetric}), where $\rho$ is the radial direction and the
indices $1,\cdots,5$ correspond to the $g^{1},\cdots,g^5$ directions. The projection
operators were defined similarly to those in the Maldacena-Nu\~nez
example. Thus, we find 2 massless fermions, while the other 6
fermions all have the same mass, $\frac{m^2_F}{T}=\frac{2\pi}{3
a_0^{3/2} g_sM}$.

\section{The effective action from correlation functions}\label{sec:corfun}

In this section we compute in confining holographic backgrounds of the
form discussed above, the low-energy effective worldsheet theory for the
$X$ fields, which are the coordinates where the gauge theory lives.
The effective action will contain corrections to the
classical interactions. We expect the corrections to be a series in
powers of $\frac{m^2}{T}$ and $\frac{\partial^2}{T}$, since the loop
expansion parameter is $\frac{1}{T}$ \footnote{We will see below
that this is not precisely correct due to logarithmic divergences.}
($T$ multiplies the whole action in some normalization of the fields).
We can write the effective action as a Nambu-Goto action plus corrections :
\begin{eqnarray}\label{eq:effaction}
S=-\int
d^2\sigma\{\sqrt{\det(-T'\delta_{\alpha,0}\delta_{\beta,0}+T'\delta_{\alpha,1}\delta_{\beta,1}
+\partial_{\alpha}\tilde{X}\cdot\partial_{\beta}\tilde{X})}-{\cal
F}(\tilde{X})\},
\end{eqnarray}
where we allow some renormalization of the fields,
$\tilde{X}=X(1+O(m^2/T))$ and the tension, $T'=T+O(m^2/T)$. In order
to evaluate the effective action we compute correlation functions in
both the original and the effective theory. By comparing the two, we
determine the coefficients of the operators appearing in ${\cal
F}(\tilde{X})$, whose general form was discussed in section \ref{sec:generalities}.
Since we are
interested only in corrections depending on $m^2$, we do not compute
the contribution of massless fields to the effective action.

Our original action (\ref{eq:10daction}) is non-renormalizable,
however all the results we find are finite, up to possible quadratic
divergences which we do not calculate as they have contributions
from the massless modes. Presumably this is because the theory is
really finite in a different gauge (conformal gauge). We use a sharp
cutoff regularization with cutoff $\Lambda$. The logarithmic
convergence of our calculation is important. Such divergences would
appear as $\frac{m^2}{T}\log[m^2/\Lambda^2]$, which implies they
vanish in flat space, and a priori one may need to add counter-terms
in order to have a finite effective theory, and the renormalization
procedure will ruin our predictability regarding ${\cal
F}(\tilde{X})$. Notice that even if there are quadratic divergences,
at one loop order they appear as $\frac{\Lambda^2}{T}$ with no powers of
$m^2$, and thus renormalizing these divergences has no finite effect
on our result.

We find that  the effective action at the six-derivative order
actually shows no deviation from Nambu-Goto, namely ${\cal F}({\tilde X})=0$
to this order.

\subsection{The tension}\label{ssec:corfunten}

The tension $T'$ is the constant term in (\ref{eq:effaction}), which can
be determined from the term linear in $L$ in the ground state
energy of a closed string of length $L$. The leading order ground
state energy of a string of length $L$ is $E=TL$, where $T$ is the
classical tension of the string. There are corrections to the
energy coming from quantum fluctuations of the worldsheet fields. At
one-loop order this is given by a summation of all (on-shell) modes in
the worldsheet theory \cite{bib:tension},
\begin{eqnarray}\label{eq:tension}
E&=&T
L+\frac{\pi}{4L}\sum_{n=-\infty}^{\infty}\{(N^0_B-N^0_F)|n|+\sum_{B}\sqrt{n^2+\frac{m^2_BL^2}{\pi^2}}-\sum_{F}\sqrt{n^2+\frac{m^2_FL^2}{\pi^2}}\}
\\ \nonumber&=&T L+\frac{L}{8\pi}\{\sum_{F}m^2_F\log(m^2_F)-\sum_{B}m^2_B\log(m^2_B)\}+O(L^0).
\end{eqnarray}
Here we approximated the sum as an integral, which is correct up to
$\frac{1}{L}$ corrections.  Interpreting terms linear in $L$ as
corrections to the string tension, we find $T'=T+\Delta T$ with
\begin{eqnarray}
\Delta
T&=&\frac{1}{8\pi}\{\sum_{F}m^2_F\log(m^2_F)-\sum_{B}m^2_B\log(m^2_B)\}.
\end{eqnarray}
Note that the logs appearing in (\ref{eq:tension}) are really
$\log(\frac{m^2}{\Lambda^2})$, but the cutoff dependence cancels out
exactly when $\sum_{F}m^2_F=\sum_{B}m^2_B$, and we will assume this
from here on.

\subsection{2-point function}\label{ssec:corfun2pt}

In this subsection we compute $\Pi(k) \equiv \vev{X^i(k)X^j(-k)}$ to see if we need to
perform a wave-function renormalization of $X$ in order to obtain the
quadratic term in (\ref{eq:effaction}).

\subsubsection{Fermion diagrams}

There are two fermionic diagrams contributing to the propagator
$\Pi(k)$. One should be careful about the Wick contractions, paying
attention to various minus signs. For example, a 4-point diagram
involving two vertices of the form
$XX\theta^1\tilde{\gamma}\theta^2$ is evaluated with the following
fermionic trace:
\begin{eqnarray}
\nonumber
<X^iX^j\theta^1_a\tilde{\gamma}^{ab}\theta^2_bX^kX^l\theta^1_c\tilde{\gamma}^{cd}\theta^2_d>
&=&<X^iX^jX^kX^l\theta^1_a(k-p)\tilde{\gamma}^{ab}\theta^2_b(p)\theta^1_c(-\tilde{p})\tilde{\gamma}^{cd}\theta^2_d(\tilde{p}-k)>
\\&\propto&{\rm tr}[-G_{21}(p-k)\tilde{\gamma}G_{21}(p)\tilde{\gamma}+G_{11}(p-k)\tilde{\gamma}G_{22}(p)\tilde{\gamma}^T]\textrm{ ,}
\end{eqnarray}
where $G_{ab}(p)$ are the entries in the fermion propagator matrix
(\ref{eq:ferprop}).

\begin{figure}[!htb]
\centering
\includegraphics[width=5.2cm,height=2.5cm]{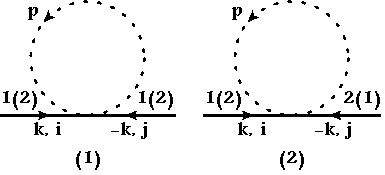}
\caption{The 2-point fermion diagrams: (1) $\Delta \Pi_1$, (2)
$\Delta \Pi_2$. The fermionic propagator is marked by a dashed line.
The numbers indicate the propagator indices; e.g. in (2) there are
two contributions coming from $G_{12}$ and $G_{21}$. External solid
lines mark the incoming scalars $X^{i}$ and $X^j$, with momenta $k$
and $-k$ respectively.\label{twoptferm}}
\end{figure}
We need to evaluate the diagrams of figure \ref{twoptferm},
where the indices on the fermion loops refer to the type of fermion
($\theta^{1}$ or $\theta^{2}$) in the loop.
The first diagram has no contribution,
\begin{eqnarray}
\Delta\Pi_1&=&\sum_{F}\frac{i}{4T}\delta_{ij}\int
\frac{d^2p}{(2\pi)^2}\{(-p_+k^2+2k_+k\cdot
p)(-ip_-){\rm tr}[(1+\gamma^c)\tilde{\gamma}_F^T\tilde{\gamma}_F]
\nonumber\\
\nonumber&+&(-p_-k^2+2k_-k\cdot
p)(-ip_+){\rm tr}[(1\pm\gamma^c)\tilde{\gamma}_F^T\tilde{\gamma}_F]\}\frac{1}{2}\frac{4}{p^2+m^2_F}
\\  &=&\sum_{F}\frac{1}{2T}\delta^{ij}\int \frac{d^2p}{(2\pi)^2}(\frac{1}{2}p^2k^2-(k\cdot p)^2)\frac{1}{p^2+m^2_F}=0\quad.
\end{eqnarray}

The second fermionic diagram is given by
\begin{eqnarray}
\Delta\Pi_2&=&\sum_{F}\frac{m_F}{8T}(-k^2)\frac{m_F}{2}\delta^{ij}\int
\frac{d^2p}{(2\pi)^2}\frac{4}{p^2+m_F^2}\frac{1}{2}\{-{\rm tr}[\tilde{\gamma}_F^T(1\pm\gamma^c)\tilde{\gamma}_F^T\tilde{\gamma}_F\tilde{\gamma}_F]
\nonumber \\
\nonumber&-&{\rm tr}[\tilde{\gamma}^{F}(1+\gamma^c)\tilde{\gamma}_F^T\tilde{\gamma}^{F}\tilde{\gamma}_F^T]\}
=\sum_{F}\frac{m^2_F}{2T}k^2\delta^{ij}\int
\frac{d^2p}{(2\pi)^2}\frac{1}{p^2+m_F^2}
\\ &=&\sum_{F}\frac{i m^2_F
}{8\pi T}k^2\delta^{ij}\log[\frac{\Lambda^2}{m_F^2}]\quad.
\end{eqnarray}
Here we used ${\rm tr}[\tilde{\gamma}_F^T\tilde{\gamma}_F\gamma^c]=0$ and
(\ref{eq:Projector}). The $i$ in the last line comes from computing
the integral in Euclidean space. Because of our $(-,+)$ signature choice,
$p^2=p^2_E$, and the only change is a factor of $i$ from
setting $p_0=ip_E$. Note that we evaluate all other diagrams in the
same way.

\subsubsection{Scalar diagram}
There is a single scalar diagram,
\begin{figure}[!htb]
\centering
\includegraphics[width=2.5cm,height=2cm]{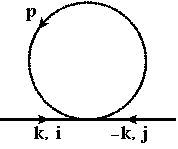}
\caption{The 2-point scalar diagram $\Delta\Pi_3$. Both massive and
massless scalars are marked by solid lines.}
\end{figure}
\begin{eqnarray}
\Delta\Pi_3&=&\sum_{B}\frac{i}{2T}\delta^{ij}\int
\frac{d^2p}{(2\pi)^2}(-m^2_Bk^2-k^2p^2+2(k\cdot
p)^2)\frac{-i}{p^2+m^2_B} \nonumber \\
&=&\sum_{B}-\frac{m^2_B
k^2}{2T}\delta^{ij}\int\frac{d^2p}{(2\pi)^2}\frac{1}{p^2+m^2_B}
=\sum_{B}-\frac{i m^2_B}{8 \pi
T}k^2\delta^{ij}\log[\frac{\Lambda^2}{m^2_B}]\quad.
\end{eqnarray}

\subsubsection{Conclusion}

The sum of all diagrams is,
\begin{eqnarray}
\Delta\Pi&=&\frac{ik^2}{8 \pi
T}\{\sum_{B}m^2_B\log(m^2_B)-\sum_{F}m^2_F\log(m^2_F)\}=-i k^2\frac{\Delta
T}{T}\quad.
\end{eqnarray}
We see that the logarithmic divergence cancels between the fermionic
and scalar diagrams and we are left with a finite contribution. We
can now obtain the two-point function:
\begin{eqnarray}
\nonumber<X(k)X(-k)>&=&-\frac{i}{k^2}+\Delta
\Pi(-\frac{i}{k^2})^2=-\frac{i}{k^2}\left(1-\frac{ i\Delta\Pi}{k^2}\right)
\\&=&-\frac{i}{k^2(1+\frac{i
\Delta\Pi}{k^2})}=-\frac{i}{k^2(1+\frac{\Delta T}{T})}.
\end{eqnarray}

This result, together with the tension correction found in section
\ref{ssec:corfunten}, are obtained from the following Minkowskian
effective action,
\begin{eqnarray}
S&=&-\int d^2\sigma \{T+\Delta
T+\frac{1}{2}\left(1+\frac{\Delta
T}{T}\right)\partial_{\alpha}X\partial^{\alpha}X\}
\\ \nonumber&=&-\int d^2\sigma \{\sqrt{-\det[(T+\Delta
T)(-\delta_{\alpha,0}\delta_{\beta,0}+\delta_{\alpha,1}\delta_{\beta,1})+\partial_{\alpha}
\tilde{X}\cdot\partial_{\beta} \tilde{X}]}+O(k^4/T^2)\}\quad.
\end{eqnarray}
We find that to this order the effective action is the NG action,
with an effective tension $T'=T+\Delta T$, and rescaled fields
$\tilde{X}^i=(1+\frac{\Delta T}{2T})X^i$ (to leading order in
$\Delta T$).

\subsection{4-point function}\label{ssec:corfun4pt}

In this subsection we compute the 4-point function $\vev{X^i(k_1)X^j(k_2)X^k(k_3)X^l(k_4)}$.
We perform the calculation first at leading order in the external momenta, to extract the leading term
$O(\frac{k^4}{T^2})$. For further simplicity, we only consider terms
proportional to $\delta_{ij}\delta_{kl}$, since the other terms follow by permutations;
in most diagrams this means that only vertices where $X^i$ is paired with $X^j$ and $X^k$ is paired with $X^l$ contribute, so we only evaluate these.

\subsubsection{Fermion diagrams}

We start from diagrams with a fermion loop.
We compute in turn the contribution of each diagram to the 4-point function.
In some cases we write the expressions for momentum integrals for arbitrary dimension $d$, to help keep track of numerical factors.
At the end of the computation we always set $d=2$. The first diagram is
\begin{eqnarray}
\nonumber
M^4_{F1}&=&(-2)\times\delta_{ij}\delta_{kl}\sum_{F}\int\frac{d^2p}{(2\pi)^2}(\frac{i}{4T})^2\frac{(-4i)^2}{(p^2+m^2_F)^{2}}\frac{1}{4}{\rm tr}[(1+\gamma^c)^2\tilde{\gamma}_F^T\tilde{\gamma}^{F}]
\\ \nonumber &&\times \{(-k_1\cdot k_2p_++{k_1}_+k_2\cdot p+{k_2}_+k_1\cdot p)(p_-)(-k_3\cdot k_4p_++{k_3}_+k_4\cdot p+{k_4}_+k_3\cdot p)(p_-)\}
\\ \nonumber&=&\delta_{ij}\delta_{kl}\sum_{F}(\frac{-2}{T^2})\int\frac{d^2p}{(2\pi)^2}\frac{1}{(p^2+m^2_F)^{2}}
\\ \nonumber &&\times \{(p_+p_-)^2k_1\cdot
k_2 k_3\cdot k_4+{k_1}_+p_-{k_3}_+p_-p\cdot k_2 p\cdot
k_4+{k_1}_+p_-{k_4}_+p_-p\cdot k_2 p\cdot
k_3\\
\nonumber&&+{k_2}_+p_-{k_3}_+p_-p\cdot k_1 p\cdot
k_4+{k_2}_+p_-{k_4}_+p_-p\cdot k_1 p\cdot k_3
-{k_1}_+p_-p_+p_-p\cdot k_2 k_3\cdot k_4
\\ &&-{k_2}_+p_-p_+p_-p\cdot k_1 k_3 \cdot k_4-{k_3}_+p_-p_+p_-p\cdot k_4 k_1 \cdot k_2-{k_4}_+p_-p_+p_-p\cdot k_3 k_1 \cdot k_2\}\quad.
\end{eqnarray}
There is a symmetry factor of $(-2)$ for the two possible contractions in
the loop. This diagram is not Lorentz invariant by itself, but
only when we add it to $M^4_{F2}$; however it is easy to see that each diagram
 separately vanishes.

\begin{figure}[!htb]
\centering
\includegraphics[width=17cm,height=9cm]{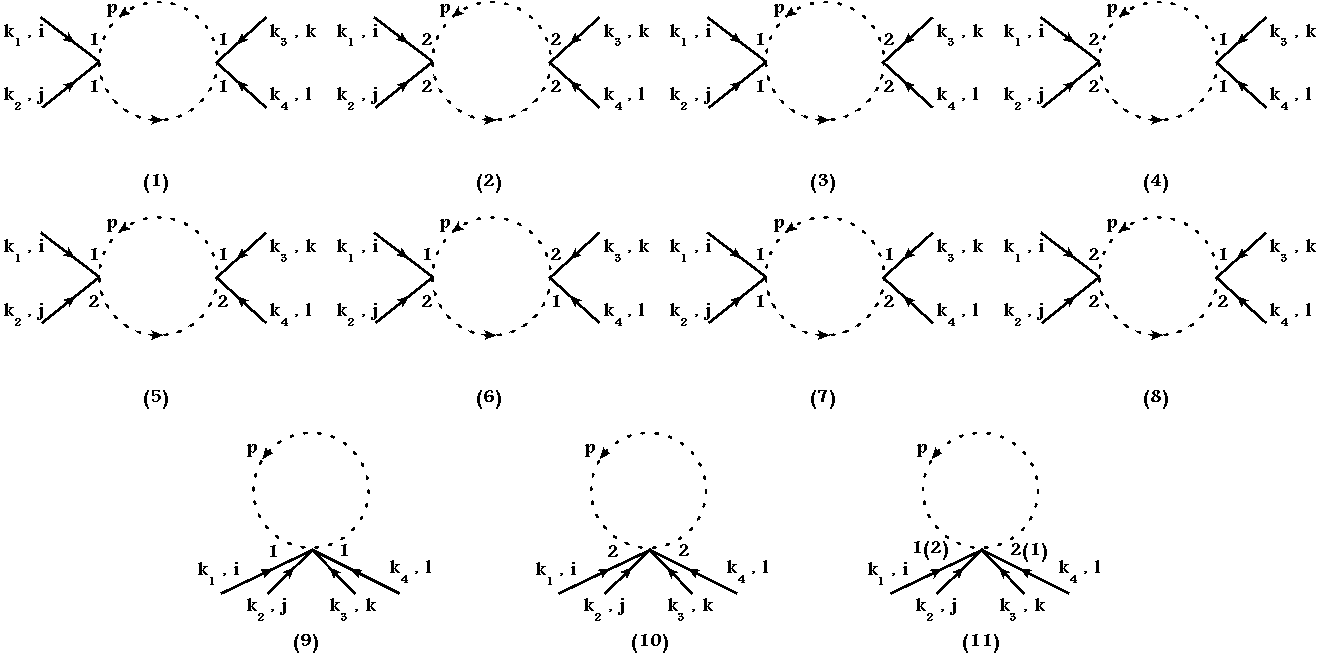}
\caption{The 4-point fermion diagrams: (1)$M^4_{F1}$, (2)$M^4_{F2}$,
(3)$M^4_{F3}$, (4)$M^4_{F4}$, (5+6)$M^4_{F5}$+$M^4_{F6}$,
(7)$M^4_{F7}$, (8)$M^4_{F8}$, (9)$M^4_{F9}$, (10)$M^4_{F10}$,
(11)$M^4_{F11}$.}
\end{figure}

The following diagram is also not Lorentz invariant by itself,
\begin{eqnarray}
\nonumber
M^4_{F3}&=&(-2)\times\sum_{F}(\frac{i}{4T})^2(\frac{m_F}{2})^2\int\frac{d^2p}{(2\pi)^2}\frac{(-4i)^2}{(p^2+m^2_F)^{2}}\frac{1}{4}{\rm tr}[(1\pm\gamma^c)(\tilde{\gamma}_F)^3(1+\gamma^c)(\tilde{\gamma}_F^T)^3]
\\ \nonumber &&\times \{(-k_1\cdot k_2p_++{k_1}_+k_2\cdot p+{k_2}_+k_1\cdot p)(-k_3\cdot k_4p_-+{k_3}_-k_4\cdot p+{k_4}_-k_3\cdot p)\}\delta_{ij}\delta_{kl}
\\ \nonumber &=&\delta_{ij}\delta_{kl}\sum_{F}(-\frac{m_F^2}{2T^2})\int\frac{d^2p}{(2\pi)^2}\frac{1}{(p^2+m_F^2)^{2}}
\times \{k_1\cdot k_2k_3\cdot k_4p_+p_-+{k_1}_+{k_3}_-k_2\cdot p
k_4\cdot p
\\ \nonumber &&+{k_2}_+{k_3}_-k_1\cdot p k_4\cdot p
+{k_1}_+{k_4}_-k_2\cdot p k_3\cdot p+{k_2}_+{k_4}_-k_1\cdot p
k_3\cdot p\\
\nonumber&&-k_1\cdot k_2p_+{k_3}_-k_4\cdot p-k_1\cdot
k_2p_+{k_4}_-k_3\cdot p-k_3\cdot k_4p_-{k_1}_+k_2\cdot p
\\ &&-k_3\cdot k_4p_-{k_2}_+k_1\cdot p\}\quad.
\end{eqnarray}
Here we used the fact that for type IIA (upper sign)
$[\tilde{\gamma},\gamma^c]=0$, while for type IIB (lower sign)
$\{\tilde{\gamma},\gamma^c\}=0$.
The next diagram is :
\begin{eqnarray}
M^4_{F4}&=&\nonumber
\delta_{ij}\delta_{kl}\sum_{F}(-\frac{m_F^2}{2T^2})\int\frac{d^2p}{(2\pi)^2}\frac{1}{(p^2+m_F^2)^{2}}
\\ \nonumber &&\times \{k_1\cdot k_2k_3\cdot k_4p_+p_-+{k_1}_-{k_3}_+k_2\cdot p k_4\cdot p+{k_2}_-{k_3}_+k_1\cdot p k_4\cdot p
+{k_1}_-{k_4}_+k_2\cdot p k_3\cdot p+\\
\nonumber&&+{k_2}_-{k_4}_+k_1\cdot p k_3\cdot p-k_1\cdot
k_2p_-{k_3}_+k_4\cdot p-k_1\cdot k_2p_-{k_4}_+k_3\cdot p-\\
&&k_3\cdot k_4p_+{k_1}_-k_2\cdot p-k_3\cdot k_4p_+{k_2}_-k_1\cdot
p\}.
\end{eqnarray}
Summing the two we obtain the Lorentz-invariant result
\begin{eqnarray}
\nonumber
M^4_{F3}+M^4_{F4}&=&\delta_{ij}\delta_{kl}\sum_{F}(-\frac{m_F^2}{2T^2})\int\frac{d^2p}{(2\pi)^2}\frac{1}{(p^2+m_F^2)^{2}}p^2
\\ \nonumber &&\times \{k_1\cdot k_2k_3\cdot k_4\left(-\frac{1}{2}+\frac{2}{d}\right)+k_1\cdot k_4k_2\cdot k_3(-\frac{1}{d})+k_1\cdot k_3k_2\cdot k_4(-\frac{1}{d})\}
\\ \nonumber
&=&\delta_{ij}\delta_{kl}\sum_{F}(-\frac{im_F^2}{16\pi
T^2})\left(-1+\log[\frac{\Lambda^2}{m_F^2}]\right)
\\ &&\times \{k_1\cdot k_2k_3\cdot k_4-k_1\cdot k_4k_2\cdot k_3-k_1\cdot k_3k_2\cdot k_4\}\quad.
\end{eqnarray}

In the evaluation of $M^4_{F5}$ the usual contraction gives a more
general index structure,
\begin{eqnarray}
M^4_{F5}&=&4\times(\frac{m_F}{16T})^2\int\frac{d^2p}{(2\pi)^2}\frac{16}{(p^2+m_F^2)^{2}}
k_1\times k_2k_3\times k_4
\\ \nonumber &&\times \frac{1}{4}
(-(\frac{m_F}{2})^2{\rm tr}[[\gamma_i,\gamma_j]\tilde{\gamma}_F(1\pm\gamma^c)\tilde{\gamma}_F^T[\gamma_k\,\gamma_l]\tilde{\gamma}_F(1\pm\gamma^c)\tilde{\gamma}_F^T]
\\ \nonumber&&+p_+p_-{\rm tr}[[\gamma_i,\gamma_j]\tilde{\gamma}_F(1+\gamma^c)[\gamma_k,\gamma_l]\tilde{\gamma}_F^T(1\pm\gamma^c)])
\\ \nonumber&&=\sum_{F}(-\frac{im_F^2}{16\pi T^2})(k_1\cdot k_4k_2\cdot k_3-k_1\cdot k_3k_2\cdot k_4) (\delta_{il}\delta_{jk}-\delta_{ik}\delta_{jl})\log[\frac{\Lambda^2}{m_F^2}]
\\ \nonumber&&\Rightarrow\sum_{F}(-\frac{im_F^2}{16\pi T^2})(2k_1\cdot k_2k_3\cdot k_4-k_1\cdot k_3k_2\cdot k_4-k_1\cdot k_4k_2\cdot k_3) \delta_{ij}\delta_{kl}\log[\frac{\Lambda^2}{m_F^2}]\quad.
\end{eqnarray}
In the last line, we applied the permutations $(l,3)\leftrightarrow (j,2)$
and $(l,3)\leftrightarrow (i,1)$ to obtain the terms proportional to
$\delta_{ij}\delta_{kl}$. There is a symmetry factor of $(4)$ due to 4
possible combinations of our 2 vertices. Each contraction in the
loop gives a different contribution as can be seen from the two
terms in the parenthesis.
We also used
\begin{eqnarray}
\nonumber&&{\rm tr}[[\gamma_i,\gamma_j][\gamma_k,\gamma_l]]=64(\delta_{il}\delta_{jk}-\delta_{ik}\delta_{jl})\quad,
\\ \nonumber &&k_1\times k_2k_3\times k_4=k_1\cdot k_4k_2\cdot k_3-k_1\cdot
k_3k_2\cdot k_4,
\\&&{\rm tr}[[\gamma_i,\gamma_j][\gamma_k,\gamma_l]\tilde{\gamma}_F^T\tilde{\gamma}_F]=\frac{1}{8}{\rm tr}[[\gamma_i,\gamma_j][\gamma_k,\gamma_l]]\quad.
\end{eqnarray}

Next, we have
\begin{eqnarray}
\nonumber
M^4_{F6}&=&4\times\delta_{ij}\delta_{kl}\sum_F(\frac{m_F}{8T})^2k_1\cdot
k_2k_3\cdot k_4\int\frac{d^2p}{(2\pi)^2}\frac{16}{(p^2+m_F^2)^{2}}
\\ \nonumber &&\times \frac{1}{4}
\{-p_+p_-{\rm tr}[\tilde{\gamma}_F(1\pm\gamma^c)\tilde{\gamma}_F^T(1+\gamma^c)]-\frac{m_F^2}{4}{\rm tr}[\tilde{\gamma}_F(1-\gamma^c)\tilde{\gamma}_F^T\tilde{\gamma}_F(1-\gamma^c)\tilde{\gamma}_F^T]\}
\\ \nonumber&=&\delta_{ij}\delta_{kl}\sum_{F}\frac{ m_F^2}{4T^2}k_1\cdot k_2k_3\cdot
k_4\int\frac{d^2p}{(2\pi)^2}\frac{p^2-m_F^2}{(p^2+m_F^2)^{2}}
\\ &=&\delta_{ij}\delta_{kl}\sum_{F}\frac{i m_F^2}{16\pi T^2}(-2+\log[\frac{\Lambda^2}{m_F^2}])k_1\cdot
k_2k_3\cdot k_4\quad.
\end{eqnarray}
There is a symmetry factor of $4$ for the four possible combinations of two
vertices.

The following diagrams vanish after setting $d=2$,
\begin{eqnarray}
\nonumber
&&M^4_{F7}+M^4_{F8}=8\times\delta_{ij}\delta_{kl}\sum_{F}(-\frac{m_F}{2})(-\frac{m_F}{8T})(\frac{i}{4T})\int
\frac{d^2p}{(2\pi)^2}\frac{(-2i)(2)}{(p^2+m_F^2)^2}
\\
\nonumber&\times&\{{\rm tr}[\tilde{\gamma}_F^T(1\pm\gamma^c)\tilde{\gamma}_F^T\tilde{\gamma}_F(1+\gamma^c)\tilde{\gamma}_F^T\tilde{\gamma}_F\tilde{\gamma}_F]-{\rm tr}[(1+\gamma^c)\tilde{\gamma}_F^T\tilde{\gamma}_F\tilde{\gamma}_F^T(1\pm\gamma^c)\tilde{\gamma}_F^T\tilde{\gamma}_F\tilde{\gamma}_F^T]\}
\\ \nonumber &\times&\{(-k_1\cdot k_2 p_++k_2\cdot pk_{1+}+k_1\cdot
pk_{2+})p_-k_3\cdot k_4+(-k_3\cdot k_4p_++k_3\cdot pk_{4+}+k_4\cdot
pk_{3+})p_-k_1\cdot k_2
\\ \nonumber&+&(-k_1\cdot k_2 p_-+k_2\cdot pk_{1-}+k_1\cdot
pk_{2-})p_+k_3\cdot k_4+(-k_3\cdot k_4p_-+k_3\cdot pk_{4-}+k_4\cdot
pk_{3-})p_+k_1\cdot k_2\}
\\ \nonumber&&\quad\quad=8\times\delta_{ij}\delta_{kl}\sum_{F}(-\frac{m^2_F}{16T^2}){\rm tr}[(\pm(\gamma^c)^2-(\gamma^c)^2)\tilde{\gamma}_F^T\tilde{\gamma}_F]\int
\frac{d^2p}{(2\pi)^2}\frac{1}{(p^2+m_F^2)^2}
\\  &&\quad\quad\times\frac{1}{2}\{(k_1\cdot k_2 p^2-2k_2\cdot p k_1\cdot p)k_3\cdot k_4+(k_3\cdot k_4p^2-2k_3\cdot
pk_4\cdot p )k_1\cdot k_2\}=0\quad.
\end{eqnarray}

The following diagrams are single vertex diagrams :
\begin{eqnarray}
\nonumber
M^4_{F9}&=&\delta_{ij}\delta_{kl}\sum_{F}\frac{i}{4T^2}\int\frac{d^2p}{(2\pi)^2}\{k_1\cdot
k_2k_3\cdot p{k_{4}}_++k_1\cdot k_2k_4\cdot p{k_{3}}_+ +k_3\cdot
k_4k_1\cdot p{k_{2}}_++k_3\cdot k_4k_2\cdot p{k_{1}}_+
\\ &-&3k_1\cdot k_2k_3\cdot k_4{p}_++k_1\cdot k_3k_2\cdot k_4{p}_++k_1\cdot k_4k_2\cdot k_3{p}_+\}\frac{4ip_-}{p^2+m_F^2}\frac{1}{2}{\rm tr}[(1+\gamma^c)\tilde{\gamma}_F^T\tilde{\gamma}_F]\quad.
\end{eqnarray}
\begin{eqnarray}
\nonumber
M^4_{F10}&=&\delta_{ij}\delta_{kl}\sum_{F}\frac{i}{4T^2}\int\frac{d^2p}{(2\pi)^2}\{k_1\cdot
k_2k_3\cdot p{k_{4}}_-+k_1\cdot k_2k_4\cdot p{k_{3}}_- +k_3\cdot
k_4k_1\cdot p{k_{2}}_-+k_3\cdot k_4k_2\cdot p{k_{1}}_-
\\ &-&3k_1\cdot k_2k_3\cdot k_4{p}_-+k_1\cdot k_3k_2\cdot k_4{p}_-+k_1\cdot k_4k_2\cdot k_3{p}_-\}\frac{4ip_+}{p^2+m_F^2}\frac{1}{2}{\rm tr}[(1\pm\gamma^c
)\tilde{\gamma}_F^T\tilde{\gamma}_F]\quad.
\end{eqnarray}
Summing the diagrams $M^4_{F9}$ and $M^4_{F10}$ we obtain
\begin{eqnarray}
\nonumber
M^4_{F9}&+&M^4_{F10}=\delta_{ij}\delta_{kl}\sum_{F}(-\frac{1}{
8})\frac{4}{T^2}(k_1\cdot k_2k_3\cdot k_4-k_1\cdot k_3k_2\cdot
k_4-k_1\cdot k_4k_2 \cdot
k_3)\int\frac{d^2p}{(2\pi)^2}\frac{p^2}{p^2+m_F^2}
\\
&=&\delta_{ij}\delta_{kl}\sum_{F}(-\frac{i m^2_F}{ 8\pi
T^2})(k_1\cdot k_2k_3\cdot k_4-k_1\cdot k_3k_2\cdot k_4-k_1\cdot
k_4k_2 \cdot
k_3)\left(\frac{\Lambda^2}{m_F^2}-\log[\frac{\Lambda^2}{m_F^2}]\right).
\end{eqnarray}

Finally,
\begin{eqnarray}
\nonumber
M^4_{F11}&=&2\times\delta_{ij}\delta_{kl}\sum_F\frac{m_F}{8T^2}(k_1\cdot
k_2k_3\cdot k_4-k_1\cdot k_3k_2\cdot k_4-k_1\cdot k_4k_2\cdot
k_3)(2m_F)
\\ \nonumber&&\times\int\frac{d^2p}{(2\pi)^2}\frac{1}{p^2+m_F^2}\frac{1}{2}{\rm tr}[\tilde{\gamma}_F^T\tilde{\gamma}_F(1+\gamma^c)]
\\ &=&\delta_{ij}\delta_{kl}\sum_{F}\frac{i m_F^2 }{8\pi T^2}(k_1\cdot k_2k_3\cdot
k_4-k_1\cdot k_3k_2\cdot k_4-k_1\cdot k_4k_2\cdot
k_3)\log[\frac{\Lambda^2}{m_F^2}].
\end{eqnarray}
There is a symmetry factor of ($2$) for $2$ vertices, $\theta^1\tilde{\gamma}\theta^2$ and $\theta^2\tilde{\gamma}\theta^1$.

\subsubsection{Scalar diagrams}

Next, we compute the diagrams with the scalar fields $Y_B$ running
in the loop. There are two diagrams at one-loop order (see figure \ref{fourptbos}) :
\begin{figure}[!htb]
\centering
\includegraphics[width=8cm,height=3cm]{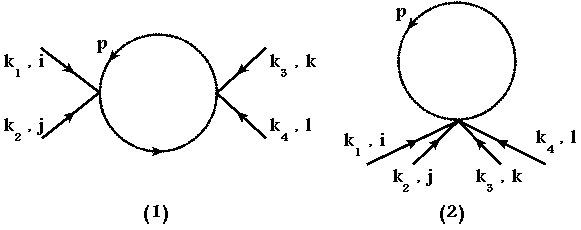}
\caption{The 4-point scalar diagrams: (1)$M^4_{B1}$, (2)$M^4_{B2}$.\label{fourptbos}}
\end{figure}
\begin{eqnarray}\label{eq:leadingbos}
\nonumber
{M^4}_{B1}&=&2\delta_{ij}\delta_{kl}\sum_{B}\int\frac{d^2p}{(2\pi)^2}\{\frac{m_B^2i}{2
T}k_1\cdot k_2+\frac{i}{2T}k_1\cdot k_2 p^2-\frac{i}{T}k_1\cdot
pk_2\cdot p\}
\\ \nonumber&&\times\{\frac{m_B^2i}{2T}k_3\cdot k_4+\frac{i}{2T}k_3\cdot k_4 p^2-\frac{i}{T}k_3\cdot pk_4\cdot p\}\left(\frac{-i}{p^2+m^2_B}\right)^2
\\ \nonumber&=&2\delta_{ij}\delta_{kl}\sum_{B}\int\frac{d^2p}{(2\pi)^2}
\left(\frac{1}{p^2+m^2_B}\right)^2\{k_1\cdot k_2 k_3\cdot
k_4(\frac{m^4_B}{4T^2}-\frac{1}{8T^2}p^4)
\\ \nonumber&&+\frac{1}{8T^2}(k_1\cdot
k_3k_2\cdot k_4+k_1\cdot k_4k_2 \cdot k_3)p^4\}
\\ \nonumber&=&\delta_{ij}\delta_{kl}\sum_{B}\frac{im_B^2}{8\pi T^2}\{k_1\cdot
k_2k_3\cdot
k_4+\left(\frac{\Lambda^2}{2m^2_B}+\frac{1}{2}-\log[\frac{\Lambda^2}{m^2_B}]\right)
\\ &\times&(-k_1\cdot
k_2 k_3\cdot k_4+k_1\cdot k_3k_2\cdot k_4 +k_1\cdot k_4k_2 \cdot
k_3)\}\quad,
\end{eqnarray}
with a symmetry factor of $(2)$ for $2$ possible contractions in
the loop, and
\begin{eqnarray}
\nonumber{M^4}_{B2}&=&\delta_{ij}\delta_{kl}\sum_{B}
\frac{i}{2T^2}\int\frac{d^2p}{(2\pi)^2}\{-2k_1\cdot k_2k_3\cdot
pk_4\cdot p-2k_3\cdot k_4k_1\cdot pk_2\cdot p
\\ \nonumber&+&p^2(3k_1\cdot k_2k_3\cdot k_4-k_1\cdot k_3k_2\cdot k_4-k_1\cdot k_4k_2\cdot k_3)
\\ \nonumber&+&m^2_B(-k_1\cdot k_2k_3\cdot k_4+k_1\cdot k_3k_2\cdot k_4+k_1\cdot k_4k_2\cdot k_3)\}\frac{-i}{p^2+m^2_B}
\\ \nonumber&=&\delta_{ij}\delta_{kl}\sum_{B} \frac{1}{2T^2}(k_1\cdot k_2k_3\cdot k_4-k_1\cdot k_3k_2\cdot k_4-k_1\cdot k_4k_2\cdot k_3)\int\frac{d^2p}{(2\pi)^2}\{p^2-m^2_B\}\frac{1}{p^2+m^2_B}
\\ &=&\delta_{ij}\delta_{kl}\sum_{B} \frac{im^2_B}{4\pi T^2}(k_1\cdot k_2k_3\cdot k_4-k_1\cdot k_3k_2\cdot k_4-k_1\cdot k_4k_2\cdot k_3)\left(\frac{\Lambda^2}{2m^2_B}-\log[\frac{\Lambda^2}{m^2_B}]\right).
\end{eqnarray}

\subsubsection{Conclusion}
If we sum our results, ignoring the quadratic divergence, we find
the finite result
\begin{eqnarray}\label{eq:4pt}
\nonumber \vev{X_i(k_1)X_j(k_2)X_k(k_3)X_l(k_4)}&=&-\delta_{ij}\delta_{kl}\frac{i\Delta{T}}{T^2}(k_1\cdot
k_2k_3\cdot k_4-k_1\cdot k_3k_2\cdot k_4-k_1\cdot k_4k_2\cdot k_3)
\\ &+&(i,1)\leftrightarrow(k,3)+(i,1)\leftrightarrow(l,4).
\end{eqnarray}
 Note that the finite contribution proportional to $\frac{1}{16\pi T^2}(\sum_B m^2_B-\sum_F m^2_F)$ vanished due to our constraint.
The quadratic divergence does not vanish in the same manner, and we believe it will vanish once we include loops of other fields (which are independent of $m$), such as the metric and the kappa gauge-fixing ghosts.
This 4-point function, with the addition of the tree-level result (\ref{eq:tree4pt}), is generated by the Minkowskian effective action:
\begin{eqnarray}\label{eq:4ptaction}
\nonumber S_4&=&-(\frac{1}{T}+\frac{\Delta
T}{T^2})\int d^2\sigma \{\frac{1}{8}(\partial^{\alpha}X\cdot
\partial_{\alpha}X)^2-\frac{1}{4}\partial_{\alpha}X\cdot\partial_{\beta}X\partial^{\alpha}X\cdot\partial^{\beta}X\}
\\ &=&-\frac{1}{T'}\int d^2\sigma \{\frac{1}{8}(\partial^{\alpha}\tilde{X}\cdot
\partial_{\alpha}\tilde{X})^2-\frac{1}{4}\partial_{\alpha}\tilde{X}\cdot\partial_{\beta}\tilde{X}\partial^{\alpha}\tilde{X}\cdot\partial^{\beta}\tilde{X}\}
\end{eqnarray}
We see that the in terms of the rescaled fields $\tilde{X}$ and tension $T'$ the effective action is precisely the NG action (\ref{eq:effaction}) expanded to fourth order in derivatives.
This is expected from the analysis of section \ref{sec:effectiveaction}, where the effective action was constrained to be the NG action to this order, but in our
computation it arises non-trivially. Note that for $D=3$ the two terms in (\ref{eq:4ptaction}) are the same, but our analysis is still valid.

\subsection{4-point function: higher derivative corrections}\label{ssec:corfun4pth}

In the previous subsection we calculated the 4-point function to lowest
order in the external momenta, so effectively we did the loop computations for zero external momenta.
In this subsection we are interested in the corrections at six-derivative order.
We will compute the 4-point function exactly as a function of the momenta, and then expand it in powers of $k$ to extract this.
To simplify our calculation, we take all the external momenta to
be on-shell ($k_i^2=0$). This is possible since contributions that are not on-shell will
create terms in the effective action that are proportional to the
equation of motion, and we know such terms can always be swallowed
by field redefinitions, and therefore do not contribute to the
partition function.
We use the Mandelstam variables, $s=(k_1+k_2)^2$, $t=(k_1+k_3)^2$, $u=(k_1+k_4)^2$,
and introduce the variable $k=k_1+k_2$ for the incoming momentum in a specific channel.
We have the on-shell relation $s+t+u=0$, and two-dimensional kinematics implies that
also $stu=0$ (we used this above in arguing that the $c_5$ term in \eqref{actionsix}
is trivial).

Apart from the UV divergences, which we expect to cancel between
fermion and scalar diagrams, we expect to find a branch-cut at
$s=-4m^2$ for diagrams with a field of mass $m$ running in the loop.
This branch cut, indicating that the fields running in the loop
become on-shell, is typical for $2\rightarrow2$ scattering.

Naively, by power counting, at one-loop the six-derivative terms
should be independent of $m$ so we are not interested in these
terms (since we are only interested in $m$-dependent contributions).
However, a dependence on $m$ can appear through logarithmic
divergences, and so we should carefully analyze the diagrams that
were quadratically divergent at four-derivative order, and thus, may
be logarithmically divergent at six-derivative order. In our case
these are the diagrams $M^4_{F1,F2}$ and $M^4_{B1}$. There are
other, single vertex diagrams, which are also quadratically divergent, but in
these the zero momentum computation was exact, and they have no
additional momentum dependence.
We note that in some of the diagrams that we do not take into account there are non-vanishing six-derivative contributions
which are not $m$-dependent, and are finite. But here we focus only on the $m$-dependent terms. Note that when discussing the $m\to 0$ limit one has to be careful, since this
limit does not commute with the small momentum limit that we analyze here.

\subsubsection{Fermion diagrams}

The diagram $M^4_{F1}$ is given by:

\begin{eqnarray}
\nonumber
M^4_{F1}&=&(-1)\times\delta_{ij}\delta_{kl}\sum_{F}\int\frac{d^2p}{(2\pi)^2}
(\frac{i}{4T})^2\frac{(-4i)^2p_- (p_- - k_-)}{(p^2+m^2_F)((p-k)^2+m^2_F)}
\frac{1}{4}{\rm tr}[(1+\gamma^c)^2\tilde{\gamma}_F^T\tilde{\gamma}_F] \frac{1}{2} \times
\\ \nonumber &\{& (k_1\cdot k_2p_+ -{k_1}_+k_2\cdot p-{k_2}_+k_1\cdot p)
(k_3\cdot k_4(2p_+-k_+)-{k_3}_+k_4\cdot (2p-k)-{k_4}_+k_3\cdot (2p-k))
\\ \nonumber &+& (k_3\cdot k_4p_+ -{k_3}_+k_4\cdot p-{k_4}_+k_3\cdot p)
(k_1\cdot k_2(2p_+-k_+)-{k_1}_+k_2\cdot (2p-k)-{k_2}_+k_1\cdot (2p-k))\}
\\ \nonumber &=&\frac{2}{T^2}\delta_{ij}\delta_{kl}\sum_{F}\int_0^1d\alpha\int\frac{d^2p}{(2\pi)^2}\frac{\alpha(1-\alpha)}{(p^2+m^2_F+k^2\alpha(1-\alpha))^2}k_- k_-\times
\\ \nonumber && \{k_1\cdot k_2k_3\cdot k_4p_+p_+-k_1\cdot k_2 (k_3\cdot pk_{4+}p_++k_4\cdot pk_{3+}p_+)-k_3\cdot k_4 (k_2\cdot pk_{1+}p_++k_1\cdot pk_{2+}p_+)
\\ &&+k_1\cdot pk_3\cdot p k_{2+}k_{4+}+k_2\cdot pk_3\cdot p k_{1+}k_{4+}+k_1\cdot pk_4\cdot p k_{2+}k_{3+}+k_2\cdot pk_4\cdot p k_{1+}k_{3+}\}=0.
\end{eqnarray}
Similarly, we find that $M^4_{F2}=0$ exactly.

\subsubsection{Scalar diagrams}

The only scalar diagram that can contribute an
$m$-dependence is $M^4_{B1}$.
We write only the term proportional to $\delta_{ij}\delta_{kl}$:
\begin{eqnarray}
\nonumber
M^4_{B1}&=&2(\frac{i}{2T})^2\sum_{B}\int\frac{d^2p}{(2\pi)^2}[m_B^2k_1\cdot
k_2 +k_1\cdot k_2p\cdot(p-k)-k_1\cdot pk_2\cdot (p-k)-k_2\cdot
pk_1\cdot (p-k)]
\\ \nonumber &\times&[m_B^2k_3\cdot k_4+ k_3\cdot k_4p\cdot(p-k)-k_3\cdot pk_4\cdot (p-k)-k_4\cdot p k_3\cdot (p-k)]\frac{-i}{p^2+m_B^2}\frac{-i}{(p-k)^2+m_B^2}
\\ \nonumber&=&\frac{1}{2T^2}\sum_{B}\int\frac{d^2p}{(2\pi)^2}[(m_B^2+p^2)k_1\cdot k_2 -2k_1\cdot pk_2\cdot p]
\\ \nonumber &\times&[(m_B^2+p^2)k_3\cdot k_4 -2k_3\cdot pk_4\cdot p]\frac{1}{p^2+m^2_B}\frac{1}{(p-k)^2+m_B^2}
\\ \nonumber &=&\frac{1}{2T^2}\sum_{B}\int_0^1d\alpha\int\frac{d^2p}{(2\pi)^2}
\frac{1}{(p^2+m^2_B+k^2\alpha(1-\alpha))^2}
\\ \nonumber&\times&[(m_B^2+(p+k(1-\alpha))^2)k_1\cdot k_2 -2k_1\cdot (p+k(1-\alpha))k_2\cdot (p+k(1-\alpha))]
\\ \nonumber &\times&[(m_B^2+(p+k(1-\alpha))^2)k_3\cdot k_4 -2k_3\cdot (p+k(1-\alpha))k_4\cdot (p+k(1-\alpha))]
\\ \nonumber &=&\frac{1}{2T^2}\sum_{B}\int_0^1d\alpha\int\frac{d^2p}{(2\pi)^2}[(m_B^2+p^2)k_1\cdot k_2 -2k_1\cdot pk_2\cdot p]
\\ \nonumber &\times&[(m_B^2+p^2)k_3\cdot k_4 -2k_3\cdot pk_4\cdot p]\frac{1}{(p^2+m^2_B+k^2\alpha(1-\alpha))^2}
\\ \nonumber&=&\frac{i}{32\pi T^2}\sum_{B}[4sm_B^4\tanh^{-1}[\sqrt{\frac{s}{s+4m_B^2}}](1+\frac{4m_B^2}{s})^{-\frac{1}{2}}
\\ \nonumber&&\qquad\qquad+\{-s^2+t^2+u^2\} \cdot \{\frac{m_B^2}{2}+\frac{(s+4m_B^2)^2}{3\sqrt{s(s+4m^2_B)}}\tanh^{-1}[\sqrt{\frac{s}{s + 4 m_B^2}}]
\\&&\qquad\qquad\qquad\qquad\qquad\qquad+\frac{1}{18}(-5s-24m_B^2-9\Lambda^2+3(s+6m_B^2)\log[\frac{m_B^2}{\Lambda^2}])\}
].
\end{eqnarray}
We can check that we reproduce the leading term (\ref{eq:leadingbos}) we computed above and we see there is a branch cut at $s=-4m_B^2$, as expected.
Taking $k\rightarrow0$, we find up to order $k^6$
\begin{eqnarray}
M^4_{B1}&=&\frac{i}{32\pi
T^2}\delta_{ij}\delta_{kl}\sum_{B}\left[m_B^2s^2-\frac{1}{6}s^3+\left\{-m_B^2-\Lambda^2+(2m_B^2+\frac{s}{3})
\log[\frac{\Lambda^2}{m_B^2}]\right\}t
u\right].
\end{eqnarray}

We see that there is apparently a logarithmic $m$-dependent term at six-derivative order,
but in fact it is proportional to $stu$ which vanishes on-shell. Thus, at one-loop order
we find no corrections to the six-derivative four-scalar term in our effective action.
This is not too surprising, since the relevant term is only non-trivial for $D>3$, while
our computation here is essentially independent of $D$.

\subsection{6-point function}\label{ssec:corfun6pt}

In this section we compute the six-point function
$\vev{X_i(k_1)X_j(k_2)X_k(k_3)X_l(k_4)X_m(k_5)X_n(k_6)}$. We write here
only the scalar contribution to six-derivative terms. We compute the
diagrams for zero external momenta, and so we expect the result
to be proportional to $\sum_{B}m^2_B\log[m^2_B/\Lambda^2]$. We know
the fermionic contribution should be
$\sum_{F}m^2_F\log[m^2_F/\Lambda^2]$, with the same proportionality
constant and opposite sign, in order to have a finite result. Thus
we skip the explicit computation of fermionic diagrams, and
calculate only the scalar contribution, from which we keep only the
term proportional to $m_B^2\log[m_B^2]$.

As in previous subsections, we focus only on the $\delta_{ij}\delta_{kl}\delta_{mn}$ terms. There are 3 scalar diagrams
(see figure \ref{fig:sixptsc}). The first is :
\begin{figure}[!htb]
\centering
\includegraphics[width=12cm,height=3cm]{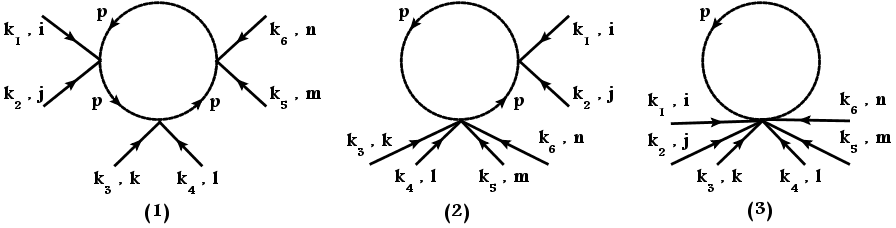}
\caption{The 6-point scalar diagrams: (1)$M^6_{B1}$, (2)$M^6_{B2}$,
(3)$M^6_{B3}$.\label{fig:sixptsc}}
\end{figure}
\begin{eqnarray}
\nonumber M^6_{B1}&=&8\times(\frac{i}{2T})^3\sum_{B}\int\frac{
d^2p}{(2\pi)^2}(\frac{-i}{m_B^2+p^2})^3(k_1\cdot
k_2(p^2+m^2)-2k_1\cdot p k_2\cdot p)
\\ \nonumber&\times&(k_3\cdot k_4(p^2+m^2)-2k_3\cdot p k_4\cdot p)(k_5\cdot k_6(p^2+m_B^2)-2k_5\cdot p k_6\cdot p)
\\ \nonumber&=&\frac{i}{2\pi T^3}\sum_{B}\{k_1\cdot k_2k_3\cdot k_4k_5\cdot k_6\left(-\frac{1}{3}\Lambda^2+\frac{13}{24}m_B^2-\frac{1}{4}m_B^2\log[\frac{\Lambda^2}{m_B^2}]\right)
\\ \nonumber&+&\left(\frac{\Lambda^2}{6}+\frac{m^2_B}{24}+
\frac{m_B^2}{4}\log[\frac{m_B^2}{\Lambda^2}]\right)[k_1\cdot k_2 k_3\cdot k_5k_4\cdot k_6+k_1\cdot k_2 k_3\cdot k_6k_4\cdot k_5
\\ \nonumber&+&(1;2)\leftrightarrow(3;4)+(1;2)\leftrightarrow(5;6)]
\\ \nonumber &-&\frac{1}{6}\left(\frac{\Lambda^2}{2}+\frac{5m^2_B}{4}+\frac{3}{2}m^2_B\log[\frac{m_B^2}{\Lambda^2}]\right)
[k_1\cdot k_3(k_2\cdot k_5 k_4\cdot k_6+k_2\cdot k_6 k_4\cdot k_5)
\\ &+&k_1\cdot k_4(k_2\cdot k_5 k_3\cdot k_6+k_2\cdot k_6 k_3\cdot k_5)+(1;2)\leftrightarrow(3;4)+(1;2)\leftrightarrow(5;6)]\}.
\end{eqnarray}
The factor of $(8)$ here is a symmetry factor for the possible contractions in the loop. Next,
\begin{eqnarray}
\nonumber
M^6_{B2}&=&2\times(\frac{i}{2T})(\frac{i}{2T^2})\sum_{B}\int
\frac{d^2p}{(2\pi)^2}\frac{(-i)^2}{(p^2+m^2_B)^2}(k_1\cdot
k_2(p^2+m^2)-2k_1\cdot pk_2\cdot p)
\\ \nonumber&\times&\{-2k_3\cdot k_4k_5\cdot pk_6\cdot p-2k_5\cdot k_6k_3\cdot pk_4\cdot p+3k_3\cdot k_4 k_5\cdot k_6 p^2
\\ \nonumber&-&p^2(k_3\cdot k_5k_4\cdot k_6+k_3\cdot k_6 k_4\cdot k_5)+m_B^2(-k_3\cdot k_4 k_5\cdot k_6+k_3\cdot k_5 k_4\cdot k_6+k_4\cdot k_5 k_3\cdot k_6)\}
\\ \nonumber&=&\frac{i}{2T^3}\sum_{B}\int \frac{d^2p}{(2\pi)^2}\frac{1}{(p^2+m^2_B)^2}
\\ \nonumber&\times&\{k_1\cdot k_2 k_3\cdot k_4 k_5\cdot k_6[(p^2+m_B^2)(p^2(3-\frac{4}{d})-m_B^2)+\frac{2}{d}p^2m_B^2-2p^4(-\frac{4}{d(d+2)}+\frac{3}{d})]
\\ \nonumber&+&k_1\cdot k_2(k_3\cdot k_5 k_4\cdot k_6+k_3\cdot k_6 k_4\cdot k_5)[(p^2+m_B^2)(-p^2+m_B^2)+\frac{2}{d}p^2(-m_B^2+p^2)]
\\ \nonumber&+&k_3\cdot k_4(k_1\cdot k_5 k_2\cdot k_6 +k_1\cdot k_6k_2\cdot k_5)(\frac{4}{d(d+2)}p^4)
\\ &+&k_5\cdot k_6(k_1\cdot k_3 k_2\cdot k_4 +k_1\cdot k_4k_2\cdot k_3)(\frac{4}{d(d+2)}p^4)\}\quad.
\end{eqnarray}
There is a symmetry factor of $(2)$ for internal contractions.
We should now sum over the two permutations $(k_1,k_2)\rightarrow(k_3,k_4)$ and $(k_1,k_2)\rightarrow(k_5,k_6)$, and then we get
\begin{eqnarray}
\nonumber M^6_{B2}&=&\frac{1}{2T^3}\sum_{B}\int
\frac{d^2p}{(2\pi)^2}\frac{1}{(p^2+m^2_B)^2}
\\ \nonumber&\times&\{3k_1\cdot k_2 k_3\cdot k_4 k_5\cdot k_6[(p^2+m_B^2)(p^2(3-\frac{4}{d})-m_B^2)+\frac{2}{d}p^2m_B^2-2p^4(-\frac{4}{d(d+2)}+\frac{3}{d})]
\\ \nonumber&+&[k_1\cdot k_2(k_3\cdot k_5 k_4\cdot k_6+k_3\cdot k_6 k_4\cdot k_5)+k_3\cdot k_4(k_1\cdot k_5 k_2\cdot k_6 +k_1\cdot k_6k_2\cdot k_5)+(1;2)\leftrightarrow(3;4)
\\ \nonumber&+&(1;2)\leftrightarrow(5;6)]\times[(p^2+m_B^2)(-p^2+m_B^2)+\frac{2}{d}p^2(-m_B^2+p^2)+\frac{8}{d(d+2)}p^4]\}
\\ \nonumber&=&\frac{i}{4\pi T^3}\sum_{B}\left[\frac{\Lambda^2}{2}+\frac{3}{2}m_B^2+\frac{3}{2}m^2_B
\log[\frac{m^2_B}{\Lambda^2}]\right]\{-3k_1\cdot k_2 k_3\cdot k_4 k_5\cdot k_6
\\ \nonumber &+&[k_1\cdot k_2(k_3\cdot k_5 k_4\cdot k_6+k_3\cdot k_6 k_4\cdot k_5)+k_3\cdot k_4(k_1\cdot k_5 k_2\cdot k_6 +k_1\cdot k_6k_2\cdot k_5)
\\ &+&(1;2)\leftrightarrow(3;4)+(1;2)\leftrightarrow(5;6)]\}\quad.
\end{eqnarray}

Finally,
\begin{eqnarray}
\nonumber M^6_{B3}&=&\frac{i}{2T^3}\sum_{B}\int
\frac{d^2p}{(2\pi)^2}\frac{-i}{p^2+m^2_B}
\\ \nonumber&\times&\{k_1\cdot k_2k_3\cdot k_4k_5\cdot k_6(9p^2-3m^2)-2k_1\cdot pk_2\cdot pk_3\cdot k_4k_5\cdot k_6
\\ \nonumber&-&2k_3\cdot pk_4\cdot pk_1\cdot k_2k_5\cdot k_6-2k_5\cdot pk_6\cdot pk_3\cdot k_4k_1\cdot k_2
\\ \nonumber&+&(m^2-p^2)[k_1\cdot k_2(k_3\cdot k_5k_4\cdot k_6+k_3\cdot k_6k_4\cdot k_5)+(1;2)\leftrightarrow(3;4)+(1;2)\leftrightarrow(5;6)]
\\ \nonumber&-&2[k_1\cdot p k_2\cdot p(k_3\cdot k_5k_4\cdot k_6+k_3\cdot k_6k_4\cdot k_5)+(1;2)\leftrightarrow(3;4)+(1;2)\leftrightarrow(5;6)]\}
\\ \nonumber&=&\frac{1}{2\pi T^3}\sum_{B}\left(\frac{1}{2}\Lambda^2+\frac{3}{4}m^2_B\log[\frac{m_B^2}{\Lambda^2}]\right)
\{3k_1\cdot k_2k_3\cdot k_4k_5\cdot k_6
\\  &-&[k_1\cdot k_2(k_3\cdot k_5k_4\cdot k_6+k_3\cdot k_6k_4\cdot k_5)+(1;2)\leftrightarrow(3;4)+(1;2)\leftrightarrow(5;6)]\}\quad.
\end{eqnarray}
Summing our results, and putting in the necessary fermionic contributions to cancel the divergences, we obtain the following 6-point function,
\begin{eqnarray}
\nonumber &&\vev{X_i(k_1)X_j(k_2)X_k(k_3)X_l(k_4)X_m(k_5)X_n(k_6)}=\frac{i\Delta T}{T^3}\delta_{ij}\delta_{kl}\delta_{mn}
\\ \nonumber&&\{ k_1\cdot k_2 k_3 \cdot k_4 k_5\cdot k_6-[k_1\cdot k_2 (k_3\cdot k_5 k_4\cdot k_6+k_4\cdot k_6 k_3\cdot k_5)+ (k_1,k_2)\rightarrow(k_3,k_4)
\\ \nonumber&&+(k_1,k_2)\rightarrow(k_5,k_6)]+[k_1\cdot k_3 (k_2\cdot k_5 k_4\cdot k_6+k_2\cdot k_6 k_4\cdot k_5)
\\ \nonumber&&+k_1\cdot k_4 (k_2\cdot k_5 k_3\cdot k_6+k_2\cdot k_6 k_3\cdot k_5)+(k_1,k_2)\rightarrow(k_3,k_4)+(k_1,k_2)\rightarrow(k_5,k_6)]\}
\\&&+\textrm{ permutations on }[(i,1),(j,2),(k,3),(l,4),(m,5),(n,6)]\}\quad.
\end{eqnarray}
However, it turns out that when we use the on-shell constraints this exactly vanishes.
Thus, the terms with six derivatives and six $X$'s in our action precisely agree with
their Nambu-Goto values, as expected from our general arguments of section \ref{sec:effectiveaction}.

\section{Conclusions}\label{sec:conclusions}

In this paper we analyzed the low-energy effective action of
confining strings. We computed its partition function using a zeta-function
regularization, argued in \cite{bib:DF} to be the unique regularization which gives
results that are independent of the UV cutoff (as we expect).
We showed that up to four-derivative order this
action must agree with the Nambu-Goto form, generalizing a result of
L\"uscher and Weisz for $D=3$. At the six-derivative order there are
three possible terms for general $D$, and we showed that our considerations do not
constrain the term $c_4$ that does not appear in the Nambu-Goto
action (the two terms appearing in
Nambu-Goto, $c_6$ and $c_7$, are uniquely determined).
Somewhat surprisingly, we found that this coefficient does
not contribute to the partition function on the torus at the first
possible order, corresponding to corrections to closed string
energies of order $1/L^5$. Thus, the corrections to energy levels
coming from this term must sum to zero
separately at each energy level. In particular we claim that the
closed string ground state energy is not corrected at order $1/L^5$
(compared to the Nambu-Goto result), so its first corrections arise
at least at order $1/L^7$. For the special case of $D=3$ we find that
there is only one coefficient in the effective action at six-derivative
order, which is uniquely determined, so that all energy levels must agree
with the Nambu-Goto results up to order $1/L^5$. This seems to be consistent with lattice
results indicating that the corrections to Nambu-Goto for the ground
state are very small \cite{bib:lat1, bib:lat2}; it is not consistent
with lattice results for percolation presented in \cite{bib:lat3},
but it is not clear if these results are reliable and if the
corresponding string theory has a weakly coupled limit where our
results should apply. It would be interesting to use lattice
simulations to measure the value of $c_4$ for
interesting confining theories with $D > 3$, such as the pure Yang-Mills theory
in $D=4$.

In the partition function on the annulus, we found that the $c_4$
term does contribute corrections to closed string energies when
$D>3$. Recall that
while the torus partition function sums over all closed string
states with weight one, the annulus partition function sums only
over specific closed string states, which have some overlap with the
boundary state, and the sum comes with different coefficients for
different states.

We then computed the specific coefficients appearing in the
effective action in a large class of holographic backgrounds, by
integrating out the massive fields on the worldsheet of strings in
confining backgrounds, to leading order in the curvature. We
verified that up to four-derivatives the Nambu-Goto action is
reproduced, and we showed that also at six-derivative order the effective
action precisely agrees with Nambu-Goto.
Somewhat surprisingly, we did
not find any $c_4$ term; it is possible that such a term only arises
at higher orders, or that it is constrained to vanish by
considerations different than the ones we used here (for instance,
by the constraints arising in the formalism of Polchinski and
Strominger). It would be interesting to understand this better. In
any case, in the backgrounds we study this means that at one-loop
order there is no correction to the partition function at
six-derivative order, both on the torus and on the annulus (at least
with the specific boundary terms we chose).

Some possible generalizations of our analysis are :
\begin{itemize}
\item It would be interesting to go to higher orders in the derivative expansion,
in particular to see at what order the effective action for $D=3$ can first deviate
from the Nambu-Goto form, and at what order corrections to the ground state energy
(for any $D$) can start appearing.
\item It would be interesting to go to two-loop order in the computation of the
effective action in holographic backgrounds, to see whether the $c_4$ term is generated
at this order or not. In particular, it would be interesting to see whether the
effective action computed in this formalism precisely agrees with the Nambu-Goto
action (to all orders in the derivative expansion), or whether deviations occur
at some order, and, if so, at what order the deviations first occur.
\item We computed the effective action and the resulting partition function, but we
did not use the effective action to compute the corrections to
specific energy levels; it would be interesting to do this.
\item We focused on closed strings, and assumed there are no boundary terms. It would
be interesting to analyze what are the possible boundary terms that
could contribute up to the order we worked in, and to compute the
corresponding corrections to open string energies. In particular, it
would be interesting to see which boundary terms appear in the
computation of the quark-anti-quark potential in holographic
backgrounds (see \cite{Kinar:1999xu} for the expansion of the action of
a holographic open confining string to quadratic order in fluctuations).
\item We assumed that the only massless fields on the worldsheet are the transverse fluctuations, but in many interesting cases (like supersymmetric gauge theories) there are
additional massless fields on the worldsheet. It would be
interesting to generalize our analysis to these cases, to see what are
the allowed terms in the effective action and whether they
contribute to the partition function or not.
\item As we mentioned in section \ref{sec:generalities}, our analysis does not apply directly to $k$-strings since they have light states on their worldsheet in the large $N$ limit; it would be interesting to understand better the form of the low-energy
effective action on $k$-strings, and to match it with recent lattice
results.
\item We considered here only orientable strings, which are relevant for
$SU(N)$ gauge theories. For $SO(N)$ or $USp(N)$ theories the confining string
is unoriented, so additional diagrams (such as a Klein bottle) are possible.
It would be interesting to check if these diagrams give additional constraints
on the effective action, and to compute the effective action for holographic
backgrounds that correspond to such theories.
\item In our analysis we wrote the effective action using only the physical
transverse fluctuations. It would be interesting to compare this, and our
constraints on the possible terms, to other formalisms, such as working in
a Poincar\'e invariant formalism and adding terms involving the extrinsic
curvature of the worldsheet (such as the ``rigidity term'' \cite{Polyakov:1986cs,Kavalov:1985di}), and the Polchinski-Strominger approach \cite{bib:constraints1}. The standard ``rigidity
term'' seems to be trivial in our long-string effective action
(in the sense that we can get rid of it by field
definitions) up to the order that we work in, but it may appear at higher orders.
\item It may also be possible to obtain inequalities on coefficients in the
effective action by using unitarity considerations, as (for instance) in \cite{Adams:2006sv}. For example, the positivity of $c_2$ (which is true) may
be argued just from these considerations. It would be interesting to incorporate
these additional constraints into our analysis.
\end{itemize}

\centerline{}
\centerline{\bf Acknowledgements}

We would like to thank Zohar Komargodski for collaboration on the
early parts of this project and for many useful discussions. We
would like to thank Mike Teper and Barak Bringoltz for many
interesting discussions which motivated this project, and for
detailed descriptions of their results. We thank Ferdinando Gliozzi
for pointing out to us a mistake in the first version of this paper.
We also thank Adi Armoni,
Ioannis Bakas, Barak Kol, David Kutasov, Julius Kuti,
Juan Maldacena, Adam Schwimmer, and especially Cobi Sonnenschein for
useful discussions. This work was supported in part by the
Israel-U.S. Binational Science Foundation, by a center of excellence
supported by the Israel Science Foundation (grant number 1468/06),
by a grant (DIP H52) of the German Israel Project Cooperation, by
the European network MRTN-CT-2004-512194, and by Minerva.

\appendix

\section{Computations for section 3}\label{sec:appendixA}

We follow the computation technique used in \cite{bib:DF}, involving
zeta function regularization, and we perform our calculations in position
space. The Feynman bubble diagrams appearing in the partition
function computations will be sums over the modes of the
worldsheet fields. These sums are typically UV divergent, and must be
regularized. Fortunately, the regularization scheme used in
\cite{bib:DF} is claimed to produce a unique finite result for any
calculation which is not cutoff-dependent. We know the partition
function is finite. Furthermore, we expect it to be finite diagram
by diagram (in the low-energy effective action),
since in our full action the divergences cancel between scalars and
fermions (while the low-energy effective action includes only scalars).
The diagrams should also
fulfill requirements such as scale invariance and
other requirements which are listed in \cite{bib:DF}. The claim is
that for such a calculation, there is a unique finite result, which
is obtained using the regularization scheme of \cite{bib:DF}.

In this scheme we analytically continue the sums
using the zeta function, so that a generic sum is written as
\begin{eqnarray}
\sum_{n,m=-\infty}^{\infty}\frac{m^a n^b}{m^2+n^2}\equiv\sum_{n,m=-\infty}^{\infty}\frac{m^{a+s} n^{b+s'}}{m^2+n^2}|_{s,s'=0}\ .
\end{eqnarray}
This can be further manipulated such that the divergent part will
always appear as a zeta function, as we will see in
subsection \ref{ssec:sumreg}. The difference of this scheme from
dimensional regularization is that there is no single parameter
which we perform the analytic continuation in. We analytically
continue and regularize each sum by itself. This will give us finite
results as long as we do not hit any poles of the zeta function,
that is as long as we do not have logarithmically divergent sums
($\zeta(1)$). This is similar to dimensional
regularization, where only logarithmic divergences are seen.

In the first two subsections we define some modular functions
and compute divergent sums which appear in our computation. We explicitly
write their regularization using the zeta function. In the following
subsections we write the details of the partition function
computation at order $O(T^{-1})$ and $O(T^{-2})$, both on the annulus
and on the torus. Finally, we explain the numerical method used
to determine the $\tilde{q}$ expansion of $F(q)$.

\subsection{Modular functions}\label{ssec:modfun}

Below is a list of functions which are related to the Dedekind eta function \eqref{etafunc},
and have nice properties under modular transformations. We define
$q, \tilde{q}, \tau, \tilde{\tau}$ as in \eqref{partdefs}.
We recall the Eisenstein series $E_{2k}(q)$,
\begin{eqnarray}\label{eq:modfun1}
E_{2k}(q)&=&1+\frac{2}{\zeta(1-2k)}\sum_{n=1}^{\infty}\frac{n^{2k-1}q^n}{1-q^n},
\end{eqnarray}
and define the functions $H_{2,k}(q)$ (for even $k$) :
\begin{eqnarray}
H_{2,k}(q)\equiv\frac{\zeta(1-k)}{2} q \frac{\partial}{\partial
q} E_{k}(q) =\sum_{n=1}^{\infty}\frac{n^k q^n}{(1-q^n)^2}.
\end{eqnarray}
For the cases we will encounter these are given by
\begin{equation}
H_{2,2}(q) = {{E_4(q) - E_2(q)^2}\over 288}, \qquad H_{2,4}(q) = {{E_2(q) E_4(q) - E_6(q)}
\over 720}.
\end{equation}

These functions obey the modular transformation properties :
\begin{eqnarray}\label{eq:modfun3}
\nonumber E_2(q)&=&-\frac{6i}{\pi} \tilde{\tau}+{\tilde\tau}^2
E_2(\tilde{q}),\quad
H_{2,2}(q)=\frac{\log(\tilde{q})^2}{4\pi^4}[-\frac{1}{8}-
\frac{1}{48}\log(\tilde{q})E_2(\tilde{q})
+\frac{1}{4}\log(\tilde{q})^2H_{2,2}(\tilde{q})],
\\ E_{k}(q)&=&\tilde{\tau}^{k}E_{k}(\tilde{q}),\quad
H_{2,k}(q)=-\frac{i k\zeta(1-k)}{4\pi} \tilde{\tau}^{k+1}E_k(\tilde{\tau})+\tilde{\tau}^{k+2} H_{2,k}(\tilde{q})
\quad \forall k>2.
\end{eqnarray}

Finally, we define the function $F(q^{ann.})$ :
\begin{eqnarray}
\nonumber F(q^{ann.})&=&\sum_{n,r=1}^{\infty}n r (n+ r)\coth(\frac{n
\pi L}{2R})\coth(\frac{(n+ r) \pi L}{2R})\coth(\frac{r \pi L}{2R}),
\\ &=&\sum_{r>n,n=1}^{\infty}n r (n- r)\coth(\frac{n \pi L}{2R})\coth(\frac{(n- r) \pi L}{2R})\coth(\frac{r \pi L}{2R})\quad.
\end{eqnarray}

\subsection{Regularization of sums}\label{ssec:sumreg}

Below we list the sums which appear in our computations. We write the
sums appearing in the annulus computation, but they are simply related
to the ones appearing for the torus. The list
includes diverging sums, which we manipulate such that the
divergence is always expressed using a zeta function.
\begin{eqnarray}
\nonumber
\sum_{m=-\infty}^{\infty}\frac{1}{\frac{n^2}{R^2}+\frac{4m^2}{L^2}}&=&\frac{\pi
R L}{2n} \coth(\frac{n \pi L}{2R}) ,\quad
\sum_{m=1}^{\infty}\frac{1}{\frac{n^2}{R^2}+\frac{4m^2}{L^2}}=\frac{\pi
R L}{4n} \coth(\frac{n \pi L}{2R})-\frac{R^2}{2n^2}
\\ \nonumber \sum_{m=-\infty}^{\infty}\frac{m^2}{\frac{n^2}{R^2}+\frac{4m^2}{L^2}}&=&\sum_{m=-\infty}^{\infty}\frac{m^{2+s}}{\frac{n^2}{R^2}+\frac{4m^2}{L^2}}|_{s=0}=\frac{L
^2}{4}\sum_{m=-\infty}^{\infty}m^s
(1-\frac{n^2}{R^2}\frac{1}{\frac{n^2m^s}{R^2}+\frac{4m^2}{L^2}})|_{s=0}
\\ \nonumber&=&\frac{L^2}{4}(1+2\zeta(0))-\frac{L^3n\pi}{8R}\coth(\frac{\pi n L}{2R})=-\frac{L^3n\pi}{8R}\coth(\frac{\pi n L}{2R}),
\\
\nonumber \sum_{m=-\infty}^{\infty}\frac{m^4}{\frac{n^2}{R^2}+\frac{4m^2}{L^2}}&=&\sum_{m=-\infty}^{\infty}\frac{m^{4+s}}{\frac{n^2}{R^2}+\frac{4m^2}{L^2}}|_{s=0}=\frac{L^2}{4}
\sum_{m=-\infty}^{\infty}m^{2+s}(1-\frac{n^2}{R^2}\frac{1}{\frac{n^2}{R^2}+\frac{4m^2}{L^2}})|_{s=0}
\\ \nonumber&=&\frac{L^2}{2}\zeta(-2)-\frac{L^2n^2}{4R^2}\sum_{m=-\infty}^{\infty}\frac{m^{2+s}}{\frac{n^2}{R^2}+\frac{4m^2}{L^2}}=\frac{L^5n^3\pi}{32R^3}\coth(\frac{n \pi L}{2R}),
\end{eqnarray}
\begin{eqnarray}\label{eq:sums}
\nonumber \sum_{n=1}^{\infty}n^s\coth(\frac{n \pi L}{2R})&=&\sum_{n=1}^{\infty}n^s\frac{e^{\frac{\pi n L}{R}}+1}{e^{\frac{\pi n L}{R}}-1}=\sum_{n=1}^{\infty}n^s(1+\frac{2}{e^{\frac{\pi n L}{R}}-1})=\zeta(-s)+2\sum_{n=1}^{\infty}\frac{n^s q_{ann.}^n}{1-q_{ann.}^n}
\\ \nonumber&=&\zeta(-s)E_{s+1}(q^{ann.}),
\\ \nonumber \sum_{n=1}^{\infty}n^s\coth^2(\frac{n \pi L
}{2R})&=&\sum_{n=1}^{\infty}n^s\left(\frac{e^{\frac{\pi n
L}{R}}+1}{e^{\frac{\pi n
L}{R}}-1}\right)^2=\sum_{n=1}^{\infty}n^s\left(1+4\frac{e^{\frac{\pi n
L}{R}}}{(e^{\frac{\pi n L}{R}}-1)^2}\right)
\\ &=&\zeta(-s)+4\sum_{n=1}^{\infty}\frac{n^s q_{ann.}^n}{(1-q_{ann.}^n)^2}=\zeta(-s)+4H_{2,s}(q^{ann.}).
\end{eqnarray}

\subsection{The annulus}

The Green's function on a cylinder worldsheet with rectangular domain $(R,L)$ is
\begin{eqnarray}
G(\sigma_1,\sigma_0;{\sigma_1}', {\sigma_0}')=\frac{2}{\pi^2 R
L}\sum_{n=1,m=-\infty}^{\infty}\frac{\sin (\frac{n \pi \sigma_1}{R})
\sin (\frac{n \pi {\sigma_1}'}{R}) e^{\frac{2\pi i
m(\sigma_0-{\sigma_0}')}{L}}}{\frac{n^2}{R^2}+4\frac{m^2}{L^2}}\quad.
\end{eqnarray}
Here $L$ is the periodic direction, and $0\leq\sigma_1\leq R$.

\subsubsection{Partition function at $O(L^{-3})$}

This computation was already performed in \cite{bib:DF}, and we reproduce it here for completeness.
The diagrams at this order are (see figure \ref{fig:Z2})
\begin{eqnarray}
\nonumber I_1^{ann.}&=&\int d^2\sigma\partial_{\alpha}\partial^{\alpha'}G\partial_{\beta}\partial^{\beta'}G=\int d^2\sigma\{(\partial_0\partial_0'G+\partial_1\partial_1'G)(\partial_0\partial_0'G+\partial_1\partial_1'G)\}
\\ \nonumber &=&\frac{2\pi^2L}{R^3}H_{2,2}(q^{ann.})=-\frac{1}{ L R}+\frac{2\pi}{3L^2}E_2(\tilde{q}) +\frac{32\pi^2 R}{ L^3}H_{2,2}(\tilde{q}),
\\ \nonumber I_2^{ann.}&=&\int d^2\sigma\partial_{\alpha}\partial_{\beta'}G\partial^{\alpha}\partial^{\beta'}G=\int d^2\sigma\{\partial_0\partial_0'G\partial_0\partial_0'G+\partial_1\partial_1'G\partial_1\partial_1'G\}
\\ &=&\frac{\pi^2L}{ R^3}\left[\frac{2}{(24)^2}E_2^2(q^{ann.})+H_{2,2}(q^{ann.})\right]=\frac{16\pi^2R}{ L^3}\left[\frac{2}{(24)^2}E_2^2(\tilde{q})+H_{2,2}(\tilde{q})\right],
\end{eqnarray}
where
\begin{eqnarray}
\partial_{\alpha}&=&\frac{\partial}{\partial\sigma_{\alpha}},
\qquad \partial_{\alpha}\partial_{\alpha}'G=\lim_{\sigma'\rightarrow\sigma}\partial_{\alpha}\partial_{\alpha}'G(\sigma,\sigma')\quad.
\end{eqnarray}
In this notation, we first take the derivative with respect to
$\sigma$ or $\sigma'$ and only then take the limit
$\sigma\rightarrow\sigma'$. One should notice that an odd number of
derivatives of the propagators with respect to $\sigma^0$ gives a sum
of antisymmetric functions of $\sigma^0-{\sigma^0}'$, and therefore
vanishes as $\sigma^0\rightarrow{\sigma^0}'$.

Below are the details of the computation:
\begin{eqnarray}
\nonumber &&\int d^2\sigma\partial_1\partial_1'G\partial_1\partial_1'G
\\ \nonumber&=&\frac{4}{R^6L^2}\sum_{n,k=1}^{\infty}\sum_{m,l=-\infty}^{\infty}\frac{n^2k^2}{(\frac{n^2}{R^2}+\frac{4m^2}{L^2})(\frac{k^2}{R^2}+\frac{4l^2}{L^2})}\int d^2\sigma \cos^2(\frac{n\pi \sigma}{R})\cos^2(\frac{k\pi \sigma}{R})
\\ \nonumber&=&\frac{1}{R^5L}\{(\sum_{m,n}\frac{n^2}{\frac{n^2}{R^2}+\frac{4m^2}{L^2}})(\sum_{k,l}\frac{k^2}{\frac{k^2}{R^2}+\frac{4l^2}{L^2}})
+\frac{1}{2}\sum_n
n^4(\sum_{m}\frac{1}{\frac{n^2}{R^2}+\frac{4m^2}{L^2}})(\sum_{l}\frac{1}{\frac{n^2}{R^2}+\frac{4l^2}{L^2}})\}
\\ &=&\frac{\pi^2L}{4R^3}\left((\zeta(-1)E_2(q^{ann.}))^2+\frac{1}{2}(\zeta(-2)+4H_{2,2}(q^{ann.}))\right),
\end{eqnarray}
\begin{eqnarray}
\nonumber &&\int d^2\sigma\partial_0\partial_0'G\partial_1\partial_1'G
\\ \nonumber&=&\frac{16}{R^4L^4}\sum_{n,k=1}^{\infty}\sum_{m,l=-\infty}^{\infty}\frac{n^2l^2}{(\frac{n^2}{R^2}+\frac{4m^2}{L^2})(\frac{k^2}{R^2}+\frac{4l^2}{L^2})}\int d^2\sigma \cos^2(\frac{n\pi \sigma}{R})\sin^2(\frac{k\pi \sigma}{R})
\\ \nonumber&=&\frac{4}{R^3L^3}\{(\sum_{m,n}\frac{n^2}{\frac{n^2}{R^2}+\frac{4m^2}{L^2}})(\sum_{k,l}\frac{l^2}{\frac{k^2}{R^2}+\frac{4l^2}{L^2}})
-\frac{1}{2}\sum_n
(\sum_{m}\frac{n^2}{\frac{n^2}{R^2}+\frac{4m^2}{L^2}})(\sum_{l}\frac{l^2}{\frac{n^2}{R^2}+\frac{4l^2}{L^2}})\}
\\ &=&-\frac{\pi^2L}{4R^3}\left((\zeta(-1)E_2(q^{ann.}))^2-
\frac{1}{2}(\zeta(-2)+4H_{2,2}(q^{ann.}))\right),
\end{eqnarray}
\begin{eqnarray}
\nonumber&&\int d^2\sigma\partial_0\partial_0'G\partial_0\partial_0'G
\\ \nonumber&=&\frac{64}{R^2L^6}\sum_{n,k=1}^{\infty}\sum_{m,l=-\infty}^{\infty}\frac{m^2l^2}{(\frac{n^2}{R^2}+\frac{4m^2}{L^2})(\frac{k^2}{R^2}+\frac{4l^2}{L^2})}\int d^2\sigma \sin^2(\frac{n\pi \sigma}{R})\sin^2(\frac{k\pi \sigma}{R})
\\ \nonumber&=&\frac{16}{RL^5}\{(\sum_{m,n}\frac{m^2}{\frac{n^2}{R^2}+\frac{4m^2}{L^2}})(\sum_{k,l}\frac{l^2}{\frac{k^2}{R^2}+\frac{4l^2}{L^2}})
+\frac{1}{2}\sum_n
(\sum_{m}\frac{m^2}{\frac{n^2}{R^2}+\frac{4m^2}{L^2}})(\sum_{l}\frac{l^2}{\frac{n^2}{R^2}+\frac{4l^2}{L^2}})\}
\\&=&\frac{\pi^2L}{4R^3}\left((\zeta(-1)E_2(q^{ann.}))^2+
\frac{1}{2}(\zeta(-2)+4H_{2,2}(q^{ann.}))\right)\quad.
\end{eqnarray}

\subsubsection{Partition function at $O(L^{-5})$}

At this order there are both 2-loop and 3-loop diagram
contributions (see figures \ref{fig:Z2} and \ref{fig:Z3}). At two loops there are
two possible contractions,
\begin{eqnarray}
\nonumber I_3^{ann.}&=& \int d^2\sigma\partial_{\alpha}\partial^{\alpha'}\partial_{\beta}\partial^{\beta'}G\partial_{\gamma}\partial^{\gamma'}G
\\ \nonumber&=&\int
d^2\sigma(\partial_0\partial_0'\partial_0\partial_0'G+\partial_1\partial_1'\partial_1\partial_1'G+2\partial_1\partial_1'\partial_0\partial_0'G)(\partial_1\partial_1'G+\partial_0\partial_0'G)
\\ &=&-4\frac{\pi^4L}{R^5}H_{2,4}(q^{ann.})=-4\frac{\pi^4L}{R^5}\left(\frac{4R^5}{15\pi
L^5}E_4(\tilde{\tau})-\frac{64R^6}{L^6}H_{2,4}(\tilde{q})\right),
\\ \nonumber I_4^{ann.}&=&\int d^2\sigma\partial_{\alpha}\partial_{\beta}\partial_{\gamma'}G\partial^{\alpha}\partial^{\beta}\partial^{\gamma'}G
\\ \nonumber&=&\int
d^2\sigma\{\partial_0\partial_0\partial_0'G\partial_0\partial_0\partial_0'G+2\partial_0\partial_1\partial_1'G\partial_0\partial_1\partial_1'G+\partial_0'\partial_1\partial_1G\partial_0'\partial_1\partial_1G
\\ \nonumber&&+2\partial_1\partial_0\partial_0'G\partial_1\partial_0\partial_0'G+\partial_0\partial_0\partial_1'G\partial_0\partial_0\partial_1'G+\partial_1\partial_1\partial_1'G\partial_1\partial_1\partial_1'G\}
\\ &=&2\frac{\pi^4L}{R^5}H_{2,4}(q^{ann.}).
\end{eqnarray}

Below are the details of the computation:
\begin{eqnarray}
\nonumber&&\int
d^2\sigma\partial_1\partial_1'\partial_1\partial_1'G\partial_1\partial_1'G
\\ \nonumber&=&\frac{4\pi^2}{R^8L^2}\sum_{n,k=1}^{\infty}\sum_{m,l=-\infty}^{\infty}\frac{n^4k^2}{(\frac{n^2}{R^2}+\frac{4m^2}{L^2})(\frac{k^2}{R^2}+\frac{4l^2}{L^2})}\int d^2\sigma \sin^2(\frac{n\pi \sigma}{R})\cos^2(\frac{k\pi \sigma}{R})
\\ \nonumber&=&\frac{\pi^2}{R^7L}\{(\sum_{m,n}\frac{n^4}{\frac{n^2}{R^2}+\frac{4m^2}{L^2}})(\sum_{k,l}\frac{k^2}{\frac{k^2}{R^2}+\frac{4l^2}{L^2}})
-\frac{1}{2}\sum_n
n^6(\sum_{m}\frac{1}{\frac{n^2}{R^2}+\frac{4m^2}{L^2}})(\sum_{l}\frac{1}{\frac{n^2}{R^2}+\frac{4l^2}{L^2}})\}
\\ &=&\frac{\pi^4L}{4R^5}(\zeta(-3)\zeta(-1)E_2(q^{ann.})E_4(q^{ann.})-\frac{1}{2}(\zeta(-4)+4H_{2,4}(q^{ann.}))),
\end{eqnarray}
\begin{eqnarray}
\nonumber&&\int
d^2\sigma\partial_1\partial_1'\partial_1\partial_1G\partial_1\partial_1'G
\\ \nonumber&=&-\frac{4\pi^2}{R^8L^2}\sum_{n,k=1}^{\infty}\sum_{m,l=-\infty}^{\infty}\frac{n^4k^2}{(\frac{n^2}{R^2}+\frac{4m^2}{L^2})(\frac{k^2}{R^2}+\frac{4l^2}{L^2})}\int d^2\sigma \cos^2(\frac{n\pi \sigma}{R})\cos^2(\frac{k\pi \sigma}{R})
\\ \nonumber&=&-\frac{\pi^2}{R^7L}\{(\sum_{m,n}\frac{n^4}{\frac{n^2}{R^2}+\frac{4m^2}{L^2}})(\sum_{k,l}\frac{k^2}{\frac{n^2}{R^2}+\frac{4l^2}{L^2}})
+\frac{1}{2}\sum_n
n^6(\sum_{m}\frac{1}{\frac{n^2}{R^2}+\frac{4m^2}{L^2}})(\sum_{l}\frac{1}{\frac{n^2}{R^2}+\frac{4l^2}{L^2}})\}
\\ &=&-\frac{\pi^4L}{4R^5}(\zeta(-3)\zeta(-1)E_2(q^{ann.})E_4(q^{ann.})+\frac{1}{2}(\zeta(-4)+4H_{2,4}(q^{ann.}))),
\end{eqnarray}
\begin{eqnarray}
\nonumber&&\int
d^2\sigma\partial_1\partial_1'\partial_1\partial_1'G\partial_0\partial_0'G
\\ \nonumber&=&\frac{16\pi^2}{R^6L^4}\sum_{n,k=1}^{\infty}\sum_{m,l=-\infty}^{\infty}\frac{n^4l^2}{(\frac{n^2}{R^2}+\frac{4m^2}{L^2})(\frac{k^2}{R^2}+\frac{4l^2}{L^2})}\int d^2\sigma \sin^2(\frac{n\pi \sigma}{R})\sin^2(\frac{k\pi \sigma}{R})
\\ \nonumber&=&\frac{4\pi^2}{R^5L^3}\{(\sum_{m,n}\frac{n^4}{\frac{n^2}{R^2}+\frac{4m^2}{L^2}})(\sum_{k,l}\frac{l^2}{\frac{k^2}{R^2}+\frac{4l^2}{L^2}})
+\frac{1}{2}\sum_n
n^4(\sum_{m}\frac{1}{\frac{n^2}{R^2}+\frac{4m^2}{L^2}})(\sum_{l}\frac{l^2}{\frac{n^2}{R^2}+\frac{4l^2}{L^2}})\}
\\ &=&\frac{\pi^4L}{4R^5}\{-\zeta(-1)\zeta(-3)E_2(q^{ann.})E_4(q^{ann.})-\frac{1}{2}(\zeta(-4)+4H_{2,4}(q^{ann.}))\},
\end{eqnarray}
\begin{eqnarray}
\nonumber&&\int
d^2\sigma\partial_1\partial_1'\partial_0\partial_0'G\partial_1\partial_1'G
\\ \nonumber&=&\frac{16\pi^2}{R^6L^4}\sum_{n,k=1}^{\infty}\sum_{m,l=-\infty}^{\infty}\frac{n^2m^2k^2}{(\frac{n^2}{R^2}+\frac{4m^2}{L^2})(\frac{k^2}{R^2}+\frac{4l^2}{L^2})}\int d^2\sigma \cos^2(\frac{n\pi \sigma}{R})\cos^2(\frac{k\pi \sigma}{R})
\\ \nonumber&=&\frac{4\pi^2}{R^5L
^3}\{(\sum_{m,n}\frac{n^2m^2}{\frac{n^2}{R^2}+\frac{4m^2}{L^2}})(\sum_{k,l}\frac{k^2}{\frac{k^2}{R^2}+\frac{4l^2}{L^2}})
+\frac{1}{2}\sum_n
n^4(\sum_{m}\frac{m^2}{\frac{n^2}{R^2}+\frac{4m^2}{L^2}})(\sum_{l}\frac{1}{\frac{n^2}{R^2}+\frac{4l^2}{L^2}})\}
\\ &=&\frac{\pi^4L}{4R^5}\{-\zeta(-1)\zeta(-3)E_2(q^{ann.})E_4(q^{ann.})-\frac{1}{2}(\zeta(-4)+4H_{2,4}(q^{ann.}))\},
\end{eqnarray}
\begin{eqnarray}
\nonumber&&\int
d^2\sigma\partial_1\partial_1\partial_0\partial_0'G\partial_1\partial_1'G
\\ \nonumber&=&-\frac{16\pi^2}{R^6L^4}\sum_{n,k=1}^{\infty}\sum_{m,l=-\infty}^{\infty}\frac{n^2m^2k^2}{(\frac{n^2}{R^2}+\frac{4m^2}{L^2})(\frac{k^2}{R^2}+\frac{4l^2}{L^2})}\int d^2\sigma \sin^2(\frac{n\pi \sigma}{R})\cos^2(\frac{k\pi \sigma}{R})
\\ \nonumber&=&-\frac{4\pi^2}{R^5L^3}\{(\sum_{m,n}\frac{n^2m^2}{\frac{n^2}{R^2}+\frac{4m^2}{L^2}})(\sum_{k,l}\frac{k^2}{\frac{k^2}{R^2}+\frac{4l^2}{L^2}})
-\frac{1}{2}\sum_n
n^4(\sum_{m}\frac{m^2}{\frac{n^2}{R^2}+\frac{4m^2}{L^2}})(\sum_{l}\frac{1}{\frac{n^2}{R^2}+\frac{4l^2}{L^2}})\}
\\ &=&\frac{\pi^4L}{4R^5}\{\zeta(-1)\zeta(-3)E_2(q^{ann.})E_4(q^{ann.})-\frac{1}{2}(\zeta(-4)+4H_{2,4}(q^{ann.}))\},
\end{eqnarray}
\begin{eqnarray}
\nonumber&&\int
d^2\sigma\partial_1\partial_1'\partial_0\partial_0'G\partial_0\partial_0'G=-\int
d^2\sigma\partial_1\partial_1'\partial_0\partial_0G\partial_0\partial_0'G
\\ \nonumber&=&\frac{64\pi^2}{R^4L^6}\sum_{n,k=1}^{\infty}\sum_{m,l=-\infty}^{\infty}\frac{n^2m^2l^2}{(\frac{n^2}{R^2}+\frac{4m^2}{L^2})(\frac{k^2}{R^2}+\frac{4l^2}{L^2})}\int d^2\sigma \cos^2(\frac{n\pi \sigma}{R})\sin^2(\frac{k\pi \sigma}{R})
\\ \nonumber&=&\frac{16\pi^2}{R^3L^5}\{(\sum_{m,n}\frac{n^2m^2}{\frac{n^2}{R^2}+\frac{4m^2}{L^2}})(\sum_{k,l}\frac{l^2}{\frac{k^2}{R^2}+\frac{4l^2}{L^2}})
-\frac{1}{2}\sum_n
n^2(\sum_{m}\frac{m^2}{\frac{n^2}{R^2}+\frac{4m^2}{L^2}})(\sum_{l}\frac{l^2}{\frac{n^2}{R^2}+\frac{4l^2}{L^2}})\}
\\ &=&\frac{\pi^4L}{4R^5}\{\zeta(-1)\zeta(-3)E_2(q^{ann.})E_4(q^{ann.})-\frac{1}{2}(\zeta(-4)+4H_{2,4}(q^{ann.}))\},
\end{eqnarray}
\begin{eqnarray}
\nonumber&&\int
d^2\sigma\partial_0\partial_0'\partial_0\partial_0'G\partial_1\partial_1'G
\\ \nonumber&=&\frac{64\pi^2}{R^4L^6}\sum_{n,k=1}^{\infty}\sum_{m,l=-\infty}^{\infty}\frac{m^4k^2}{(\frac{n^2}{R^2}+\frac{4m^2}{L^2})(\frac{k^2}{R^2}+\frac{4l^2}{L^2})}\int d^2\sigma \cos^2(\frac{n\pi \sigma}{R})\sin^2(\frac{k\pi \sigma}{R})
\\ \nonumber&=&\frac{16\pi^2}{R^3L^5}\{(\sum_{m,n}\frac{m^4}{\frac{n^2}{R^2}+\frac{4m^2}{L^2}})(\sum_{k,l}\frac{k^2}{\frac{k^2}{R^2}+\frac{4l^2}{L^2}})
-\frac{1}{2}\sum_n
n^2(\sum_{m}\frac{m^4}{\frac{n^2}{R^2}+\frac{4m^2}{L^2}})(\sum_{l}\frac{1}{\frac{n^2}{R^2}+\frac{4l^2}{L^2}})\}
\\ &=&\frac{\pi^4L}{4R^5}\{\zeta(-1)\zeta(-3)E_2(q^{ann.})E_4(q^{ann.})-\frac{1}{2}(\zeta(-4)+4H_{2,4}(q^{ann.}))\},
\end{eqnarray}
\begin{eqnarray}
\nonumber&&\int
d^2\sigma\partial_0\partial_0'\partial_0\partial_0'G\partial_0\partial_0'G=-\int
d^2\sigma\partial_0\partial_0'\partial_0\partial_0G\partial_0\partial_0'G
\\ \nonumber&=&\frac{256\pi^2}{R^2L^8}\sum_{n,k=1}^{\infty}\sum_{m,l=-\infty}^{\infty}\frac{m^4l^2}{(\frac{n^2}{R^2}+\frac{4m^2}{L^2})(\frac{k^2}{R^2}+\frac{4l^2}{L^2})}\int d^2\sigma \sin^2(\frac{n\pi \sigma}{R})\sin^2(\frac{k\pi \sigma}{R})
\\ \nonumber&=&\frac{64\pi^2}{RL^7}\{(\sum_{m,n}\frac{m^4}{\frac{n^2}{R^2}+\frac{4m^2}{L^2}})(\sum_{k,l}\frac{l^2}{\frac{k^2}{R^2}+\frac{4l^2}{L^2}})
+\frac{1}{2}\sum_n
(\sum_{m}\frac{m^4}{\frac{n^2}{R^2}+\frac{4m^2}{L^2}})(\sum_{l}\frac{l^2}{\frac{n^2}{R^2}+\frac{4l^2}{L^2}})\}
\\ &=&\frac{\pi^4L}{4R^5}\{-\zeta(-1)\zeta(-3)E_2(q^{ann.})E_4(q^{ann.})-\frac{1}{2}(\zeta(-4)+4H_{2,4}(q^{ann.}))\},
\end{eqnarray}
\begin{eqnarray}
&&\int
d^2\sigma\partial_1\partial_1\partial_1G\partial_1'\partial_1'\partial_1'G=\int
d^2\sigma\partial_1'\partial_1\partial_1G\partial_1'\partial_1\partial_1G
\\ \nonumber&=&\frac{4\pi^2}{R^8L^2}\sum_{n,k=1}^{\infty}\sum_{m,l=-\infty}^{\infty}\frac{n^3k^3}{(\frac{n^2}{R^2}+\frac{4m^2}{L^2})(\frac{k^2}{R^2}+\frac{4l^2}{L^2})}\frac{1}{4}\int d^2\sigma \sin(\frac{2n\pi \sigma}{R})\sin(\frac{2k\pi \sigma}{R})
\\ \nonumber&=&\frac{\pi^2}{2R^7L}\{\sum_n n^6(\sum_{m}\frac{1}{\frac{n^2}{R^2}+\frac{4m^2}{L^2}})(\sum_{l}\frac{1}{\frac{n^2}{R^2}+\frac{4l^2}{L^2}})\}
=\frac{\pi^4L}{8R^5}(\zeta(-4)+4H_{2,4}(q^{ann.})),
\end{eqnarray}
\begin{eqnarray}
\nonumber &&\int
d^2\sigma\partial_0\partial_0\partial_1G\partial_0'\partial_0'\partial_1'G=\int
d^2\sigma\partial_0\partial_0'\partial_1G\partial_0\partial_0'\partial_1G=-\int
d^2\sigma\partial_0\partial_0'\partial_1G\partial_0\partial_0\partial_1'G
\\ &=&\frac{64\pi^2}{R^4L^6}\sum_{n,k=1}^{\infty}\sum_{m,l=-\infty}^{\infty}\frac{nkm^2l^2}{(\frac{n^2}{R^2}+\frac{4m^2}{L^2})(\frac{k^2}{R^2}+\frac{4l^2}{L^2})}\frac{1}{4}\int d^2\sigma \sin(\frac{2n\pi \sigma}{R})\sin(\frac{2k\pi \sigma}{R})
\\ \nonumber&=&\frac{8\pi^2}{R^3L^5}\{\sum_n n^2(\sum_{m}\frac{m^2}{\frac{n^2}{R^2}+\frac{4m^2}{L^2}})(\sum_{l}\frac{l^2}{\frac{n^2}{R^2}+\frac{4l^2}{L^2}})\}
=\frac{\pi^4L}{8R^5}(\zeta(-4)+4H_{2,4}(q^{ann.})),
\end{eqnarray}
\begin{eqnarray}
&&\int
d^2\sigma\partial_1\partial_1'\partial_1G\partial_1'\partial_0\partial_0'G=\int
d^2\sigma\partial_1\partial_1'\partial_1G\partial_1\partial_0\partial_0'G
\\ \nonumber&=&\frac{16\pi^2}{R^6L^4}\sum_{n,k=1}^{\infty}\sum_{m,l=-\infty}^{\infty}\frac{n^3kl^2}{(\frac{n^2}{R^2}+\frac{4m^2}{L^2})(\frac{k^2}{R^2}+\frac{4l^2}{L^2})}(-\frac{1}{4})\int d^2\sigma \sin(\frac{2n\pi \sigma}{R})\sin(\frac{2k\pi \sigma}{R})
\\ \nonumber&=&\frac{2\pi^2}{R^5L^3}\{-\sum_n n^4(\sum_{m}\frac{1}{\frac{n^2}{R^2}+\frac{4m^2}{L^2}})(\sum_{l}\frac{l^2}{\frac{n^2}{R^2}+\frac{4l^2}{L^2}})\}
=\frac{\pi^4L}{8R^5}(\zeta(-4)+4H_{2,4}(q^{ann.}))\quad.
\end{eqnarray}

At three loops there are 3 possible contractions,
\begin{eqnarray}
\nonumber I_6^{ann.}&=&\int
d^2\sigma\{\partial_0\partial_0'G\partial_0\partial_0'G\partial_0\partial_0'G+\partial_1\partial_1'G\partial_1\partial_1'G\partial_1\partial_1'G
\\ && + 3\partial_0\partial_0'G\partial_0\partial_0'G\partial_1\partial_1'G+3\partial_1\partial_1'G\partial_1\partial_1'G\partial_0\partial_0'G\}=\frac{3\pi^3L}{2R^5}F(q^{ann.}),
\\ \nonumber I_7^{ann.}&=&\int d^2\sigma\partial^{\alpha}\partial'_{\alpha}G\partial^{\beta}\partial_{\gamma}'G\partial_{\beta}'\partial^{\gamma}G=\int d^2\sigma\{\partial_0\partial_0'G\partial_0\partial_0'G\partial_0\partial_0'G+\partial_1\partial_1'G\partial_1\partial_1'G\partial_1\partial_1'G
\\ && + \partial_0\partial_0'G\partial_0\partial_0'G\partial_1\partial_1'G+\partial_1\partial_1'G\partial_1\partial_1'G\partial_0\partial_0'G\}=\frac{3\pi^3L}{4R^5}F(q^{ann.}),
\\ \nonumber I_8^{ann.}&=&\int d^2\sigma\partial^{\alpha}\partial'_{\beta}G\partial^{\beta}\partial_{\gamma}'G\partial^{\gamma}\partial_{\alpha'}G=\int d^2\sigma\{\partial_0\partial_0'G\partial_0\partial_0'G\partial_0\partial_0'G+\partial_1\partial_1'G\partial_1\partial_1'G\partial_1\partial_1'G\}
\\ &=&\frac{3\pi^3L
}{8R^5}F(q^{ann.}).
\end{eqnarray}
The details are :
\begin{eqnarray}
\nonumber&&\int
d^2\sigma\partial_1\partial_1'G\partial_1\partial_1'G\partial_1\partial_1'G
\\ \nonumber&=&\frac{8}{R^9L^3}\sum_{n,k,r=1}^{\infty}\sum_{m,l,s=-\infty}^{\infty}\frac{n^2k^2r^2}{(\frac{n^2}{R^2}+\frac{4m^2}{L^2})(\frac{k^2}{R^2}+\frac{4l^2}{L^2})(\frac{r^2}{R^2}+\frac{4s^2}{L^2})}\int d^2\sigma \cos^2(\frac{n\pi \sigma}{R})\cos^2(\frac{k\pi \sigma}{R})\cos^2(\frac{r\pi \sigma}{R})
\\ \nonumber&=&\frac{1}{R^8L^2}\{(\sum_{m,n}\frac{n^2}{\frac{n^2}{R^2}+\frac{4m^2}{L^2}})(\sum_{k,l}\frac{k^2}{\frac{k^2}{R^2}+\frac{4l^2}{L^2}})(\sum_{r,s}\frac{r^2}{\frac{r^2}{R^2}+\frac{4s^2}{L^2}})
\\ \nonumber&&+\frac{3}{2}[\sum_n n^4(\sum_{m}\frac{1}{\frac{n^2}{R^2}+\frac{4m^2}{L^2}})(\sum_{l}\frac{1}{\frac{n^2}{R^2}+\frac{4l^2}{L^2}})](\sum_{r,s}\frac{r^2}{\frac{r^2}{R^2}+\frac{4s^2}{L^2}})
\\ \nonumber&&+\frac{1}{4}\sum_{n,r} n^2r^2(n+r)^2(\sum_{m}\frac{1}{\frac{n^2}{R^2}+\frac{4m^2}{L^2}})(\sum_{l}\frac{1}{\frac{(n+r)^2}{R^2}+\frac{4l^2}{L^2}})(\sum_{s}\frac{1}{\frac{r^2}{R^2}+\frac{4s^2}{L^2}})
\\ \nonumber&&+\frac{1}{4}\sum_{n>r} n^2r^2(n-r)^2(\sum_{m}\frac{1}{\frac{n^2}{R^2}+\frac{4m^2}{L^2}})(\sum_{l}\frac{1}{\frac{(n-r)^2}{R^2}+\frac{4l^2}{L^2}})(\sum_{s}\frac{1}{\frac{r^2}{R^2}+\frac{4s^2}{L^2}})
\\ \nonumber&&+\frac{1}{4}\sum_{n<r} n^2r^2(r-n)^2(\sum_{m}\frac{1}{\frac{n^2}{R^2}+\frac{4m^2}{L^2}})(\sum_{l}\frac{1}{\frac{(r-n)^2}{R^2}+\frac{4l^2}{L^2}})(\sum_{s}\frac{1}{\frac{r^2}{R^2}+\frac{4s^2}{L^2}})\}
\\ &=&\frac{\pi^3L}{8R^5}\{(\zeta(-1)E_2(q^{ann.}))^3+\frac{3}{2}\zeta(-1)E_2(q^{ann.})(\zeta(-2)+4H_{2,2}(q^{ann.}))+\frac{3}{4}F(q^{ann.})\},
\end{eqnarray}
\begin{eqnarray}
\nonumber&&\int
d^2\sigma\partial_1\partial_1'G\partial_1\partial_1'G\partial_0\partial_0'G
\\ \nonumber&=&\frac{32}{R^7L^5}\sum_{n,k,r=1}^{\infty}\sum_{m,l,s=-\infty}^{\infty}\frac{n^2k^2s^2}{(\frac{n^2}{R^2}+\frac{4m^2}{L^2})(\frac{k^2}{R^2}+\frac{4l^2}{L^2})(\frac{r^2}{R^2}+\frac{4s^2}{L^2})}\int d^2\sigma \cos^2(\frac{n\pi \sigma}{R})\cos^2(\frac{k\pi \sigma}{R})\sin^2(\frac{r\pi \sigma}{R})
\\ \nonumber&=&\frac{4}{R^6L^4}\{(\sum_{m,n}\frac{n^2}{\frac{n^2}{R^2}+\frac{4m^2}{L^2}})(\sum_{k,l}\frac{k^2}{\frac{k^2}{R^2}+\frac{4l^2}{L^2}})(\sum_{r,s}\frac{s^2}{\frac{r^2}{R^2}+\frac{4s^2}{L^2}})
\\ \nonumber&&-[\sum_n n^2(\sum_{m}\frac{1}{\frac{n^2}{R^2}+\frac{4m^2}{L^2}})(\sum_{s}\frac{s^2}{\frac{n^2}{R^2}+\frac{4s^2}{L^2}})](\sum_{l}\frac{k^2}{\frac{k^2}{R^2}+\frac{4l^2}{L^2}})
\\ \nonumber&&+\frac{1}{2}[\sum_n n^4(\sum_{m}\frac{1}{\frac{n^2}{R^2}+\frac{4m^2}{L^2}})(\sum_{l}\frac{1}{\frac{n^2}{R^2}+\frac{4l^2}{L^2}})](\sum_{r,s}\frac{s^2}{\frac{r^2}{R^2}+\frac{4s^2}{L^2}})
\\ \nonumber&&-\frac{1}{4}\sum_{n,k} n^2k^2(\sum_{m}\frac{1}{\frac{n^2}{R^2}+\frac{4m^2}{L^2}})(\sum_{l}\frac{1}{\frac{n^2}{R^2}+\frac{4l^2}{L^2}})(\sum_{r,s}\frac{s^2}{\frac{(n+k)^2}{R^2}+\frac{4s^2}{L^2}})
\\ \nonumber&&-\frac{1}{4}\sum_{n,k} n^2k^2(\sum_{m}\frac{1}{\frac{n^2}{R^2}+\frac{4m^2}{L^2}})(\sum_{l}\frac{1}{\frac{n^2}{R^2}+\frac{4l^2}{L^2}})(\sum_{r,s}\frac{s^2}{\frac{(n-k)^2}{R^2}+\frac{4s^2}{L^2}})\}
\\ &=&\frac{\pi^3L}{8R^5}\{-(\zeta(-1)E_2(q^{ann.}))^3+\frac{1}{2}\zeta(-1)E_2(q^{ann.})(\zeta(-2)+4H_{2,2}(q^{ann.}))+\frac{3}{4}F(q^{ann.})\},
\end{eqnarray}
\begin{eqnarray}
\nonumber&&\int
d^2\sigma\partial_1\partial_1'G\partial_0\partial_0'G\partial_0\partial_0'G
\\ \nonumber&=&\frac{128}{R^5L^7}\sum_{n,k,r=1}^{\infty}\sum_{m,l,s=-\infty}^{\infty}\frac{n^2l^2s^2}{(\frac{n^2}{R^2}+\frac{4m^2}{L^2})(\frac{k^2}{R^2}+\frac{4l^2}{L^2})(\frac{r^2}{R^2}+\frac{4s^2}{L^2})}\int d^2\sigma \cos^2(\frac{n\pi \sigma}{R})\sin^2(\frac{k\pi \sigma}{R})\sin^2(\frac{r\pi \sigma}{R})
\\ \nonumber&=&\frac{16}{R^4L^6}\{(\sum_{m,n}\frac{n^2}{\frac{n^2}{R^2}+\frac{4m^2}{L^2}})(\sum_{k,l}\frac{l^2}{\frac{k^2}{R^2}+\frac{4l^2}{L^2}})(\sum_{r,s}\frac{s^2}{\frac{r^2}{R^2}+\frac{4s^2}{L^2}})
\\ \nonumber&&-[\sum_n n^2(\sum_{m}\frac{1}{\frac{n^2}{R^2}+\frac{4m^2}{L^2}})(\sum_{s}\frac{s^2}{\frac{n^2}{R^2}+\frac{4s^2}{L^2}})](\sum_{l}\frac{l^2}{\frac{k^2}{R^2}+\frac{4l^2}{L^2}})
\\ \nonumber&&+\frac{1}{2}[\sum_k (\sum_{s}\frac{s^2}{\frac{k^2}{R^2}+\frac{4s^2}{L^2}})(\sum_{l}\frac{l^2}{\frac{k^2}{R^2}+\frac{4l^2}{L^2}})](\sum_{n,m}\frac{n^2}{\frac{n^2}{R^2}+\frac{4m^2}{L^2}})
\\ \nonumber&&+\frac{1}{4}\sum_{r,k} (r+k)^2 (\sum_{s}\frac{s^2}{\frac{r^2}{R^2}+\frac{4s^2}{L^2}})(\sum_{l}\frac{l^2}{\frac{k^2}{R^2}+\frac{4l^2}{L^2}})(\sum_{m}\frac{1}{\frac{(k+r)^2}{R^2}+\frac{4m^2}{L^2}})
\\ \nonumber&&+\frac{1}{4}\sum_{r,k} (r-k)^2 (\sum_{s}\frac{s^2}{\frac{r^2}{R^2}+\frac{4s^2}{L^2}})(\sum_{l}\frac{l^2}{\frac{k^2}{R^2}+\frac{4l^2}{L^2}})(\sum_{m}\frac{1}{\frac{(k-r)^2}{R^2}+\frac{4m^2}{L^2}})\}
\\ &=&\frac{\pi^3L}{8R^5}\{(\zeta(-1)E_2(q^{ann.}))^3-\frac{1}{2}\zeta(-1)E_2(q^{ann.})(\zeta(-2)+4H_{2,2}(q^{ann.})+\frac{3}{4}F(q^{ann.})\},
\end{eqnarray}
\begin{eqnarray}
\nonumber&&\int
d^2\sigma\partial_0\partial_0'G\partial_0\partial_0'G\partial_0\partial_0'G
\\ \nonumber&=&\frac{512}{R^3L^9}\sum_{n,k,r=1}^{\infty}\sum_{m,l,s=-\infty}^{\infty}\frac{m^2l^2s^2}{(\frac{n^2}{R^2}+\frac{4m^2}{L^2})(\frac{k^2}{R^2}+\frac{4l^2}{L^2})(\frac{r^2}{R^2}+\frac{4s^2}{L^2})}\int d^2\sigma \sin^2(\frac{n\pi \sigma}{R})\sin^2(\frac{k\pi \sigma}{R})\sin^2(\frac{r\pi \sigma}{R})
\\ \nonumber&=&\frac{64}{R^2L^8}\{(\sum_{m,n}\frac{m^2}{\frac{m^2}{R^2}+\frac{4m^2}{L^2}})(\sum_{k,l}\frac{l^2}{\frac{k^2}{R^2}+\frac{4l^2}{L^2}})(\sum_{r,s}\frac{s^2}{\frac{r^2}{R^2}+\frac{4s^2}{L^2}})
\\ \nonumber&&+\frac{3}{2}[\sum_n (\sum_{m}\frac{m^2}{\frac{n^2}{R^2}+\frac{4m^2}{L^2}})(\sum_{s}\frac{s^2}{\frac{n^2}{R^2}+\frac{4s^2}{L^2}})](\sum_{k,l}\frac{l^2}{\frac{k^2}{R^2}+\frac{4l^2}{L^2}})
\\ \nonumber&&-\frac{1}{4}\sum_{n,k} (\sum_{m}\frac{m^2}{\frac{n^2}{R^2}+\frac{4m^2}{L^2}})(\sum_{s}\frac{s^2}{\frac{(n+k)^2}{R^2}+\frac{4s^2}{L^2}})(\sum_{l}\frac{l^2}{\frac{k^2}{R^2}+\frac{4l^2}{L^2}})\}
\\ &=&\frac{\pi^3L}{8R^5}\{-(\zeta(-1)E_2(q^{ann.}))^3-\frac{3}{2}\zeta(-1)E_2(q^{ann.})(\zeta(-2)+4H_{2,2}(q^{ann.}))+\frac{3}{4}F(q^{ann.})\}\quad.
\end{eqnarray}

\subsection{The torus}

The Green's function on the torus with periodicities $L,R$ is given by
\begin{eqnarray}
G(\sigma_0,\sigma_1;\sigma_0',\sigma_1')=\frac{1}{4\pi^2RL}\sum_{(n,m)\neq(0,0)}\frac{e^{\frac{2\pi i
m}{L}(\sigma_0-\sigma_0')}e^{\frac{2\pi i
n}{R}(\sigma_1-\sigma_1')}}{\frac{n^2}{R^2}+\frac{m^2}{L^2}}\quad.
\end{eqnarray}

\subsubsection{Partition function at $O(L^{-3})$}

This was computed already in \cite{bib:DF}, and we reproduce it here for completeness.
The diagrams at this order are,
\begin{eqnarray}
\nonumber I_1^{tor.}&=&\int d^2\sigma\partial_{\alpha}\partial^{\alpha'}G\partial_{\beta}\partial^{\beta'}G=\int d^2\sigma\{(\partial_0\partial_0'G+\partial_1\partial_1'G)(\partial_0\partial_0'G+\partial_1\partial_1'G)\}
=\frac{1}{RL},
\\ \nonumber I_2^{tor.}&=&\int d^2\sigma\partial_{\alpha}\partial_{\beta'}G\partial^{\alpha}\partial^{\beta'}G=\int d^2\sigma\{\partial_0\partial_0'G\partial_0\partial_0'G+\partial_1\partial_1'G\partial_1\partial_1'G\}
\\ &=&\frac{\pi^2L}{18R^3}E_2^2(q^{tor.})-\frac{\pi}{3R^2}E_2(q^{tor.})+\frac{1}{RL},
\end{eqnarray}
and it is easy to check that they are invariant under $R\leftrightarrow L$  as they must be.

Below are the details of the computation :
\begin{eqnarray}
\nonumber \int d^2\sigma\partial_1\partial_1'G\partial_1\partial_1'G
&=&\frac{1}{R^5L}(2\sum_{n=1}^{\infty}\sum_{m=-\infty}^{\infty}\frac{n^2}{\frac{m^2}{L^2}+\frac{n^2}{R^2}}+\sum_{m\neq0}0)^2
\\ &=&\frac{1}{R^5L}(2\pi R L\zeta(-1)E_2(q^{tor.}))^2
=\frac{4\pi^2L}{R^3}(\zeta(-1)E_2(q^{tor.}))^2,
\end{eqnarray}
\begin{eqnarray}
\nonumber\int d^2\sigma\partial_0\partial_0'G\partial_1\partial_1'G
&=&\frac{1}{R^5L}(2\sum_{n=1}^{\infty}\sum_{m=-\infty}^{\infty}\frac{n^2}{\frac{m^2}{L^2}+\frac{n^2}{R^2}}+\sum_{m\neq0}0)(2\sum_{n=1}^{\infty}\sum_{m=-\infty}^{\infty}\frac{m^2}{\frac{m^2}{L^2}+\frac{n^2}{R^2}}+\sum_{m\neq0}L^2)
\\ \nonumber&=&\frac{1}{R^3L^3}(2\pi R L\zeta(-1)E_2(q^{tor.}))\left(-2\frac{\pi L^3}{R}\zeta(-1)E_2(q^{tor.})+2L^2\zeta(0)\right)
\\ &=&-\frac{4\pi^2L}{R^3}(\zeta(-1)E_2(q^{tor.}))^2+\frac{4\pi}{R^2}\zeta(0)\zeta(-1)E_2(q^{tor.}),
\end{eqnarray}
\begin{eqnarray}
\nonumber\int d^2\sigma\partial_0\partial_0'G\partial_0\partial_0'G
&=&\frac{1}{RL^5}(2\sum_{n=1}^{\infty}\sum_{m=-\infty}^{\infty}\frac{m^2}{\frac{m^2}{L^2}+\frac{n^2}{R^2}}+\sum_{m\neq0}L^2)^2
\\ &=&\frac{1}{RL^5}\left(-2\frac{\pi L^3}{R}\zeta(-1)E_2(q^{tor.})+2L^2\zeta(0)\right)^2
\\ \nonumber &=&\frac{4\pi^2L}{R^3}(\zeta(-1)E_2(q^{tor.}))^2-\frac{8\pi}{R^2}\zeta(0)\zeta(-1)E_2(q^{tor.})+\frac{4}{RL}(\zeta(0))^2,
\end{eqnarray}

\subsubsection{Partition function at $O(L^{-5})$}

At this order there are both 2-loop and 3-loop contributions (see figures \ref{fig:Z2}
and \ref{fig:Z3}) .
At two loops there are two possible contractions, which both turn out to vanish:
\begin{eqnarray}
\nonumber I_3^{tor.}&=& \int d^2\sigma\partial_{\alpha}\partial^{\alpha'}\partial_{\beta}\partial^{\beta'}G\partial_{\gamma}\partial^{\gamma'}G
\\ &=&\int
d^2\sigma(\partial_0\partial_0'\partial_0\partial_0'G+\partial_1\partial_1'\partial_1\partial_1'G+2\partial_1\partial_1'\partial_0\partial_0'G)(\partial_1\partial_1'G+\partial_0\partial_0'G)
=0,
\\  I_4^{tor.}&=&\int d^2\sigma\partial_{\alpha}\partial_{\beta}\partial_{\gamma'}G\partial^{\alpha}\partial^{\beta}\partial^{\gamma'}G=0.
\end{eqnarray}

Below are the details of the computation :
\begin{eqnarray}
\nonumber\int
d^2\sigma\partial_1\partial_1'\partial_1\partial_1'G\partial_1\partial_1'G&=&-\nonumber\int
d^2\sigma\partial_1\partial_1'\partial_1\partial_1G\partial_1\partial_1'G
\\ \nonumber &=&\frac{4\pi^2}{R^7L}(2\sum_{n=1}^{\infty}\sum_{m=-\infty}^{\infty}\frac{n^4}{\frac{m^2}{L^2}+\frac{n^2}{R^2}}+\sum_{m\neq0}0)(2\sum_{n=1}^{\infty}\sum_{m=-\infty}^{\infty}\frac{n^2}{\frac{m^2}{L^2}+\frac{n^2}{R^2}}+\sum_{m\neq0}0)
\\ &=&\frac{16\pi^4L}{R^5}\zeta(-1)\zeta(-3)E_2(q^{tor.})E_4(q^{tor.}),
\end{eqnarray}
\begin{eqnarray}
\nonumber \int
d^2\sigma\partial_1\partial_1'\partial_1\partial_1'G\partial_0\partial_0'G
&=&\frac{4\pi^2}{R^5L^3}(2\sum_{n=1}^{\infty}\sum_{m=-\infty}^{\infty}\frac{n^4}{\frac{m^2}{L^2}+\frac{n^2}{R^2}}+\sum_{m\neq0}0)(2\sum_{n=1}^{\infty}\sum_{m=-\infty}^{\infty}\frac{m^2}{\frac{m^2}{L^2}+\frac{n^2}{R^2}}+\sum_{m\neq0}L^2)
\\ &=&\frac{16\pi^3}{R^4}\left(-\frac{\pi L}{R}\zeta(-1)\zeta(-3)E_2(q^{tor.})E_4(q^{tor.})+\zeta(0)\zeta(-3)E_4(q^{tor.})\right),
\end{eqnarray}
\begin{eqnarray}
\nonumber \int
d^2\sigma\partial_1\partial_1'\partial_0\partial_0'G\partial_1\partial_1'G&=&-\nonumber \int
d^2\sigma\partial_1\partial_1\partial_0\partial_0'G\partial_1\partial_1'G
\\ \nonumber &=&\frac{4\pi^2}{R^5L^3}(2\sum_{n=1}^{\infty}\sum_{m=-\infty}^{\infty}\frac{m^2n^2}{\frac{m^2}{L^2}+\frac{n^2}{R^2}}+\sum_{m\neq0}0)(2\sum_{n=1}^{\infty}\sum_{m=-\infty}^{\infty}\frac{n^2}{\frac{m^2}{L^2}+\frac{n^2}{R^2}}+\sum_{m\neq0}0)
\\ &=&-\frac{16\pi^4L}{R^5}\zeta(-1)\zeta(-3)E_2(q^{tor.})E_4(q^{tor.}),
\end{eqnarray}
\begin{eqnarray}
\nonumber\int
d^2\sigma\partial_1\partial_1'\partial_0\partial_0'G\partial_0\partial_0'G&=&-\nonumber\int
d^2\sigma\partial_1\partial_1'\partial_0\partial_0G\partial_0\partial_0'G
\\ \nonumber &=&\frac{4\pi^2}{R^3L^5}(2\sum_{n=1}^{\infty}\sum_{m=-\infty}^{\infty}\frac{m^2n^2}{\frac{m^2}{L^2}+\frac{n^2}{R^2}}+\sum_{m\neq0}0)(2\sum_{n=1}^{\infty}\sum_{m=-\infty}^{\infty}\frac{m^2}{\frac{m^2}{L^2}+\frac{n^2}{R^2}}+\sum_{m\neq0}L^2)
\\ &=&\frac{16\pi^3}{R^4}\left(\frac{\pi L}{R}\zeta(-1)\zeta(-3)E_2(q^{tor.})E_4(q^{tor.})-\zeta(0)\zeta(-3)E_4(q^{tor.})\right),
\end{eqnarray}
\begin{eqnarray}
\nonumber \int
d^2\sigma\partial_0\partial_0'\partial_0\partial_0'G\partial_1\partial_1'G
&=&\frac{4\pi^2}{RL^7}(2\sum_{n=1}^{\infty}\sum_{m=-\infty}^{\infty}\frac{m^4}{\frac{m^2}{L^2}+\frac{n^2}{R^2}}+\sum_{m\neq0}m^2L^2)(2\sum_{n=1}^{\infty}\sum_{m=-\infty}^{\infty}\frac{n^2}{\frac{m^2}{L^2}+\frac{n^2}{R^2}}+\sum_{m\neq0}0)
\\ &=&\frac{16\pi^4L}{R^5}\zeta(-1)\zeta(-3)E_2(q^{tor.})E_4(q^{tor.}),
\end{eqnarray}
\begin{eqnarray}
\nonumber\int
d^2\sigma\partial_0\partial_0'\partial_0\partial_0'G\partial_0\partial_0'G&=&-\nonumber\int
d^2\sigma\partial_0\partial_0\partial_0\partial_0'G\partial_0\partial_0'G
\\ \nonumber &=&\frac{4\pi^2}{RL^7}(2\sum_{n=1}^{\infty}\sum_{m=-\infty}^{\infty}\frac{m^4}{\frac{m^2}{L^2}+\frac{n^2}{R^2}}+\sum_{m\neq0}m^2L^2)(2\sum_{n=1}^{\infty}\sum_{m=-\infty}^{\infty}\frac{m^2}{\frac{m^2}{L^2}+\frac{n^2}{R^2}}+\sum_{m\neq0}L^2)
\\ &=&-\frac{16\pi^4L}{R^5}\zeta(-1)\zeta(-3)E_2(q^{tor.})E_4(q^{tor.})+\frac{16\pi^3}{R^4}\zeta(0)\zeta(-3)E_4(q^{tor.}).
\end{eqnarray}

At three loops there are 3 possible contractions :
\begin{eqnarray}
\nonumber I_6^{tor.}&=&\int d^2\sigma
\partial_{\alpha}\partial^{\alpha'}G\partial_{\beta}\partial^{\beta'}G\partial_{\gamma}\partial^{\gamma'}G
=\int
d^2\sigma\{\partial_0\partial_0'G\partial_0\partial_0'G\partial_0\partial_0'G+\partial_1\partial_1'G\partial_1\partial_1'G\partial_1\partial_1'G
\\ &&+3\partial_0\partial_0'G\partial_0\partial_0'G\partial_1\partial_1'G+3\partial_1\partial_1'G\partial_1\partial_1'G\partial_0\partial_0'G\}
=\frac{8}{L^2R^2}\zeta(0)^3,
\\ \nonumber I_7^{tor.}&=&\int d^2\sigma\partial^{\alpha}\partial'_{\alpha}G\partial^{\beta}\partial_{\gamma}'G\partial_{\beta}'\partial^{\gamma}G=\int d^2\sigma\{\partial_0\partial_0'G\partial_0\partial_0'G\partial_0\partial_0'G+\partial_1\partial_1'G\partial_1\partial_1'G\partial_1\partial_1'G
\\ \nonumber&&+\partial_0\partial_0'G\partial_0\partial_0'G\partial_1\partial_1'G+\partial_1\partial_1'G\partial_1\partial_1'G\partial_0\partial_0'G\}
\\ &=&-\frac{16\pi}{LR^3}\zeta(-1)E_2(q^{tor.})\zeta(0)+\frac{16\pi^2}{R^4}(\zeta(-1)E_2(q^{tor.}))^2\zeta(0)+\frac{8}{L^2R^2}\zeta(0)^3,
\\ \nonumber I_8^{tor.}&=&\int d^2\sigma\partial^{\alpha}\partial'_{\beta}G\partial^{\beta}\partial_{\gamma}'G\partial^{\gamma}\partial_{\alpha'}G
=\int d^2\sigma\{\partial_0\partial_0'G\partial_0\partial_0'G\partial_0\partial_0'G+\partial_1\partial_1'G\partial_1\partial_1'G\partial_1\partial_1'G\}
\\ &=&-\frac{24\pi}{L R^3}\zeta(-1)E_2(q^{tor.})\zeta(0)+\frac{24\pi^2}{R^4}(\zeta(-1)E_2(q^{tor.}))^2\zeta(0)+\frac{8}{L^2R^2}\zeta(0)^3\quad.
\end{eqnarray}

Below are the details of the computation:
\begin{eqnarray}
\nonumber &&\int
d^2\sigma\partial_0\partial_0'G\partial_0\partial_0'G\partial_0\partial_0'G
=\frac{1}{R^2L^8}\left(2\sum_{n=1}^{\infty}\sum_{m=-\infty}^{\infty}
\frac{m^2}{\frac{n^2}{R^2}+\frac{m^2}{L^2}}+\sum_{m\neq0}L^2\right)^3
\\ \nonumber&=&-\frac{8\pi^3L}{R^5}(\zeta(-1)E_2(q^{tor.}))^3+\frac{24\pi^2}{R^4}(\zeta(-1)E_2(q^{tor.}))^2\zeta(0)-\frac{24\pi}{LR^3}\zeta(-1)E_2(q^{tor.})\zeta(0)^2
\\ &&+\frac{8}{L^2R^2}\zeta(0)^3,
\end{eqnarray}
\begin{eqnarray}
\nonumber &&\int
d^2\sigma\partial_0\partial_0'G\partial_0\partial_0'G\partial_1\partial_1'G
\\ &=&\frac{1}{R^4L^6}\left(2\sum_{n=1}^{\infty}\sum_{m=-\infty}^{\infty}
\frac{m^2}{\frac{n^2}{R^2}+\frac{m^2}{L^2}}+\sum_{m\neq0}L^2\right)^2
\left(2\sum_{n=1}^{\infty}\sum_{m=-\infty}^{\infty}
\frac{n^2}{\frac{n^2}{R^2}+\frac{m^2}{L^2}}+\sum_{m\neq0}0\right)
\\ \nonumber &=&\frac{8\pi^3L}{R^5}(\zeta(-1)E_2(q^{tor.}))^3-\frac{16\pi^2}{R^4}(\zeta(-1)E_2(q^{tor.}))^2\zeta(0)+\frac{8\pi}{LR^3}\zeta(-1)E_2(q^{tor.})\zeta(0)^2,
\end{eqnarray}
\begin{eqnarray}
\nonumber &&\int
d^2\sigma\partial_0\partial_0'G\partial_1\partial_1'G\partial_1\partial_1'G
\\ \nonumber &=&\frac{1}{R^4L^6}\left(2\sum_{n=1}^{\infty}\sum_{m=-\infty}^{\infty}
\frac{m^2}{\frac{n^2}{R^2}+\frac{m^2}{L^2}}+\sum_{m\neq0}L^2\right)
\left(2\sum_{n=1}^{\infty}\sum_{m=-\infty}^{\infty}
\frac{n^2}{\frac{n^2}{R^2}+\frac{m^2}{L^2}}+\sum_{m\neq0}0\right)^2
\\ &=&-\frac{8\pi^3L}{R^5}(\zeta(-1)E_2(q^{tor.}))^3+\frac{8\pi^2}{R^4}(\zeta(-1)E_2(q^{tor.}))^2\zeta(0),
\end{eqnarray}
\begin{eqnarray}
\int
d^2\sigma \partial_1\partial_1'G\partial_1\partial_1'G\partial_1\partial_1'G
&=&\frac{1}{R^4L^6}\left(2\sum_{n=1}^{\infty}\sum_{m=-\infty}^{\infty}
\frac{n^2}{\frac{n^2}{R^2}+\frac{m^2}{L^2}}+\sum_{m\neq0}0\right)^3
\\ \nonumber &=&\frac{1}{R^8L^2}(2\pi R L \zeta(-1)E_2(q^{tor.}))^3=\frac{8\pi^3L}{R^5}(\zeta(-1)E_2(q^{tor.}))^3 \quad.
\end{eqnarray}

\subsection{$F(q)$ : numerical evaluation}
\label{numericalF}

Due to technical difficulties in the evaluation of $F(q)$ as a
finite sum of Eisenstein series, we study its modular properties
using a numerical method. We are able to extract the
coefficients in the series,
$F(\tilde{q})=\sum_{n} a_n(\frac{R}{L})^n+O(\tilde{q})$, which is a good
approximation for
$\frac{R}{L}\rightarrow\infty$.
We first extract the divergent part which we compute with a zeta function
regularization :
\begin{eqnarray}
F(q)&=&\sum_{n,r=1}^{\infty}(n^2r+nr^2)\frac{1+q^n}{1-q^n}\frac{1+q^r}{1-q^r}\frac{1+q^{n+r}}{1-q^{n+r}}
\\ \nonumber&=&\sum_{n,r=1}^{\infty}(n^2r+nr^2)\left(1+2\frac{q^n}{1-q^n}\right)\left(1+2\frac{q^r}{1-q^r}\right)
\left(1+2\frac{q^{n+r}}{1-q^{n+r}}\right)
\\ \nonumber&=&4\left[\sum_{n=1}^{\infty}n^2\left(\frac{q^n}{1-q^n}\right)\right]
\left[-\frac{1}{12}+2\sum_{r=1}^{\infty}r\left(\frac{q^r}{1-q^r}\right)\right]
\\ \nonumber&&+2\sum_{n,r=1}^{\infty}(n^2r+nr^2)\left(\frac{1+q^n}{1-q^n}\right)
\left(\frac{1+q^r}{1-q^r}\right)\left(\frac{q^{n+r}}{1-q^{n+r}}\right).
\end{eqnarray}
We then sum $F$ using this expression up to the $n,r=1000$
term, and perform a fit for small $\tilde q$ of the form
$F(\tilde{q})=-a_5\frac{\pi^4 }{\log(q)^5}-a_4\frac{\pi^2 }{\log(q)^4}-a_3\frac{1}{\log(q)^3}+O(\tilde{q})$.
Our result (expressed as rational numbers times $\pi$'s as expected) is
\begin{eqnarray}
\nonumber
F(\tilde{q})&=&-\frac{\pi^4}{3\log(q)^5}-\frac{4\pi^2}{3\log(q)^4}-\frac{4}{3\log(q)^3}
+O(\tilde q) \\&=&\frac{ R^5}{3\pi L^5}-\frac{4R^4}{3\pi^2 L^4}+\frac{4R^3}{3\pi^3L^3}+O(\tilde{q}).
\end{eqnarray}
%

\section{Conventions for sections 4-6}\label{sec:appendixB}

\subsection{General conventions}\label{ssec:appendixBgen}

The coordinates we use are $Z^{\mu}=\{X^{\alpha},X^i,Y^B,e^a\}$,
where the indices are arranged in the following way (unless written
otherwise) :
\begin{eqnarray}
\nonumber &&\mu,\nu,\rho,\cdots=0,\cdots,9  ,\quad \alpha, \beta, \gamma,\cdots=0,1 ,\quad  i,j,k,\cdots=2,\cdots,D-1
\\ && \xi=0,\cdots,D-1 ,\quad B=D,\cdots,D+N_B-1 ,\quad a=D+N_B,\cdots,9,
\end{eqnarray}
where $N_B$ is the number of massive scalars, $N_F$ is the number of massive fermions, $N_B^0$ is the number of massless transverse scalars
and $N_F^0$ is the number of massless fermions.

We use the following notation to sum over the scalar fields,
\begin{eqnarray}
X\cdot X=\delta_{ij}X^{i}X^{j},\quad Y\cdot Y=\delta_{ab}Y^{a}Y^{b}\quad.
\end{eqnarray}

On the worldsheet we have the following metric and antisymmetric tensor :
\begin{eqnarray}
\eta_{\alpha \beta}&=&\eta^{\alpha
\beta}=\left(\begin{array}{cc}
-1 & 0 \\
0 & 1
\end{array}\right) ,\quad \epsilon_{\alpha \beta}=\left(\begin{array}{cc}
0 & 1 \\
-1 & 0
\end{array}\right) ,\quad \epsilon^{\alpha \beta}=\left(\begin{array}{cc}
0 & -1 \\
1 & 0
\end{array}\right).
\end{eqnarray}
We define
\begin{eqnarray}
k_1\times k_2&\equiv&\epsilon^{\alpha\beta}k_{1\alpha}k_{2\beta}.
\end{eqnarray}
We also use the lightcone coordinates $\tilde{\alpha}=(+,-)$,
defined by the relation
$\sigma^{\pm}=\sigma^0\pm \sigma^1$. In these coordinates,
\begin{eqnarray}
\nonumber \eta_{\tilde{\alpha} \tilde{\beta}}&=&\left(\begin{array}{cc}
0 & -\frac{1}{2} \\
-\frac{1}{2} & 0
\end{array}\right) ,\quad \eta^{\tilde{\alpha}
\tilde{\beta}}=\left(\begin{array}{cc}
0 & -2 \\
-2 & 0
\end{array}\right) ,\quad \epsilon^{\tilde{\alpha} \tilde{\beta}}=\left(\begin{array}{cc}
0 & -2 \\
2 & 0
\end{array}\right),
\\ \frac{1}{4}k^2&=&\frac{1}{4}(k^{\sigma}k^{\sigma}-k^{\tau}k^{\tau})=-k_+k_-
,\quad ik_{\pm}=\partial_{\pm}
,\quad  \partial_{\pm}=\frac{1}{2}(\partial_0\pm\partial_1).
\end{eqnarray}

\subsection{Spinor conventions}\label{ssec:appendixBfer}

Our spinor notation is almost identical to the one used in \cite{bib:action2}.
We choose the Majorana condition such that the fermions are real
variables. This is consistent with choosing the conjugation
operation to be $\bar{\theta}=\theta^{T}\Gamma^0$. We introduce the
10 space-time gamma matrices $\Gamma_{\mu}$ which satisfy
$\{\Gamma_{\mu},\Gamma_{\nu}\}=2g_{\mu\nu}$, and the chirality
operator $\Gamma^{11}=\Gamma^0\cdots\Gamma^9$. We use the notation
$\Gamma_{\mu_1\cdots\mu_n}=\frac{1}{n!}\Gamma_{[\mu_1}\cdots\Gamma_{\mu_n]}$
where the brackets indicate anti-symmetrization of the gamma
matrices; e.g.
$\Gamma_{01}=\frac{1}{2}(\Gamma_0\Gamma_1-\Gamma_1\Gamma_0)$. The
matrices are real and can be broken into blocks in the following way
(using the metric (\ref{eq:metric})),
\begin{eqnarray}
\Gamma_{\alpha}=\sqrt{2\pi\alpha'} \rho_{\alpha}\otimes I ,\quad \Gamma_{i(B)}=\sqrt{2\pi\alpha'}\rho\otimes\gamma_{i (B)}, \quad \Gamma_{a}=\rho\otimes\gamma_{a},
\end{eqnarray}
with the chirality operators,
\begin{eqnarray}
\Gamma^{11}=\rho\otimes\gamma^c ,\quad \rho=\rho_0\rho_1, \quad
\gamma^c=\gamma^2\cdots\gamma^9\quad.
\end{eqnarray}
The worldsheet gamma matrices obey the following anti-commutation relations,
\begin{eqnarray}
\{\rho_{\alpha},\rho_{\beta}\}&=&2\eta_{\alpha\beta}, \quad
\{\rho_{\alpha},\rho\}=0\quad.
\end{eqnarray}
We explicitly write the matrices we will use :
\begin{eqnarray}
\begin{array}{l}
\rho_0=\left(\begin{array}{cc} 0 & 1 \\ -1 & 0 \end{array}\right) ,\quad \rho^+=-2\rho_-=\left(\begin{array}{cc} 0 & 0 \\ 2 & 0
\end{array}\right)\quad,\\\\
\rho_1=\left(\begin{array}{cc} 0 & 1 \\ 1 & 0 \end{array}\right) ,\quad \rho^-=-2\rho_+=\left(\begin{array}{cc} 0 & -2 \\ 0 & 0 \end{array}\right),\quad
\rho=\rho_0 \rho_1=\left(\begin{array}{cc} 1 & 0 \\ 0 & -1 \end{array}\right)\quad,
\end{array}
\end{eqnarray}
and the following relations :
\begin{eqnarray}
\nonumber  \rho^+ &=&\rho^{0}+\rho^{1} ,\quad \rho^-=\rho^{0}-\rho^{1},
\\ \epsilon_{\alpha\beta}\rho^{\alpha}\partial^{\beta}&=&\rho^-\partial_--\rho^+\partial_+, \quad
\eta_{\alpha\beta}\rho^{\alpha}\partial^{\beta}=\rho^+\partial_++\rho^-\partial_-\quad.
\end{eqnarray}
The 8 dimensional gamma matrices $\gamma_i$ obey flat space anti-commutation relations :
\begin{eqnarray}
\{\gamma_{i},\gamma_{j}\}=2\delta_{i j}, \quad
\{\gamma_{i},\gamma^c\}=0 \quad (i=2,\cdots,9)\quad.
\end{eqnarray}

\end{document}